\documentclass[aps,prd,twocolumn,superscriptaddress,groupedaddress,floatfix,nofootinbib]{revtex4-1}

\usepackage{filecontents} 

\usepackage{graphicx}  
\graphicspath{{./plots/}}
\usepackage{bm}        
\usepackage{amsmath}
\usepackage{amssymb}   
\usepackage{enumitem}
\usepackage{usebib}
\newbibfield{author}
\newbibfield{journal}
\newbibfield{volume}
\newbibfield{pages}
\newbibfield{year}
\newbibfield{eprint}
\bibinput{pcm_wc_mspace}
\usepackage{pdfcomment}
\usepackage{etoolbox}
\newcommand{\ubauthortitle}[1]{\usebibentry{#1}{author}, ``\usebibentry{#1}{title}''}
\newcommand{\ubjref}[1]{\usebibentry{#1}{journal} \usebibentry{#1}{volume} (\usebibentry{#1}{year}) \usebibentry{#1}{pages}}
\newcommand{\ubeprint}[1]{[ArXiv:\usebibentry{#1}{eprint}]}
\newcommand{\ubref}[1]{\ubauthortitle{#1}, \ubjref{#1}, \ubeprint{#1}}
\let\oldcite\cite
\renewcommand{\cite}[1]{\oldcite{#1}%
\renewcommand{\do}[1]{ + \ubref{##1}\textCR}%
\pdftooltip{${}^{>}$}{\docsvlist{#1}}%
}

\providecommand{\texorpdfstring}[2]{#1}

\newcommand{\lr}[1]{ \left( #1 \right) }
\newcommand{\lrs}[1]{ \left[ #1 \right] }
\newcommand{\lrc}[1]{ \left\{ #1 \right\} }
\newcommand{\vev}[1]{ \langle \, #1 \, \rangle }

\newcommand{\Tr}{ {\rm Tr} \, }
\newcommand{\tr}{ {\rm Tr} \, }
\newcommand{\re}{ {\rm Re} \, }

\renewcommand{\Re}{ {\rm Re} \, }
\renewcommand{\Im}{ {\rm Im} \, }

\newcommand{\const}{ {\rm const}}
\renewcommand{\det}[1]{ {\rm det} \left( #1 \right) }
\newcommand{\expa}[1]{ \exp{\left( #1 \right)} }

\newcommand{\sign}[1]{ {\rm sign}\left( #1 \right)}

\newcommand{\comment}[1]{}

\newcommand{\D}{\Delta}
\newcommand{\T}{\mathcal{T}}
\newcommand{\N}{\mathcal{N}}

\newcommand{\partsumI}{\N_{\T} = \sum_{X_{n+1} \in \T} |A\lr{X_{n+1} | X_n}|}
\newcommand{\partsum}{\N_{\T}}

\usepackage{xcolor}

\newtoggle{activecolor}
\toggletrue{activecolor}

\newcommand{\activeseq}[1]{\iftoggle{activecolor}{\textcolor{violet}{#1}}{#1}}
\newcommand{\activeidx}[1]{\iftoggle{activecolor}{\textcolor{red}{#1}}{#1}}
\newcommand{\myremark}[1]{\iftoggle{activecolor}{\textcolor{olive}{#1}}{#1}}

\newcounter{desccount}[subsection]

\newcommand{\desclabel}[1]{\\{}\\\refstepcounter{desccount}\textbf{#1 \thesection.\thesubsection.\arabic{desccount}}}

\begin{document}
\sloppy

\title{Diagrammatic Monte-Carlo for weak-coupling expansion of non-Abelian lattice field theories: \texorpdfstring{large-$N$ $U\lr{N} \times U\lr{N}$}{large-N U(N)xU(N)} principal chiral model}
\thanks{This paper includes PDF tooltips for citations. Please hover the mouse over the $>$ symbol after citations to see a tooltip with bibliographic information.}

\author{P.~V.~Buividovich}
\email{Pavel.Buividovich@physik.uni-regensburg.de}
\affiliation{Institute of Theoretical Physics, University of Regensburg,
D-93053 Germany, Regensburg, Universit\"{a}tsstrasse 31}

\author{A.~Davody}
\email{Ali.Davody@physik.uni-regensburg.de}
\altaffiliation[On leave of absence from ]{IPM, Teheran, Iran}
\affiliation{Institute of Theoretical Physics, University of Regensburg,
D-93053 Germany, Regensburg, Universit\"{a}tsstrasse 31}

\date{November 1st, 2017}

\begin{abstract}
We develop numerical tools for Diagrammatic Monte-Carlo simulations of non-Abelian lattice field theories in the t'Hooft large-$N$ limit based on the weak-coupling expansion. First we note that the path integral measure of such theories contributes a bare mass term in the effective action which is proportional to the bare coupling constant. This mass term renders the perturbative expansion infrared-finite and allows to study it directly in the large-$N$ and infinite-volume limits using the Diagrammatic Monte-Carlo approach. On the exactly solvable example of a large-$N$ $O\lr{N}$ sigma model in $D=2$ dimensions we show that this infrared-finite weak-coupling expansion contains, in addition to powers of bare coupling, also powers of its logarithm, reminiscent of re-summed perturbation theory in thermal field theory and resurgent trans-series without exponential terms. We numerically demonstrate the convergence of these double series to the manifestly non-perturbative dynamical mass gap. We then develop a Diagrammatic Monte-Carlo algorithm for sampling planar diagrams in the large-$N$ matrix field theory, and apply it to study this infrared-finite weak-coupling expansion for large-$N$ $U\lr{N} \times U\lr{N}$ nonlinear sigma model (principal chiral model) in $D = 2$. We sample up to $12$ leading orders of the weak-coupling expansion, which is the practical limit set by the increasingly strong sign problem at high orders. Comparing Diagrammatic Monte-Carlo with conventional Monte-Carlo simulations extrapolated to infinite $N$, we find a good agreement for the energy density as well as for the critical temperature of the ``deconfinement'' transition. Finally, we comment on the applicability of our approach to planar QCD at zero and finite density.
\end{abstract}

\maketitle

\section{Introduction}
\label{sec:intro}

An infamous fermionic sign problem in Monte-Carlo simulations based on sampling field configurations in the path integral is currently one of the main obstacles for systematic first-principle studies of the phase diagram and equation of state of dense strongly interacting matter as described by Quantum Chromodynamics (QCD). This problem has motivated the search for alternative simulation algorithms for non-Abelian gauge theories, which would hopefully avoid the sign problem or make it milder.

One of the alternative first-principle simulation strategies is the so-called Diagrammatic Monte-Carlo approach \cite{Prokofev:0802.2923}, conventionally abbreviated as DiagMC, which stochastically samples the diagrams of the weak- or strong-coupling expansions, rather than field configurations. DiagMC algorithms, complemented with the ``worm'' algorithm techniques \cite{Prokofev:01:1}, turned out to be very helpful for eliminating sign problem in physical models with relatively simple weak- or strong-coupling expansions. The physical models and theories which were successfully studied using DiagMC range from strongly interacting fermionic systems \cite{Prokofev:cond-mat/0605350,Prokofev:cond-mat/0702555,Prokofev:1110.3747,Rossi:1703.10141,Rossi:1612.05184,Prokofev:1608.00133,Chandrasekharan:1709.06048} relevant in condensed matter physics, via scalar field theories \cite{Wolff:09:2,Wolff:11:1,Gattringer:12:1,Davody:1307.7699,Pollet:1701.02680}, to $O\lr{N}$ and $\mathbb{CP}^{N-1}$ nonlinear sigma models \cite{Wolff:09:1,Wolff:10:1,Sulejmanpasic:15:1,deForcrand:17:1} and Abelian gauge theories with fermionic and bosonic matter fields \cite{Wolff:10:3,Gattringer:15:1,Gattringer:13:1}, which are more similar to QCD. This selection of references is by no means exhaustive.

This success has motivated various attempts to apply DiagMC to simulations of non-Abelian lattice gauge theories at finite density \cite{Miura:11:1,deForcrand:11:1,deForcrand:14:1,Vairinhos:14:1,Unger:14:1,Gattringer:1609.00124,Unger:17:1} or the effective models thereof \cite{Philipsen:13:1,Gattringer:11:1,Gattringer:11:2}. Despite significant progress, these attempts have not so far resulted in a first-principle, systematically improvable simulation strategy. At the qualitative level, one of the most successful strategies is based on the strong-coupling expansion of lattice QCD, which captures such non-perturbative features of QCD as confinement, chiral symmetry breaking and the baryon and meson bound states already at the leading order \cite{Rossi:84:1,Wolff:85:1}. DiagMC simulations of lattice QCD with staggered fermions at infinitely strong coupling are free of the sign problem in the massless limit, and allow one to reproduce the expected qualitative features of the QCD phase diagram \cite{deForcrand:11:1,deForcrand:14:1}. Unfortunately, an extension of this approach to the finite values of gauge coupling meets conceptual difficulties, as the next orders of the strong-coupling expansion coming from the bosonic part of the action should be incorporated in the DiagMC algorithm manually \cite{deForcrand:14:1}, which becomes more and more complicated for higher orders \cite{Unger:14:1}. Ideally, one would also like to sample the strong-coupling expansion in powers of inverse gauge coupling using DiagMC algorithm. Since on finite-volume lattices the strong-coupling expansion is convergent for all values of coupling, simulating sufficiently high orders should allow to access even the weak-coupling scaling region where lattice theory approaches continuum QCD. Recent attempts to extend the DiagMC simulations to finite gauge coupling using the method of auxiliary Hubbard-Stratonovich variables \cite{Vairinhos:14:1} or the Abelian color cycles \cite{Gattringer:1609.00124} are very promising, but still not quite practical. Development of practical tools for generating high orders of strong-coupling expansion is thus currently an important unsolved problem.

However, in the infinite-volume limit strong coupling expansion is known to have only a finite radius of convergence (see e.g.~\cite{Philipsen:0805.1163}). One can thus expect that as one approaches the weak-coupling regime, the convergence of the strong-coupling series will increasingly stronger depend on the lattice size, which might well make the continuum and infinite-volume extrapolation difficult in practice, especially for the infrared-sensitive physical observables such as hadron masses and transport coefficients. Moreover, in the t'Hooft large-$N$ limit one even expects a phase transition separating the strong-coupling and the weak-coupling phases \cite{Gross:80:1,Samuel:81:1,Rossi:94:2} even on the finite-volume lattice. These properties of the strong-coupling expansion might significantly limit the ability of DiagMC to work directly in the thermodynamic limit, which is one of the most important advantages of DiagMC over conventional Monte-Carlo methods \cite{Rossi:1612.05184,Rossi:1703.10141}.

These considerations suggest that the extrapolation to the continuum and thermodynamic limit might be easier for DiagMC simulations based on the weak-coupling expansion. However, the weak-coupling expansion for asymptotically free non-Abelian field theories, such as four-dimensional non-Abelian gauge theories and two-dimensional nonlinear sigma models, has two closely related problems, which make a direct application of a DiagMC approach conceptually difficult at first sight. The first problem are infrared divergences due to massless gluon propagators, which cancel only in the final expressions for physical observables. The second problem is the non-sign-alternating factorial growth of the weak-coupling expansion coefficients, which is commonly referred to as the infrared renormalon \cite{Beneke:99:1} and cannot be dealt with using the Borel resummation techniques. These factorial divergences in the perturbative expansions of physical observables originate in the renormalization-group running of the coupling constant, and have nothing to do with the purely combinatorial factorial growth of the number of Feynman diagrams with diagram order\footnote{The recent work \cite{Chandrasekharan:1709.06048} suggests that renormalon divergences might be absent in two-dimensional asymptotically free fermionic theories.}. In particular, infrared renormalon behaviour persists also in the large-$N$ limit \cite{David:81:1,David:82:1,Wiegmann:94:1,Wiegmann:94:2,Braun:98:1}, in which the number of Feynman diagrams grows only exponentially with order \cite{Brezin:78:1}. It is known that if one stops the running of the coupling at some artificially introduced infrared cutoff scale (e.g. by considering the theory in a finite volume), this factorial growth disappears \cite{Pineda:14:1,Pineda:15:1}, thus in a sense it is a remainder of IR divergences in perturbative expansion.

Recently infrared renormalon divergences have been given a physical interpretation in terms of the action of unstable or complex-valued saddles of the path integral, along with a resummation prescription based on the mathematical notion of resurgent trans-series \cite{Unsal:12:1,Unsal:14:1,Unsal:14:2,Dunne:14:1,Unsal:15:1,Unsal:15:2}. Trans-series is a generalization of the ordinary power series to a more general form
\begin{eqnarray}
\label{transseries_example}
 \vev{\mathcal{O}}\lr{\lambda} = \sum\limits_i e^{-S_i/\lambda} \sum\limits_{l = 0}^{m_i} \log\lr{\lambda}^l \sum\limits_{p=0}^{+\infty} c_{i,l,p} \lambda^p ,
\end{eqnarray}
where $\vev{\mathcal{O}}\lr{\lambda}$ is the expectation value of some physical observable as a function of the coupling constant $\lambda$, $c_{i,l,p}$ are the coefficients of perturbative expansion of path integral around the $i$-th saddle point of the path integral with the action $S_i$ and $l$ labels flat directions of the action in the vicinity of this saddle. It turns out that non-Borel-summable factorial divergences cancel between perturbative expansions around different saddle points $i$ in (\ref{transseries_example}). Unfortunately, currently no method is available to generate the trans-series (\ref{transseries_example}) for strongly coupled field theories in a systematic way. The resurgent analysis of perturbative expansions has been done so far almost exclusively in the regime where gauge theories or sigma models are artificially driven to the weak-coupling limit by a special compactification of space-time to $D = 1 + 0$ dimensions with twisted boundary conditions \cite{Unsal:12:1,Unsal:14:1,Unsal:14:2,Dunne:14:1,Unsal:15:1,Unsal:15:2}.

The aim of this work is to develop DiagMC methods suitable for the bosonic sector of non-Abelian lattice field theories, which is currently a challenge for DiagMC approach. We develop two practically independent tools, which are finally combined together to perform practical simulations.

The first tool is the resummation prescription which renders the bare weak-coupling expansion in non-Abelian field theories infrared-finite and thus suitable for DiagMC simulations which sample Feynman diagrams. This prescription is introduced in Section~\ref{sec:lattpt} and tested in Section~\ref{sec:solvable_examples} on some exactly solvable examples. The basic idea is to work with massive bare propagators, where the bare mass term comes from the nontrivial path integral measure for non-Abelian groups and is proportional to the bare coupling. By analogy with Fujikawa's approach to axial anomaly \cite{Fujikawa:79:1}, this bare mass serves as a seed for dynamical mass generation and conformal anomaly in non-Abelian field theories which are scale invariant at the classical level. A similar partial resummation of the perturbative series with the help of the bare mass term is also often used in finite-temperature quantum field theory \cite{Espinosa:hep-ph/9212248,Arnold:hep-ph/9408276}, in particular, in the contexts of hard thermal loop approximation \cite{Pisarski:92:1} and screened perturbation theory \cite{Karsch:hep-ph/9702376}. Upon this resummation, the weak-coupling expansion becomes formally similar to the trans-series (\ref{transseries_example}), but contains only the powers of coupling and of the logs of coupling only, while the exponential factors $e^{-S_i/\lambda}$ seem to be absent. This expansion appears to be manifestly infrared-finite and free of factorial divergences at least in the large-$N$ limit. An exactly solvable example of the $O\lr{N}$ nonlinear sigma model demonstrates that our expansion reproduces non-perturbative quantities such as the dynamically generated mass gap (see Subsection~\ref{sec:solvable_examples:subsec:2D}).

The second tool, described in Section~\ref{sec:diagmc}, is the general method for devising DiagMC algorithms from the full untruncated hierarchy of Schwinger-Dyson equations \cite{Buividovich:10:2,Buividovich:11:1}, bypassing any explicit construction of a diagrammatic representation. Here the basic idea is to generate a series expansion by stochastic iterations of Schwinger-Dyson equations. We believe the method to be advantageous for sampling expansions for which the complicated form of diagrammatic rules makes the straightforward application of e.g. worm algorithm hardly possible, with strong-coupling expansion in non-Abelian gauge theories being one particular example. In this work we apply the method to sample the (re-summed) weak-coupling expansion of Section~\ref{sec:lattpt}, which for non-Abelian lattice field theories also has complicated form due to the multitude of higher-order interaction vertices.

As a practical application which takes advantage of both tools, in this work we consider the large-$N$ $U\lr{N} \times U\lr{N}$ principal chiral model on the lattice, which is defined by the following partition function:
\begin{eqnarray}
\label{pcm_partition0}
 \mathcal{Z} = \int\limits_{U\lr{N}} dg_x \expa{-\frac{N}{\lambda} \sum\limits_{<x,y>} 2\re\tr\lr{g_x^{\dag} g_y}} ,
\end{eqnarray}
where integration in the path integral is performed over unitary $N \times N$ matrices $g_x$ living on the sites of the square two-dimensional lattice, $\lambda$ is the t'Hooft coupling constant which is kept fixed when taking the large-$N$ limit, and $\sum\limits_{<x,y>}$ denotes summation over all pairs of neighboring lattice sites $x$ and $y$. Numerical results for this model obtained by combining the methods of Sections~\ref{sec:lattpt}~and~\ref{sec:diagmc} are presented in Section~\ref{sec:pcm_numres}.

While we expect that our approach can be extended to large-$N$ pure gauge theory without conceptual difficulties, in this proof-of-concept study we prefer to work with the principal chiral model (\ref{pcm_partition0}). The reason is that while this model exhibits non-perturbative features very similar to those of non-Abelian gauge theories, including dynamical mass gap generation, dimensional transmutation and infrared renormalons \cite{Wiegmann:94:1,Wiegmann:94:2}, it has a much simpler structure of the perturbative expansion than non-Abelian gauge theories. Correspondingly, the DiagMC algorithm is also simpler to implement. We discuss the extension of our approach to large-$N$ gauge theories and illustrate the expected structure of infrared-finite weak-coupling expansion for this case in Section~\ref{sec:gaugeext}.

\section{Infrared-finite weak-coupling expansion for the \texorpdfstring{$U\lr{N} \times U\lr{N}$}{U(N) x U(N)} principal chiral model}
\label{sec:lattpt}

The first step in the construction of lattice perturbation theory is to parameterize the small fluctuations of lattice fields around some perturbative vacuum state. For non-Abelian lattice fields $g_x \in SU\lr{N}$ the most popular parametrization is the exponential mapping from the space of traceless Hermitian matrices $\phi_x$ to $SU\lr{N}$ group: $g_x = \expa{i \alpha \phi_x}$, where $\alpha$ is related to the coupling constant and $g_x = 1$ is the perturbative vacuum. While all the results in this paper could be also obtained for exponential mapping, for our purposes it will be more convenient to use the Cayley map
\begin{eqnarray}
\label{cayley_map}
 g_x
 =
 \frac{1 + i \alpha \phi_x}{1 - i \alpha \phi_x}
 =
 1 + 2 \sum\limits_{k=1}^{+\infty} \lr{i \alpha}^k \phi_x^k
,
\end{eqnarray}
where $\phi_x$ are again Hermitian matrices and the value of $\alpha$ will be specified later. Since in this work we will be working in the large-$N$ limit in which the $U\lr{N}$ and $SU\lr{N}$ groups are indistinguishable, we omit here the zero trace condition for $\phi_x$ and work with $U\lr{N}$ group. This will significantly simplify the derivation of Schwinger-Dyson equations in terms of $\phi_x$ fields, see Subsection~\ref{sec:diagmc:subsec:sdeqs}.

One of the advantages of the Cayley map is the particularly simple form of the $U\lr{N}$ Haar measure (see Appendix \ref{apdx:integration_measures:subsec:cayley} for the derivation):
\begin{eqnarray}
\label{cayley_jacobian}
 \int\limits_{U\lr{N}} dg_x
 =
 \int d\phi_x \det{1 + \alpha^2 \phi_x^2}^{-N}
 = \nonumber \\ =
 \int d\phi_x \expa{-N \tr\ln\lr{1 + \alpha^2 \phi_x^2}}
 = \nonumber \\ =
 \int d\phi_x \expa{-N\lr{\alpha^2 \tr{\phi_x^2} + \frac{\alpha^4}{2} \tr{\phi_x^4} + \ldots} } ,
\end{eqnarray}
where the integral on the right-hand side is over all Hermitian matrices $\phi_x$ and $\ldots$ represent the terms of order $O\lr{\alpha^6 \phi_x^6}$ or higher. Thus the Cayley map provides a unique one-to-one mapping between the $U\lr{N}$ group manifold and the space of all Hermitian matrices. Strictly speaking, one should exclude a single point $g_x = -1$ from this mapping, which, however, has zero measure in the path integral. In contrast, for exponential mapping the integration over $\phi_x$ should be restricted to a certain region within the space of Hermitian matrices in order to avoid multiple covering of $U\lr{N}$ group. In addition, for exponential mapping the exponentiated Jacobian contains double-trace terms \cite{Buividovich:15:3}, which make the planar perturbation theory technically somewhat more complicated (see Appendix \ref{apdx:integration_measures:subsec:expmap}). The Cayley map was also used in the context of lattice QCD in the Landau gauge \cite{Smekal:07:1,Smekal:08:1,Smekal:10:1,Smekal:12:1} and for studying unitary matrix models \cite{Mizoguchi:05:1}.

The classical action of the principal chiral model (\ref{pcm_partition0}) can be rewritten in terms of the fields $\phi_x$ as
\begin{eqnarray}
\label{pcm_action0}
 S_0 = \lambda^{-1} \sum\limits_{x,y} \D_{x y} \tr\lr{g^{\dag}_x g_y}
 = \nonumber\\ =
 4 \lambda^{-1} \sum\limits_{k,l=1}^{+\infty} \lr{-i \alpha}^k \lr{i \alpha}^l
      \sum\limits_{x,y} \D_{xy} \tr\lr{\phi_x^k \, \phi_y^l}
 = \nonumber\\ =
 4 \lambda^{-1} \sum\limits_{\substack{k,l=1\\k+l=2 n}}^{+\infty}
  \lr{-1}^{\frac{k-l}{2}} \alpha^{k+l}
       \sum\limits_{x,y} \D_{xy} \tr\lr{\phi_x^k \, \phi_y^l} ,
\end{eqnarray}
where we have introduced the lattice Laplacian operator in $D$-dimensional Euclidean space
\begin{eqnarray}
\label{lattice_laplacian_def}
 \D_{x,y} = 2 D \delta_{x,y} - \sum\limits_{\mu=0}^{D-1} \delta_{x,y-\hat{\mu}}  - \sum\limits_{\mu=0}^{D-1} \delta_{x,y+\hat{\mu}} ,
\end{eqnarray}
and used the unitarity of $g_x$ to add the diagonal terms $\tr\lr{g^{\dag}_x g_x} = N$ to the action. Here $\mu = 0, 1$ labels the two lattice coordinates, and $\hat{\mu}$ denotes the unit lattice vector in the direction $\mu$. We see now that in order to get the canonical normalization of the kinetic term in the action $S_0 = \frac{1}{2} \sum\limits_{x,y} \phi_x \D_{x,y} \phi_y + \ldots$, where $\ldots$ denotes the terms with more than two fields $\phi$, we have to set $\alpha^2 = \frac{\lambda}{8}$. We also note that the classical action (\ref{pcm_action0}) is invariant under the shifts of the field variable $\phi_x \rightarrow \phi_x + c I$, where $I$ is the $N \times N$ identity matrix. This symmetry is a consequence of the scale invariance of the classical action of the principal chiral model.

Rewriting the path integral in the partition function (\ref{pcm_partition0}) in terms of the new fields $\phi_x$, we also need to include the Jacobian (\ref{cayley_jacobian}). Adding the expansion of the Jacobian (\ref{cayley_jacobian}) in powers of $\lambda$ and $\phi$ to the classical action (\ref{pcm_action0}), we obtain the full action for the fields $\phi_x$:
\begin{eqnarray}
\label{pcm_action}
 S\lrs{\phi} =
 \frac{1}{2} \sum\limits_{x,y} \lr{\D_{xy} + \frac{\lambda}{4} \delta_{x y}} \tr\lr{\phi_x \phi_y}
 + \nonumber \\ +
 \sum\limits_{k=2}^{+\infty} \lr{-1}^{k-1} \frac{\lambda^{k}}{k \, 8^k} \sum\limits_x \tr \phi_x^{2 k}
 + \nonumber \\ +
 \sum\limits_{\substack{k,l=1\\k+l=2 n,k+l>2}}^{+\infty}
  \lr{-1}^{\frac{k-l}{2}} \frac{4 \, \lambda^{\frac{k+l-2}{2}}}{8^{\frac{k+l}{2}}}
       \sum\limits_{x,y} \D_{xy} \tr\lr{\phi_x^k \, \phi_y^l} ,
\end{eqnarray}
so that the partition function (\ref{pcm_partition0}) reads
\begin{eqnarray}
\label{pcm_partition_phi}
 \mathcal{Z} = \int d\phi_x \expa{-N S\lrs{\phi}} .
\end{eqnarray}

The observation which is central for this work is that the Jacobian (\ref{cayley_jacobian}) contributes a term
\begin{eqnarray}
\label{mass_term1}
 \frac{\lambda}{8} \sum\limits_x \phi_x^2 = \frac{m_0^2}{2} \sum\limits_x \phi_x^2
\end{eqnarray}
to the quadratic part
\begin{eqnarray}
\label{pcm_action_free}
 S_F\lrs{\phi} =
 \frac{1}{2} \sum\limits_{x,y} \lr{\D_{xy} + \frac{\lambda}{4} \delta_{x y}} \tr\lr{\phi_x \phi_y}
 \equiv \nonumber \\ \equiv
 \frac{1}{2} \sum\limits_{x,y} \lr{\D_{xy} + m_0^2 \delta_{x y}} \tr\lr{\phi_x \phi_y}
 \equiv \nonumber \\ \equiv
 \frac{1}{2} \sum\limits_{x,y} \lr{\lr{G^0}^{-1}}_{xy} \tr\lr{\phi_x \phi_y}
\end{eqnarray}
of the action (\ref{pcm_action}), which endows the field $\phi_x$ with the bare mass $m_0^2 = \frac{\lambda}{4}$. Thus the effect of the Jacobian (\ref{cayley_jacobian}) is to break the shift symmetry $\phi_x \rightarrow \phi_x + c \, I$ of the classical action, serving as a seed for dynamical scale generation and conformal anomaly in the principal chiral model (\ref{pcm_partition0}).

A conventional approach in lattice perturbation theory, which constructs weak-coupling expansion as formal series in powers of coupling $\lambda$, is to treat the quadratic term (\ref{mass_term1}) in the action (\ref{pcm_action}) as a perturbation \cite{Rossi:84:2} on top of the kinetic term. Perturbative expansion then involves massless propagators $\lr{\D^{-1}}_{xy}$ coming from the scale-invariant classical part of the action, which leads to infrared divergences in the contributions of individual diagrams. These infrared divergences, however, cancel in the final results for physical observables \cite{Rossi:84:2}. Furthermore, the resulting power series in the coupling constant $\lambda$ contain factorially divergent terms, some of which have non-alternating signs and thus cannot be removed by Borel resummation techniques. The only recently proposed way to deal with these divergences is to use the mathematical apparatus of resurgent trans-series \cite{Unsal:12:1,Unsal:14:1,Unsal:14:2,Dunne:14:1,Unsal:15:1,Unsal:15:2}.

In this work we consider a different approach to the weak-coupling expansion based on the action (\ref{pcm_action}), somewhat similar in spirit to the screened perturbation theory \cite{Espinosa:hep-ph/9212248,Arnold:hep-ph/9408276,Karsch:hep-ph/9702376} and to hard thermal loop perturbation theory \cite{Pisarski:92:1,Rebhan:94:1}. Namely, we include the mass $m_0^2 = \frac{\lambda}{4}$ coming from (\ref{mass_term1}) into the bare lattice propagators $G^0_{xy} = \lr{\Delta + m_0^2}^{-1}_{xy}$ which are used to construct the weights of Feynman diagrams. This corresponds to a trivial resummation of an infinite number of chain-like diagrams with ``two-leg'' vertices. Since this prescription puts the coupling $\lambda$ both in the propagators and vertices, we need to define the formal power counting scheme for our expansion. To this end we group together the terms with the products of $\lr{2 v + 2}$ fields, $v = 1, 2, \ldots +\infty$, which correspond to interaction vertices with $\lr{2 v + 2}$ legs, and multiply them with the powers $\xi^v$ of an auxiliary parameter $\xi$ which should be set to $\xi = 1$ to obtain the physical result:
\begin{eqnarray}
\label{pcm_action_int_general}
 S_I\lrs{\phi_x}
 = \sum\limits_{v=1}^{+\infty} \lr{- \frac{\lambda}{8}}^v \, \xi^v \, S_I^{\lr{2 v + 2}}\lrs{\phi_x} ,
\\
\label{pcm_action_int_vertices}
 S_I^{\lr{2 v + 2}}\lrs{\phi_x} =
 \frac{m_0^2}{2 \lr{v + 1}} \sum\limits_x \tr \phi_x^{2 v + 2}
 + \nonumber \\ +
 \frac{1}{2} \sum\limits_{l=1}^{2 v + 1} \lr{-1}^{l-1}
 \sum\limits_{x,y} \D_{x y} \tr\lr{\phi_x^{2 v + 2 - l} \phi_y^l } ,
\end{eqnarray}
where in the last equation we have used our definition $m_0^2 = \frac{\lambda}{4}$ in order to formally eliminate the coupling $\lambda$ from interaction vertices $S_I^{\lr{2 v + 2}}\lrs{\phi_x}$. This construction is similar to the ``shifted action'' of \cite{Rossi:1508.03654}.

Perturbative expansion of any physical observable $\vev{\mathcal{O}\lrs{\phi_x}}$ is then organized as a formal power series expansion in the auxiliary parameter $\xi$ in (\ref{pcm_action_int_general}):
\begin{eqnarray}
\label{expansion_general}
 \vev{\mathcal{O}\lrs{\phi_x}} =
 \lim\limits_{\xi \rightarrow 1}
 \lim\limits_{M \rightarrow \infty}
 \sum\limits_{m = 0}^{M}
 \xi^m \, \lr{-\frac{\lambda}{8}}^m \, \mathcal{O}_m\lr{m_0} ,
\end{eqnarray}
where $M$ is the maximal order of the expansion. This means that we ignore the relation between the mass $m_0^2 = \frac{\lambda}{4}$ and the coupling constant $\lambda$ and treat the terms $S_I^{\lr{2 v + 2}}\lrs{\phi_x}$ as bare vertices with $2 v + 2$ legs, counting only those powers $\lambda^v$ of the t'Hooft coupling constant which come from the vertex pre-factors in (\ref{pcm_action_int_general}). The bare mass $m_0^2$ in the expansion coefficients in (\ref{expansion_general}) is substituted with $\lambda/4$ at the very end of the calculation.

Let us discuss the general structure of an expansion (\ref{expansion_general}) for the free energy of the two-dimensional principal chiral model (\ref{pcm_partition0}) in the large-$N$ limit, in which only connected planar Feynman diagrams with no external legs contribute. Consider such a planar diagram with $f-1$ independent internal loop momenta $q_1, \ldots, q_{f-1}$, and with $l$ bare propagators and $v$ vertices. The kinematic weight of such a diagram can be written as
\begin{eqnarray}
\label{kinematic_weight_estimate}
  W_K \sim \int d^2 q_1 \ldots d^2 q_{f-1}
  \nonumber \\
  \frac{V_1 \ldots V_v}{\lr{\Delta\lr{Q_1} + m_0^2} \ldots \lr{\Delta\lr{Q_l} + m_0^2}} ,
\end{eqnarray}
where $Q_1 \ldots Q_l$ are momenta flowing through each diagram line,
\begin{eqnarray}
\label{lattice_laplacian_def_momentum}
 \Delta\lr{Q} = \sum\limits_{\mu=0}^{D-1} 4 \sin^2\lr{\frac{Q_{\mu}}{2}}
\end{eqnarray}
is the lattice Laplacian (\ref{lattice_laplacian_def}) in momentum space which behaves as $\Delta\lr{Q} \sim Q^2$ at small momenta and $V_1 \ldots V_v$ are the weights of the vertices which in general depend on $q_1 \ldots q_{f-1}$. The momenta $Q_1 \ldots Q_l$ are in general not independent and can be expressed as some linear combinations of $q_1, \ldots, q_{f-1}$. All momenta belong to the Brillouin zone $Q_{\mu} \in \lrs{-\pi, \pi}$ of the square lattice. Thus any possible ultraviolet divergences in the integral (\ref{kinematic_weight_estimate}) are regulated by the lattice ultraviolet cutoff $\Lambda_{UV} \sim 1$, and we only have to care about infrared divergences. Each vertex in (\ref{kinematic_weight_estimate}) contributes either a power of $m_0^2$ or a square of some combination of momenta $q_1 \ldots q_{f-1}$, coming from the first or the second terms in (\ref{pcm_action_int_vertices}), respectively. If our planar Feynman diagram is considered as a planar graph drawn on the sphere, the number $f$ is the number of faces of this planar graph. We now take into account the identity $f - l + v = 2$ for the Euler characteristic of a planar graph. Applying now the standard dimensional analysis, we see that the above relation between $f$, $l$ and $v$ implies that $W_K$ can only contain the terms proportional to $\Lambda_{UV}^2$, $m_0^2$, $m_0^4/\Lambda_{UV}^2$ and so on, probably times logarithmic terms $\sim \log\lr{m_0^2}$. In Subsection \ref{sec:solvable_examples:subsec:2D} we will explicitly
illustrate the importance of these logarithmic terms. No negative powers of $m_0$ can ever appear.

Remembering now the relation $m_0^2 = \frac{\lambda}{4}$, and taking into account the powers of coupling associated with vertices, we immediately conclude that the weights of Feynman diagrams in our perturbation theory with the bare mass $m_0 \sim \lambda$ are proportional to positive powers of $\lambda$, times possible logarithmic terms $\log\lr{\lambda}^m$, $\log\lr{\log\lr{\lambda}}^m$ ($m>0$) and so on. From (\ref{pcm_action_int_vertices}) we see that the combinatorial pre-factor of vertex with $2 v + 2$ legs grows at most linearly with $v$. Since the number of planar diagrams which contribute to the expansion (\ref{expansion_general}) at order $m$ grows at most exponentially with $m$, and the weight of any diagram is both UV and IR finite, contributions which grow factorially with $m$ cannot appear in the expansion (\ref{expansion_general}) in $D = 2$ dimensions.

The numerical results which we present below suggest that (\ref{expansion_general}) is a convergent weak-coupling expansion. At first sight this contradicts the common lore that perturbative expansion cannot capture the non-perturbative physics of the model (\ref{pcm_partition0}). Let us remember, however, that our expansion is no longer a strict power series in $\lambda$, which is known to be factorially divergent. Rather, the mass term $m_0^2 \sim \lambda$ in bare propagators allows for logarithms of coupling in the series, in close analogy with resummation of infrared divergences in hard thermal loop perturbation theory \cite{Rebhan:94:1}. It is easy to convince oneself that the formal expansion which contains powers of logs of coupling along with the powers of coupling can incorporate the non-perturbative scaling of the dynamical mass gap
\begin{eqnarray}
\label{nonpert_mass_gap_def}
 m^2 \sim \expa{-\frac{1}{\beta_0 \lambda}} ,
\end{eqnarray}
where $\beta_0$ is the first coefficient in the perturbative $\beta$-function. Indeed, we can rewrite the function (\ref{nonpert_mass_gap_def}) as a convergent expansion in powers of $z = \log\lr{\lambda}$:
\begin{eqnarray}
\label{explog_example}
 \expa{-\frac{1}{\beta_0 \lambda}}
 =
 \expa{-\beta_0^{-1} e^{-z}}
 = \nonumber \\ =
 \sum\limits_{k=0}^{+\infty} \frac{\lr{-\beta_0}^{-k}}{k!} \, e^{- k z}
 =
 \sum\limits_{l=0}^{+\infty} \frac{\lr{-z}^l}{l!}
 \sum\limits_{k=0}^{+\infty} \frac{\beta_0^{-k} \, k^l}{k!} .
\end{eqnarray}

In what follows we apply this resummation prescription to construct the infrared-finite weak-coupling expansion for several models: exactly solvable large-$N$ $O\lr{N}$ sigma model in $D = 1, 2, 3$ dimensions, exactly solvable large-$N$ principal chiral model on one-dimensional lattices with $L_0 = 2, \, 3, \, 4$ lattice sites, and, most importantly, the principal chiral model in $D = 2$ dimensions at zero and finite temperature.

\section{Exactly solvable examples and counterexamples}
\label{sec:solvable_examples}

\subsection{\texorpdfstring{Large-$N$ $O\lr{N}$}{Large-N O(N)} sigma model in \texorpdfstring{$D = 1, \, 2, \, 3$}{D = 1, 2, 3} dimensions}
\label{sec:solvable_examples:subsec:on_vardim}

\newlength{\mfwa}
\setlength{\mfwa}{0.33\textwidth}
\begin{figure}[h!tpb]
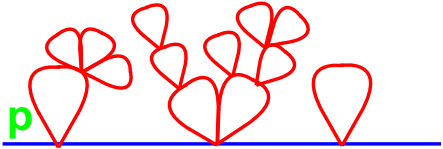

  \centering
  \includegraphics[width=\mfwa]{{{cactus_diagrams}}}\\
  \caption{Feynman diagrams which contribute to the two-point correlator in the large-$N$ limit of the $O\lr{N}$ sigma model.}
  \label{fig:cactus_diagrams}
\end{figure}

Large-$N$ $O\lr{N}$ sigma model provides one of the simplest examples of a non-perturbative quantum field theory which can be exactly solved by using saddle point approximation. In $D = 2$ dimensions this model has dynamically generated non-perturbative mass gap, similarly to the more complicated principal chiral models and non-Abelian gauge theories. On the other hand, for this model one can also explicitly construct the formal weak-coupling perturbative expansion, which is dominated by the ``cactus''-type diagrams, schematically shown on Fig.~\ref{fig:cactus_diagrams}. These properties make this model an ideal testing ground for the infrared-finite weak-coupling expansion described in Section~\ref{sec:lattpt}.

The partition function of the $O\lr{N}$ sigma model is given by the path integral over $N$-component unit vectors $n_{a \, x}$, $a = 0 \ldots N-1$ attached to the sites of $D$-dimensional square lattice, which we label by $x$:
\begin{eqnarray}
\label{on_partition1}
 \mathcal{Z} = \int\mathcal{D}n_x \expa{-\frac{N}{2 \lambda} \sum\limits_{x,y,a} \D_{xy} n_{a \, x} \, n_{a \, y} } ,
\end{eqnarray}
where $\D_{x,y}$ is again the lattice Laplacian (\ref{lattice_laplacian_def}). The well-known saddle-point solution shows that the two-point correlation function $G_{xy} = \vev{n_{a \, x} \, n_{a \, y}}$ is proportional to the propagator of a free massive scalar field, with the mass $m$ being the solution of the gap equation:
\begin{eqnarray}
\label{on_gap_equation}
 \lambda I_0\lr{m} = 1, \quad I_0\lr{m} \equiv \int \frac{d^D p}{\lr{2 \pi}^D} \frac{1}{\D\lr{p} + m^2} .
\end{eqnarray}
In $D = 1, \, 2, \, 3$ dimensions the function $I_0\lr{m}$ is
\begin{eqnarray}
\label{on_I0}
I_0\lr{m} =
\left\{
  \begin{array}{ll}
   \frac{1}{2 m} \, \lr{1 + \frac{m^2}{4}}^{-1/2}, & \hbox{D = 1;} \\
   - \frac{1}{4 \pi} \, \log\lr{\frac{m^2}{32}}\lr{1 + O\lr{m^2}} , & \hbox{D = 2;} \\
  A + B m + \cdots, & \hbox{D = 3 ,}
  \end{array}
 \right.
\end{eqnarray}
where in the last line $A=0.252731$ and $B=-0.0795775$. If the mass gap $m$ is small enough, the corresponding solutions for $m$ read
\begin{eqnarray}
\label{on_mass_exact}
m^2\lr{\lambda} =
\left\{
  \begin{array}{ll}
   \sqrt{\lambda^2 + 4} - 2, & \hbox{D = 1;} \\
   32 \expa{-\frac{4 \pi}{\lambda}}, & \hbox{D = 2;} \\
   m = |B^{-1}|\lr{A - 1/\lambda}, & \hbox{D = 3.}
  \end{array}
 \right.
\end{eqnarray}

The goal of this Section is to construct the infrared-finite weak-coupling expansion following the prescription of Section~\ref{sec:lattpt}. In order to parameterize the fluctuations of the $n_{a \, x}$ field around the classical vacuum with $n_{0 \, x} = 1$, we use the stereographic projection from the unit sphere in $N$-dimensions to $N-1$-dimensional real space of fields $\phi_{i \, x}$, $i = 1, \ldots, N-1$, which is the analogue of the Cayley map (\ref{cayley_map}):
\begin{eqnarray}
\label{stereographic_map}
 n_{0 \, x} = \frac{1 - \frac{\lambda}{4} \phi_x^2}{1 + \frac{\lambda}{4} \phi_x^2},
 \quad
 n_{i \, x} = \frac{\sqrt{\lambda} \, \phi_{i \, x}}{1 + \frac{\lambda}{4} \phi_x^2} ,
\end{eqnarray}
where $\phi_x^2 \equiv \sum\limits_{i} \phi_{i \, x} \phi_{i \, x}$. This is a unique map between the unit sphere embedded in $N$-dimensional space and the $N-1$-dimensional real space, in which only a single point with $n_{0 \, x} = -1$, $n_{i \, x} = 0$ is excluded. The integration measure on the unit sphere can be written in terms of the stereographic coordinates $\phi_{i \, x}$ as \cite{Buividovich:15:3} (see also Appendix~\ref{apdx:integration_measures:subsec:stereo}):
\begin{eqnarray}
\label{stereographic_jacobian}
 \mathcal{D}n_x = \mathcal{D}\phi_x \lr{1 + \frac{\lambda}{4} \phi_x^2}^{-N} .
\end{eqnarray}
Following the construction of Section~\ref{sec:lattpt}, we now include this Jacobian in the action for the fields $\phi_{i \, x}$, which reads:
\begin{eqnarray}
\label{on_action}
 S\lrs{\phi} = \frac{1}{2} \sum\limits_{x,y} \lr{\D_{x y} + \frac{\lambda}{2} \delta_{x y} } \lr{\phi_x \cdot \phi_y}
 + \nonumber \\ +
 \sum\limits_{\substack{k,l=0\\k+l\neq 0}}^{+\infty} \frac{\lr{-1}^{k+l} \lambda^{k+l}}{2 \cdot 4^{k+l}}
   \sum\limits_{x, y} \D_{x y} \lr{\phi_x^2}^k \lr{\phi_y^2}^l \lr{\phi_x \cdot \phi_y}
 + \nonumber \\ +
\sum\limits_{k=2}^{+\infty} \frac{\lr{-1}^{k-1} \lambda^k}{4^k \, k} \sum\limits_x \lr{\phi_x^2}^k
 , \hspace{1cm}
\end{eqnarray}
where $\phi_x \cdot \phi_y \equiv \sum\limits_i \phi_{i\,x} \cdot \phi_{i\,y}$. We have also taken into account that the terms of the form $\sum\limits_{x,y} \D_{xy} \lr{\phi_x^2}^k \lr{\phi_y^2}^l$ vanish at large $N$ due to factorization property $\vev{\lr{\phi_x \cdot \phi_y} \, \lr{\phi_z \cdot \phi_t}} = \vev{\phi_x \cdot \phi_y} \vev{\phi_z \cdot \phi_t}$ and translational invariance. In full analogy with the derivation of the action (\ref{pcm_action}) in Section~\ref{sec:lattpt}, the Jacobian of the transformation from compact to non-compact field variables results in the bare ``mass term'' $m_0^2 = \lambda/2$.

A convenient way to arrive at the perturbative expansion for the action (\ref{on_action}) is to iterate the corresponding Schwinger-Dyson equations
\begin{eqnarray}
\label{on_SD0}
 G^{-1}\lr{p} = \D\lr{p} + m_0^2
 -
 \D\lr{p} \frac{\eta \lr{2 + \eta}}{\lr{1 + \eta}^2}
 - \nonumber \\ -
 \frac{\lambda}{2} \frac{\eta}{1 + \eta}
 -
 \frac{\lambda}{2} \frac{1}{\lr{1 + \eta}^3} \int\frac{d^D q}{\lr{2 \pi}^D} \, \D\lr{q} \, G\lr{q} ,
 \nonumber \\
 \eta \equiv \frac{\lambda}{4} \, \vev{\phi_x^2} = \frac{\lambda}{4} \, \int\frac{d^D p}{\lr{2 \pi}^D} \, G\lr{p} ,
\end{eqnarray}
which in the large-$N$ limit involve only the two-point function
\begin{eqnarray}
\label{on_two_point_ansatz}
 \vev{\phi_x \cdot \phi_y} = \int\frac{d^D p}{\lr{2 \pi}^D} e^{i p_{\mu} \lr{x - y}_{\mu}} \, G\lr{p} .
\end{eqnarray}
The form of equation (\ref{on_SD0}) suggests that the solution has the form of the free scalar field propagator with the mass $m$, times some wave function renormalization factor $z$:
\begin{eqnarray}
\label{on_two_point_ansatz2}
 G\lr{p} = \frac{z}{\Delta\lr{p} + m^2} .
\end{eqnarray}
Substituting this ansatz in (\ref{on_SD0}), we obtain the following equations for $m$ and $z$:
\begin{eqnarray}
\label{on_SD}
 m^2 = m_0^2 z^2 - \frac{\lambda \xi}{2} z^2 + \frac{\lambda \xi}{2} z m^2 I_0\lr{m},
 \nonumber \\
 z = 1 + \frac{\lambda \xi}{4} z^2 I_0\lr{m} ,
\end{eqnarray}
where we have again introduced the auxiliary parameter $\xi$, to be set to $\xi = 1$, in order to define the formal power counting scheme. Substituting $\xi = 1$ and $m_0^2 = \lambda/2$, we immediately arrive at the exact solutions (\ref{on_mass_exact}) for the mass $m$. For $z$ the exact solution is simply $z = 2$.

\begin{figure}[h!tpb]
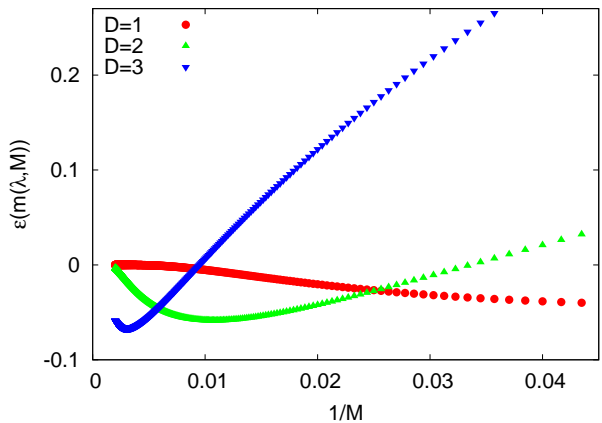

  \centering
  \includegraphics[width=\mfwa,angle=-90]{{{compare_mass_series_.4}}}\\
  \caption{Relative error (\ref{mass_relative_error_def}) of the mass gap for the $O\lr{N}$ sigma model from the truncated expansion (\ref{on_formal_series}) as a function of truncation order $M$ for $D = 1, \, 2, \, 3$. The coupling is fixed to yield the mass $m\lr{\lambda} = 0.4$ in (\ref{on_mass_exact}) for all $D$.}
  \label{fig:on_convergence_vardim}
\end{figure}

On the other hand, we can use the equations (\ref{on_SD}) for an iterative calculation of the coefficients of the formal weak-coupling expansion
\begin{eqnarray}
\label{on_formal_series}
 m^2 = m_0^2 + \lambda \, \xi \, \sigma_1\lr{m_0} + \lambda^2 \, \xi^2 \, \sigma_2\lr{m_0} + \ldots,
 \nonumber \\
 z = 1 + \lambda \, \xi \, z_1\lr{m_0} + \lambda^2 \, \xi^2 \, z_2\lr{m_0} + \ldots ,
\end{eqnarray}
which is also the formal power series in the auxiliary parameter $\xi$ in (\ref{on_SD}). In practice, we continue the iterations up to some finite order $M$, and calculate $m^2$ and $z$ by summing up all the terms in the series (\ref{on_formal_series}) with powers of $\xi$ less than or equal to $M$. After such a truncation of series, we substitute $\xi = 1$ and $m_0 \rightarrow \lambda/2$.

In $D = 1$ dimensions, our prescriptions yields the series with summands of the form $\lambda^k \lr{8 + \lambda}^{- M - 2}$ and $\lambda^{k+1/2} \lr{8 + \lambda}^{- M - 3/2}$, $k = 1, 2, \ldots$, times some integer-valued coefficients which depend on $M$. The expansion thus behaves regularly in the limit $\lambda \rightarrow 0$, where the mass goes to zero linearly in $\lambda$.

In $D = 2$ dimensions, we get the expansion which involves both powers of $\lambda$ and $\log\lr{\lambda}$ (see also \cite{Buividovich:15:3}):
\begin{eqnarray}
\label{on_series_2D}
 m^2\lr{\lambda, M} =
 \sum\limits_{l = 0}^{M} \lr{-\log\lr{\frac{\lambda}{64}}}^l
 \times \nonumber \\ \times
 \sum\limits_{k = \min\lr{l+2,M}}^{M + \min\lr{l,1}} \sigma_{l,k} \lambda^k ,
\end{eqnarray}
and similarly for $z\lr{\lambda, M}$. As discussed above, the expansion involving the logs of $\lambda$ can reproduce the non-perturbative scaling (\ref{on_mass_exact}) of the mass gap in $D = 2$. Finally, in $D = 3$ dimensions we obtain the expansion in positive powers of $\sqrt{\lambda}$.

On Fig.~\ref{fig:on_convergence_vardim} we compare the convergence of the expansions in $D = 1, \, 2, \, 3$ dimensions, plotting the relative error
\begin{eqnarray}
\label{mass_relative_error_def}
 \epsilon\lr{m\lr{\lambda, M}} = \frac{m\lr{\lambda, M} - m\lr{\lambda}}{m\lr{\lambda}}
\end{eqnarray}
of the truncated series (\ref{on_formal_series}) with respect to the exact result $m\lr{\lambda}$ given by (\ref{on_mass_exact}). The values of the coupling $\lambda$ are fixed in such a way that the exact non-perturbative mass gap is equal to $m = 0.4$ for all dimensions $D = 1,\, 2,\, 3$. The series (\ref{on_formal_series}) exhibit the fastest convergence at $D = 1$. At $D = 2$, the convergence is not monotonic, and also slower than in $D = 1$. In $D = 3$ the convergence is slowest, and from the convergence plot in Fig.~\ref{fig:on_convergence_vardim} it is difficult to say whether the series converge or not, even with up to $500$ terms in the expansion. These results suggest that for the large-$N$ $O\lr{N}$ sigma model $D = 2$ seems to be the critical dimension separating the convergent and non-convergent expansions.

\subsection{Non-perturbative mass gap in the 2D \texorpdfstring{$O\lr{N}$}{O(N)} sigma model}
\label{sec:solvable_examples:subsec:2D}

\begin{figure*}[h!tpb]
  \centering
  \includegraphics[width=\mfwa,angle=-90]{{{mass_series}}}\includegraphics[width=\mfwa,angle=-90]{{{z_series}}}\\
  \caption{Convergence of the truncated series (\ref{on_formal_series}) to the exact values (\ref{on_mass_exact}) for the effective mass term $m$ (on the left) and the renormalization factor $z$ (on the right), which are plotted as functions of $1/M$. The data points at $1/M = 0$ correspond to the exact results (\ref{on_mass_exact}).}
  \label{fig:on_model_series}
\end{figure*}

\begin{figure}[h!tpb]
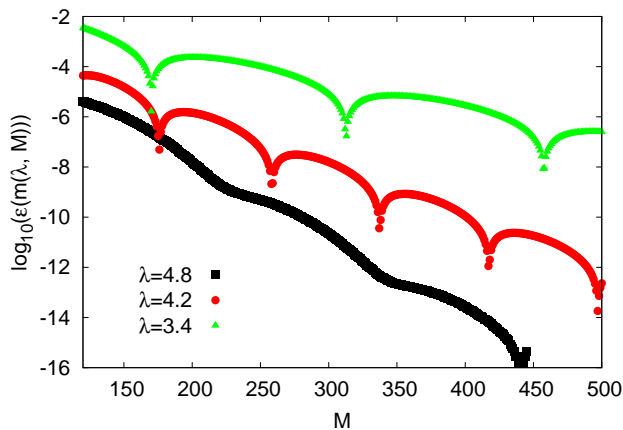

  \centering
  \includegraphics[width=\mfwa,angle=-90]{{{delta_mass_n}}}\\
  \caption{Relative error $\epsilon\lr{m\lr{\lambda, M}}$ of the effective mass $m$ obtained from the truncated series (\ref{on_formal_series}) at large values of $M$.}
  \label{fig:on_model_series_enlarged}
\end{figure}

We now turn to the physically most interesting example of the two-dimensional $O\lr{N}$ sigma model in the large-$N$ limit, where the mass gap (\ref{on_mass_exact}) exhibits non-perturbative scaling $m^2 \sim \expa{-\frac{1}{\beta_0 \lambda}}$ similar to the one found also in two-dimensional principal chiral model and in non-Abelian gauge theories. On Fig.~\ref{fig:on_model_series} we illustrate the convergence of the series (\ref{on_series_2D}) for the mass $m$ and the renormalization factor $z$ to the exact results (\ref{on_mass_exact}) and $z = 2$ for different values of the coupling constant $\lambda$. To make the convergence at large truncation orders $M$ more obvious, on Fig.~\ref{fig:on_model_series_enlarged} we also show the relative error (\ref{mass_relative_error_def}) of the truncated series (\ref{on_series_2D}) as a function of $M$ in logarithmic scale for several values of $\lambda$. While at large values of $\lambda$ the series converge with the precision of $10^{-16}$ at $M \sim 500$, the convergence becomes slower at smaller values of $\lambda$ - and, correspondingly, at smaller values of the mass $m$.

\begin{figure}[h!tpb]
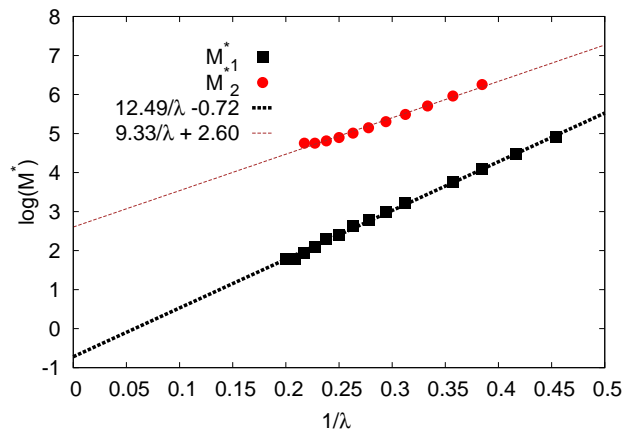

  \centering
  \includegraphics[width=\mfwa,angle=-90]{{{turning_order}}}\\
  \caption{The expansion truncation orders $M^{\star}_{1,2}\lr{\lambda}$ at which the truncated series (\ref{on_series_2D}) have local extrema, plotted as functions of the inverse coupling constant $\lambda^{-1}$. Solid lines are the fits of the form $\log\lr{M^{\star}\lr{\lambda}} = \const - \beta/\lambda$.}
  \label{fig:on_turning_point}
\end{figure}

An interesting feature of the dependence of $m\lr{\lambda, M}$ on $M$ is that it is not monotonous, but rather has several extrema with respect to $M$. We denote the smallest value of $M$ at which the dependence of $m\lr{\lambda; M}$ on $M$ has a minimum as $M^{\star}_1$, the position of the next maximum of this dependence - as $M^{\star}_2$, and so on. These positions depend on $\lambda$ and shift towards larger $M$ at smaller $\lambda$. We can think of the extrema positions $M^{\star}_1, M^{\star}_2, \ldots$ as of some characteristic scales which determine the convergence rate of the truncated series. In particular, at $M > M^{\star}_1$ the convergence to the exact value is rather fast. Remarkably, we have found that $M^{\star}_1$ as a function of $\lambda$ with good precision appears to be proportional to the square of correlation length (in lattice units):
\begin{eqnarray}
\label{mmax_scaling_2D_ON}
 M^{\star}_1 \sim l_c^2 \sim m^{-2} \sim \expa{\frac{4 \pi}{\lambda}} .
\end{eqnarray}
For illustration, on Fig.~\ref{fig:on_turning_point} we plot $\log\lr{M^{\star}_{1,2}\lr{\lambda}}$ versus $1/\lambda$, together with the linear fits of the form $\log\lr{M^{\star}\lr{\lambda}} = \const - \beta/\lambda$. For $M^{\star}_1$ the best fit yields the coefficient $\beta = 12.49$ which is very close to $4 \pi = 12.57$. For $M^{\star}_2$ we get the best fit value $\beta = 9.33$, which is close to $3 \pi = 9.42$ and suggests the scaling $M^{\star}_2 \sim l_c^{3/2} \sim m^{-3/2}$. This might imply that at sufficiently small $\lambda$ the two extrema might merge. However, for $M^{\star}_2$ we have fewer data points, and those exhibit some tendency for faster growth at larger values of $1/\lambda$, thus the scaling exponent for $M^{\star}_2$ might be different at asymptotically small $\lambda$.

The data presented on Figs.~\ref{fig:on_model_series_enlarged}~and~\ref{fig:on_turning_point} provide a clear numerical evidence that the infrared-finite weak-coupling expansion (\ref{on_series_2D}) indeed incorporates the non-perturbative dynamically generated mass gap $m^2 = 32 \expa{-\frac{4 \pi}{\lambda}}$. If we were to sample the series (\ref{on_series_2D}) using DiagMC algorithm, the linear scaling of $M^{\star}_1\lr{\lambda}$ with correlation length would also result in the expectable critical slowing down of simulations near the continuum limit.

The series which we obtain from the field-theoretical weak-coupling expansion (\ref{on_formal_series}) based on Schwinger-Dyson equations (\ref{on_SD}) are in this respect drastically different from the simplest formal expansion (\ref{explog_example}) of the exact answer (\ref{on_mass_exact}) in powers of logs. We have explicitly checked that for this case the positions of the extrema of $m$ with respect to $M$ are independent of $\lambda$. In this respect our infrared-finite weak-coupling expansion (\ref{on_formal_series}) seems to be closer to the notion of trans-series.

Let us stress, however, that the series (\ref{on_series_2D}) are not trans-series yet, despite formal similarity with (\ref{transseries_example}). In a mathematically strict sense, resurgent trans-series (\ref{transseries_example}) should give a ``minimal'' representation of a function with optimal convergence. For the large-$N$ $O\lr{N}$ sigma model, this optimal trans-series representation for the dynamically generated mass is given by a single term $m^2 = 32 \expa{-\frac{4 \pi}{\lambda}}$ (see the exact solution (\ref{on_mass_exact})). Thus the true trans-series for this model consist of only one term, which is enough for convergence! The expansion (\ref{on_series_2D}) also converges to the exact result (\ref{on_mass_exact}), but at a non-optimal rate \footnote{We thank Ovidiu~Costin for pointing out these properties of trans-series to us at the ``Resurgence 2016'' workshop in Lisbon.}.

\subsection{Exactly solvable counter-example: finite chiral chain models}
\label{sec:solvable_examples:subsec:finite_LS}

\begin{figure}[h!tpb]
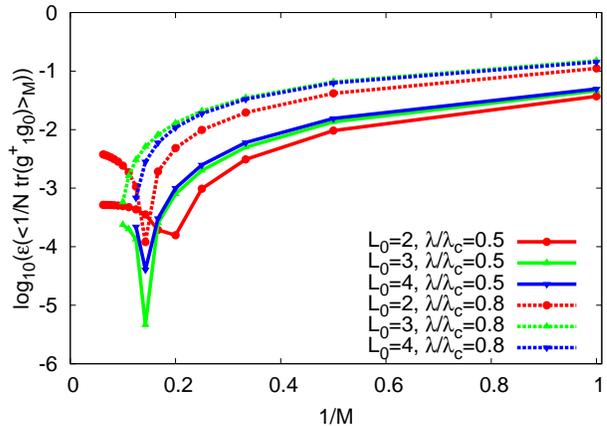

  \centering
  \includegraphics[width=\mfwa,angle=-90]{{{link_error_vs_order}}}\\
  \caption{Relative error of the weak-coupling expansion (\ref{expansion_general}) for the mean link $\vev{\frac{1}{N} \tr\lr{g^{\dag}_1 g_0}}_M$ of the principal chiral model (\ref{pcm_partition0}) on the one-dimensional lattices with $L_0 = 2, \, 3, \, 4$ sites, as a function of truncation order $M$ for fixed $\lambda/\lambda_c = 0.5$ and $\lambda/\lambda_c = 0.8$.}
  \label{fig:chiral_chains_convergence}
\end{figure}

The $U\lr{N} \times U\lr{N}$ principal chiral model (\ref{pcm_partition0}) can be solved exactly in $D = 1$ dimensions for lattice sizes $L_0 = 2, \, 3, \, 4$ and $L_0 = +\infty$ \cite{Rossi:98:1}. The first and the last cases can be reduced to the Gross-Witten unitary matrix model \cite{Gross:80:1}. For other two cases, solutions can be found in \cite{Rossi:98:1}. For all values of $L_0$ the one-dimensional principal chiral model exhibits a third-order Gross-Witten phase transition at some critical coupling $\lambda = \lambda_c\lr{L_0}$, which separates the strong-coupling and the weak-coupling regimes. In particular, for the mean link $\vev{\frac{1}{N} \tr\lr{g^{\dag}_1 g_0}}$ the exact results in the weak-coupling regime $\lambda < \lambda_c$ are:
\begin{eqnarray}
\label{exact_link_LS2}
 \vev{\frac{1}{N} \tr\lr{ g^{\dag}_1 g_0 }} = 1 - \frac{\lambda}{8}, \, (L_0 = 2, \lambda_c = 4)
\\
\label{exact_link_LS3}
 \vev{\frac{1}{N} \tr\lr{ g^{\dag}_1 g_0 }}
 =
 \frac{1}{\lambda} + \frac{1}{2} - \frac{\lambda}{8}
 - \nonumber \\ -
 \frac{1}{\lambda} \lr{1-\frac{\lambda}{3}}^{\frac{3}{2}}, \, (L_0 = 3, \lambda_c = 3) .
\end{eqnarray}
For $L_0 = 4$ the weak-coupling solution is only available in the implicit form:
\begin{eqnarray}
\label{exact_link_LS4}
 \vev{\frac{1}{N} \tr\lr{g^{\dag}_1 g_0 }}
 =
 \frac{2}{\lambda} - \frac{\lambda}{8} - \frac{2}{\lambda} \delta^2, \, (\lambda_c = \frac{8}{\pi}) ,
\end{eqnarray}
where $\delta$ is the solution of the equation $\frac{8}{\pi \lambda}\lr{E\lr{1-\delta^2} - \delta^2 F\lr{1-\delta^2}} = 1$ and $E$ and $F$ are the complete elliptic integrals of second and first type, respectively.

These analytic expressions offer the possibility to check the convergence of the infrared-finite weak-coupling expansion of Section~\ref{sec:lattpt} for small lattice sizes. Of course, for finite lattices the dimensional analysis following the equation (\ref{kinematic_weight_estimate}) would not work, and loop integrals will only produce some rational functions of $\lambda$ instead of logs. Nevertheless it is interesting to check whether these rational functions provide a good approximation to the exact results (\ref{exact_link_LS2}), (\ref{exact_link_LS3}) and (\ref{exact_link_LS4}). To this end we have performed exact calculations of the coefficients of the weak-coupling expansion (\ref{expansion_general}), generating them recursively up to some maximal order $M$ from the Schwinger-Dyson equations (\ref{sd_eqs_mom_vertex}), to be presented in Subsection~\ref{sec:diagmc:subsec:sdeqs}. Such an exact calculation is possible for $L_0 = 2$ up to $M = 16$ and for $L_0 = 4$ up to $M = 8$. Explicit expression for the mean link $\vev{\frac{1}{N} \tr\lr{g^{\dag}_1 g_0 }}$ in terms of the correlators of $\phi_x$ matrices and the prescription for truncating the expansion at some finite order $M$ are given in Subsection~\ref{sec:pcm_numres:subsec:observables}.

On Fig.~\ref{fig:chiral_chains_convergence} we illustrate the relative error of the truncated weak-coupling expansion (\ref{expansion_general}) for the mean link $\vev{\frac{1}{N} \tr\lr{g^{\dag}_1 g_0 }}_M$ as a function of inverse truncation order $1/M$ for $L_0 = 2, \, 3, \, 4$, fixing $\lambda/\lambda_c = 0.5$ and $\lambda/\lambda_c = 0.8$. The relative error is defined in full analogy with (\ref{mass_relative_error_def}). At the first sight the convergence seems to be rather fast, however, a detailed inspection of numerical results reveals a tiny discrepancy of order of $0.1\%$ which cannot be ascribed to numerical round-off errors, and which seems to stabilize at some finite value as $M$ is increased (at least for $L_0 = 2$). An analytic calculation of the few first orders revealed that the weak-coupling expansion (\ref{expansion_general}) correctly reproduces only the three leading coefficients of the standard perturbative expansion in powers of $\lambda$, which is convergent for finite $L_0$.

It seems thus that for finite-size one-dimensional lattices we observe the misleading convergence of the re-summed perturbative expansion to some unphysical branch of the solutions of Schwinger-Dyson equations. This phenomenon might happen even in zero-dimensional models \cite{Rossi:1508.03654} and was first noticed in Bold DiagMC simulations of interacting fermions \cite{Kozik:1407.5687}. Remembering the fact that for the large-$N$ $O\lr{N}$ sigma model in the infinite space the expansion (\ref{expansion_general}) converged with much higher precision than the discrepancy which we observe here for finite-size chiral chains (compare Figs.~\ref{fig:chiral_chains_convergence}~and~\ref{fig:on_model_series_enlarged}), we conclude that infinite lattice size and the presence of $\log\lr{\lambda}$ terms seem to be crucial for the convergence to the correct physical result. An important extension of our analysis, which we relegate for future work, would be to extend the sufficient conditions of \cite{Rossi:1508.03654} for the convergence to the physical result to the weak-coupling expansion (\ref{expansion_general}).

\section{Schwinger-Dyson equations and DiagMC algorithm for \texorpdfstring{$U\lr{N} \times U\lr{N}$}{U(N) x U(N)} principal chiral model}
\label{sec:diagmc}

The aim of this Section is to construct a DiagMC algorithm for sampling the weak-coupling expansion (\ref{expansion_general}). While in this work we will only consider the weak-coupling expansion (\ref{expansion_general}) based on the action (\ref{pcm_action}) of the large-$N$ principal chiral model (\ref{pcm_partition0}), our construction of the DiagMC algorithm is quite general and can be applied to both to the weak and strong-coupling expansions in general large-$N$ matrix field theories. We will briefly outline the extension to non-Abelian gauge theories in Section~\ref{sec:gaugeext}.

\subsection{Schwinger-Dyson equations for the \texorpdfstring{large-$N$ $U\lr{N} \times U\lr{N}$}{large-N U(N) x U(N)} principal chiral model with Cayley parameterization of \texorpdfstring{$U\lr{N}$}{U(N)} group}
\label{sec:diagmc:subsec:sdeqs}

As in \cite{Prokofev:cond-mat/0702555,Prokofev:1110.3747,Prokofev:1608.00133,Davody:1307.7699,Pollet:1701.02680}, the starting point for the construction of our DiagMC algorithm are the Schwinger-Dyson equations, which we derive for the action (\ref{pcm_action}) of the principal chiral model (\ref{pcm_partition0}). In contrast to the works \cite{Davody:1307.7699,Pollet:1701.02680} on the bosonic DiagMC, here we consider the full untruncated hierarchy of Schwinger-Dyson equations for multi-trace, in general disconnected, correlators of the $\phi_x$ fields defined in (\ref{cayley_map}):
\begin{eqnarray}
\label{multitrace_observables_def}
 \vev{\lrs{x^1_1, \ldots, x^1_{n_1}} \lrs{x^2_1, \ldots, x^2_{n_2}} \ldots \lrs{x^r_1, \ldots, x^r_{n_r}} }
 \equiv \nonumber \\ \equiv
 \left\langle
 \frac{1}{N}\, \tr\lr{\phi_{x^1_1} \ldots \phi_{x^1_{n_1}} }
 \frac{1}{N}\, \tr\lr{\phi_{x^2_1} \ldots \phi_{x^2_{n_2}} }
 \ldots \right. \nonumber \\ \left. \ldots
 \frac{1}{N}\, \tr\lr{\phi_{x^r_1} \ldots \phi_{x^r_{n_r}} }
 \right\rangle .
\end{eqnarray}

\begin{widetext}
To derive the Schwinger-Dyson equations for the field correlators (\ref{multitrace_observables_def}), let's consider the full derivative of the form
\begin{eqnarray}
\label{matrix_sd_general_derivation}
 \frac{\delta_{ik} \delta_{jl}}{N^r} \, \int \mathcal{D}\phi \frac{\partial}{\partial \phi_{x \, ij} }
 \left(
 \lr{\phi_{x^1_2} \ldots \phi_{x^1_{n_1}} }_{kl}
 \tr\lr{\phi_{x^2_1} \ldots \phi_{x^2_{n_2}} }
 \ldots
 \tr\lr{\phi_{x^r_1} \ldots \phi_{x^r_{n_r}} }
 \expa{- S_F\lrs{\phi} - S_I\lrs{\phi} }
 \right) = 0 ,
\end{eqnarray}
where we have explicitly separated the action (\ref{pcm_action}) into the quadratic and interacting parts given by (\ref{pcm_action_free}) and (\ref{pcm_action_int_general}), (\ref{pcm_action_int_vertices}) to simplify the derivation and notation. We now expand the derivative according to the Leibniz product rule, contract matrix indices and finally contract the result with the bare propagator $G^0_{x^1_1 \, x} = \lr{\Delta + m_0^2}^{-1}_{xy}$ coming from the quadratic part $S_F\lrs{\phi_x}$ of the action (\ref{pcm_action}). This yields the following infinite system of equations for the multi-trace observables (\ref{multitrace_observables_def}):
\begin{eqnarray}
\label{matrix_sd_general_n2}
\vev{
  \activeseq{\lrs{x_1, x_2}}
 }
 =
 G^0_{\activeidx{x_1 x_2}}
 +
 \sum\limits_x G^0_{\activeidx{x_1 x}}
 \vev{
  \activeseq{\lrs{ \activeidx{\frac{\partial S_I}{\partial \phi_x},} \,x_2}}
 } ,
\end{eqnarray}
\begin{eqnarray}
\label{matrix_sd_general_n2_mtr}
 \vev{
  \activeseq{\lrs{x_1^1, x_2^1}}
  \lrs{x_1^2, \ldots, x_{n_2}^2}
  \ldots
  \lrs{x_1^r, \ldots, x_{n_r}^r}
 }
 = 
 G^0_{\activeidx{x_1^1 x_2^1}}
 \vev{
  \lrs{x_1^2, \ldots, x_{n_2}^2}
  \ldots
  \lrs{x_1^r, \ldots, x_{n_r}^r}
 }
 + \nonumber \\ +
 \sum\limits_x G^0_{\activeidx{x_1^1 x}}
 \vev{
  \activeseq{\lrs{ \activeidx{\frac{\partial S_I}{\partial \phi_x},} \,x_2^1}}
  \lrs{x_1^2, \ldots, x_{n_2}^2}
  \ldots
  \lrs{x_1^r, \ldots, x_{n_r}^r}
 }
 + \nonumber \\ +
 \frac{1}{N^2} \,
 \sum\limits_{s = 2}^r
 \sum\limits_{p = 1}^{n_s}
 G^0_{\activeidx{x_1^1 x_p^s}}
 \vev{
  \lrs{x_1^2, \ldots, x_{n_2}^2}
  \ldots
  \activeseq{\lrs{x_1^s, \ldots, x^s_{p-1}, \, \activeidx{x^1_2,} \, x^s_{p+1}, \ldots, x^s_{n_s}} }
  \ldots
  \lrs{x_1^r, \ldots, x_{n_r}^r}
 }
\end{eqnarray}
\begin{eqnarray}
\label{matrix_sd_general}
 \vev{
  \activeseq{\lrs{x_1^1, \ldots, x_{n_1}^1}}
  \lrs{x_1^2, \ldots, x_{n_2}^2}
  \ldots
  \lrs{x_1^r, \ldots, x_{n_r}^r}
 }
 = \nonumber \\ =
 G^0_{\activeidx{x_1^1 x_2^1}}
 \vev{
  \activeseq{\lrs{x_3^1, \ldots, x_{n_1}^1}}
  \lrs{x_1^2, \ldots, x_{n_2}^2}
  \ldots
  \lrs{x_1^r, \ldots, x_{n_r}^r}
 }
 + \nonumber \\ +
 G^0_{\activeidx{x_1^1 x_{n_1}^1}}
 \vev{
  \activeseq{\lrs{x_2^1, \ldots, x_{n_1-1}^1}}
  \lrs{x_1^2, \ldots, x_{n_2}^2}
  \ldots
  \lrs{x_1^r, \ldots, x_{n_r}^r}
 }
 + \nonumber \\ +
 \sum\limits_{u = 3}^{n-2}
 G^0_{\activeidx{x_1^1 x_u^1}}
 \vev{
  \activeseq{\lrs{x_2^1, \ldots, x_{u-1}^1}
  \lrs{x_{u+1}^1, \ldots, x_{n_1}^1}}
  \lrs{x_1^2, \ldots, x_{n_2}^2}
  \ldots
  \lrs{x_1^r, \ldots, x_{n_r}^r}
 }
 + \nonumber \\ +
 \sum\limits_x G^0_{\activeidx{x_1^1 x}}
 \vev{
  \activeseq{\lrs{ \activeidx{
  \frac{\partial S_I}{\partial \phi_x},} \,x_2^1, \ldots, x_{n_1}^1}}
  \lrs{x_1^2, \ldots, x_{n_2}^2}
  \ldots
  \lrs{x_1^r, \ldots, x_{n_r}^r}
 }
 + \nonumber \\ +
 \frac{1}{N^2} \,
 \sum\limits_{s = 2}^r
 \sum\limits_{p = 1}^{n_s}
 G^0_{\activeidx{x_1^1 x_p^s}}
 \vev{
  \lrs{x_1^2, \ldots, x_{n_2}^2}
  \ldots
  \activeseq{\lrs{x_1^s, \ldots, x^s_{p-1}, \, \activeidx{x^1_2, \ldots, x^1_{n_1},} \, x^s_{p+1}, \ldots, x^s_{n_s} }}
  \ldots
  \lrs{x_1^r, \ldots, x_{n_r}^r}
 } \iftoggle{activecolor}{,}{.}
\end{eqnarray}
\iftoggle{activecolor}{where we have used \textcolor{red}{red} and \textcolor{violet}{violet} colors to highlight the field arguments $x$ and the single-trace terms which are different on the left- and the right-hand sides. By}{Here by} $\frac{\partial S_I}{\partial \phi_x}$ we have schematically denoted the terms which are produced by applying the derivative operator in (\ref{matrix_sd_general_derivation}) to the interacting part (\ref{pcm_action_int_general}) of the action (\ref{pcm_action}):
\begin{eqnarray}
\label{dSdphi_coord}
 \frac{\partial S_I\lrs{\phi}}{\partial \phi_x}
 =
 \sum\limits_{v = 1}^{+\infty} \xi^v \, \lr{-\frac{\lambda}{8}}^v \,
 \frac{\partial S_I^{\lr{2 v + 2}}\lrs{\phi}}{\partial \phi_x} ,
 \nonumber \\
 \frac{\partial S_I^{\lr{2 v + 2}}\lrs{\phi}}{\partial \phi_x}
 =
 m_0^2 \phi_x^{2 v + 1}
 + 
 \sum\limits_{l=1}^{2 v + 1} \lr{-1}^{l-1}
 \sum\limits_{m=0}^{2 v + 1 - l}
 \sum\limits_y \D_{x y}
 \phi_x^m \phi_y^l \phi_x^{2 v + 1 - l - m}  .
\end{eqnarray}
\end{widetext}
These terms contain three or more $\phi$ fields and should be inserted into the single-trace expectation values (\ref{multitrace_observables_def}) within the equations (\ref{matrix_sd_general_n2}), (\ref{matrix_sd_general_n2_mtr}) and (\ref{matrix_sd_general}): $\vev{\lrs{\frac{\partial S_I\lrs{\phi}}{\partial \phi_x} \, y}} = \vev{\frac{1}{N} \tr\lr{\frac{\partial S_I\lrs{\phi}}{\partial \phi_x} \phi_y } }$, and so on. For the sake of brevity, we omit the matrix indices of $\phi_x$, so that all products of $\phi_x$ are understood as matrix products. In particular, this implies that $\phi_x$ and $\phi_y$ do not commute for $x \neq y$.

In the t'Hooft large-$N$ limit, the last summands in the equations (\ref{matrix_sd_general_n2_mtr}) and (\ref{matrix_sd_general}) can be omitted, since they are suppressed by a factor $1/N^2$ and all multi-trace correlators (\ref{multitrace_observables_def}) are of order $N^0$. Since in this limit the correlators (\ref{multitrace_observables_def}) with odd numbers of $\phi_x$ fields within some of the traces, such as $\vev{\lrs{x} \lrs{y}} \equiv \vev{\frac{1}{N} \tr\lr{\phi_x} \frac{1}{N} \tr\lr{\phi_y}}$, are suppressed as $1/N^2$, we can also limit the equations to the space of multi-trace correlator which contain an even number of $\phi$ fields within each trace. Furthermore, in the t'Hooft large-$N$ limit the equations (\ref{matrix_sd_general_n2}), (\ref{matrix_sd_general_n2_mtr}) and (\ref{matrix_sd_general}) without the $1/N^2$-suppressed terms admit the large-$N$ factorized solution
\begin{eqnarray}
\label{factorization_largeN}
\vev{\lrs{x^1_1, \ldots, x^1_{n_1}} \ldots \lrs{x^r_1, \ldots, x^r_{n_r}} }
 = \nonumber \\ =
\vev{\lrs{x^1_1, \ldots, x^1_{n_1}}} \ldots \vev{\lrs{x^r_1, \ldots, x^r_{n_r}} }
\end{eqnarray}
with single-trace correlators $\vev{\lrs{x_1, \ldots, x_n}}$ obeying the much simpler non-linear equation
\begin{eqnarray}
\label{matrix_sd_factorized}
 \vev{\lrs{x_1, \ldots, x_n}}
 = \nonumber \\ =
 G^0_{x_1 x_2} \vev{\lrs{x_3, \ldots, x_n}}
 +
 G^0_{x_1 x_n} \vev{\lrs{x_2, \ldots, x_{n-1}}}
 + \nonumber \\ +
 \sum\limits_{u = 3}^{n-2}
 G^0_{x_1 x_u}
 \vev{\lrs{x_2, \ldots, x_{u-1}}}
 \vev{\lrs{x_{u+1}, \ldots, x_n}}
 + \nonumber \\ +
 \sum\limits_x G^0_{x_1 x}
 \vev{\lrs{ \frac{\partial S_I}{\partial \phi_x}, \, x_2, \ldots, x_n}}  .
\end{eqnarray}
The form of the Schwinger-Dyson equation (\ref{matrix_sd_general_n2}) for the two-point correlator does not change.

Schwinger-Dyson equations for large-$N$ quantum field theories are almost always considered in the nonlinear form (\ref{matrix_sd_factorized}), which is much more compact than the linear equations (\ref{matrix_sd_general}). One of the well-known examples of such non-linear Schwinger-Dyson equations in the large-$N$ limit are the Migdal-Makeenko loop equations for non-Abelian gauge theories \cite{Migdal:81:1,Eguchi:82:1}. Likewise, even at finite $N$ one can also arrive at a more compact \emph{non-linear} representations of the Schwinger-Dyson equations (\ref{matrix_sd_general_n2}), (\ref{matrix_sd_general_n2_mtr}) and (\ref{matrix_sd_general}) by rewriting them in terms of connected or one-particle-irreducible correlators. This strategy is numerically very efficient for approximate solutions based on some truncations of the full set of equations (especially for correlators with small numbers of fields), and is commonly used for numerical solutions of QCD Schwinger-Dyson equations \cite{Alkofer:00:1,Berges:hep-ph/0401172} and in condensed matter physics \cite{Prokofev:98:1,Prokofev:0802.2923,Pollet:1701.02680}. However, Schwinger-Dyson equations are always linear equations when written in terms of disconnected correlators of the form (\ref{multitrace_observables_def}). This linear form is very convenient for the construction of DiagMC algorithms, which we consider in the next Subsection.

To proceed towards the DiagMC algorithm, we go to the momentum space, and define the momentum-space field operators and correlators as
\begin{eqnarray}
\label{momentum_space_fields}
 \phi_p = \frac{1}{V} \sum\limits_x e^{i p x} \phi_x,
 \quad
 \phi_x = \sum\limits_p e^{-i p x} \phi_p .
\end{eqnarray}
Note that we have assumed the finite lattice volume $V$, which will be convenient for further analysis. In particular, this allows us to interpret all the delta-functions with momentum-valued arguments as Kronecker deltas on the discrete set of lattice momenta. For brevity, we will denote $\delta\lr{p} \equiv \delta_{p,0}$. We will see in what follows that the algorithm performance is practically volume-independent, thus one can make a smooth transition from discrete to continuous momenta if necessary.

It is also convenient to introduce the momentum-space vertex functions
\begin{eqnarray}
\label{vertex_func_def}
 V\lr{q_1, \ldots, q_{2 v + 1}}
 =
 m_0^2
 + \nonumber \\ +
 \sum\limits_{l=1}^{2 v + 1} \lr{-1}^{l-1}
 \sum\limits_{m=0}^{2 v + 1 - l}
 \D\lr{q_{m+1} + \ldots + q_{m+l}} ,
\end{eqnarray}
so that the Fourier transform of the derivative (\ref{dSdphi_coord}) of the interacting part of the action can be compactly represented as
\begin{eqnarray}
\label{dSdphi_mom_vertex}
 \frac{\partial S_I^{\lr{2 v + 2}}\lrs{\phi}}{\partial \phi_p}
 \equiv
 \frac{1}{V} \sum\limits_x e^{i p x}
 \frac{\partial S_I^{\lr{2 v + 2}}\lrs{\phi}}{\partial \phi_x}
 = \nonumber \\ =
 \sum\limits_{q_1 \ldots q_{2 v + 1}} \delta\lr{p - q_1 - \ldots - q_{2 v + 1}}
 \times \nonumber \\ \times
 V\lr{q_1, \ldots, q_{2 v + 1}} \phi_{q_1} \ldots \phi_{q_{2 v + 1}} .
\end{eqnarray}

An important property of the vertex functions (\ref{vertex_func_def}) is that they are positive for all values of $q_1, \ldots, q_{2 v + 1}$. While we don't have an explicit proof of this statement, in our Monte-Carlo simulations with over $10^{14}$ Metropolis updates in total we have explicitly checked that the values of vertex function were always positive. Furthermore, in Appendix~\ref{apdx:vertex_functions_ir} we demonstrate that if all the momenta in (\ref{vertex_func_def}) are small and the lattice Laplacian can be approximated as $\D\lr{p} = p^2$, the vertex functions take the simple form
\begin{eqnarray}
\label{vertex_func_IR}
 V\lr{q_1, \ldots, q_{2 v + 1}} = m_0^2 + \lr{q_1 + q_3 + \ldots + q_{2 v + 1}}^2 ,
\end{eqnarray}
which manifestly demonstrates their positivity.

Furthermore, in order to define the power counting scheme to be used in our DiagMC simulations, we expand the momentum-space correlators of $\phi$ fields in formal power series in our auxiliary parameter $\xi$:
\begin{eqnarray}
\label{corrs_expansion_def}
 \vev{
  \frac{1}{N} \tr\lr{\phi_{p^1_1} \ldots \phi_{p^1_{n_1}} }
  \ldots
  \frac{1}{N} \tr\lr{\phi_{p^r_1} \ldots \phi_{p^r_{n_r}} }
 }
 = \nonumber \\ =
 \sum\limits_{m = 0}^{+\infty}
 \xi^m \lr{-\frac{\lambda}{8}}^m
 \times \nonumber \\ \times
 \vev{
  \lrs{p^1_1, \ldots, p^1_{n_1}}
  \ldots
  \lrs{p^r_1, \ldots, p^r_{n_r}}
 }_m  .
\end{eqnarray}
In terms of the correlators (\ref{corrs_expansion_def}) of the momentum-space fields $\phi_p$, the large-$N$ limit of the Schwinger-Dyson equations (\ref{matrix_sd_general_n2}), (\ref{matrix_sd_general_n2_mtr}) and (\ref{matrix_sd_general}) read:
\begin{widetext}
\begin{subequations}
\label{sd_eqs_mom_vertex}
\begin{eqnarray}
\label{sd_eqs_mom_vertex_n2_contact}
 \vev{\activeseq{\lrs{p_1, p_2}}}_{\activeseq{m}}
 =
 \delta_{\activeseq{m,0}} \frac{\delta\lr{\activeidx{p_1 + p_2}}}{V} \, G_0\lr{\activeidx{p_1}}
 -
 \\ \label{sd_eqs_mom_vertex_n2_vertex}
 -
 G_0\lr{\activeidx{p_1}} \sum\limits_{v=1}^{m}
 \sum\limits_{q_1, \ldots, q_{2 v + 1}}
 \delta\lr{\activeidx{p_1 - q_1 - \ldots - q_{2 v + 1}}}
 V\lr{\activeidx{q_1, \ldots, q_{2 v + 1}}}
 \vev{\activeseq{\lrs{\activeidx{q_1, \ldots, q_{2 v + 1},} \, p_2}}}_{\activeidx{m - v}} ,
\end{eqnarray}
\begin{eqnarray}
\label{sd_eqs_mom_vertex_n2_mtr}
 \vev{
  \activeseq{\lrs{p^1_1, p^1_2}}
  \lrs{p^2_1, \ldots, p^2_{n_2}}
  \ldots
  \lrs{p^r_1, \ldots, p^r_{n_r}}
 }_{\activeseq{m}}
 = \nonumber
 \\ \label{sd_eqs_mom_vertex_n2_mtr_newseq}
 =
 \frac{\delta\lr{\activeidx{p^1_1 + p^1_2}}}{V} \, G_0\lr{\activeidx{p^1_1}}
 \vev{
  \lrs{p^2_1, \ldots, p^2_{n_2}}
  \ldots
  \lrs{p^r_1, \ldots, p^r_{n_r}}
 }_m - 
 \\ \label{sd_eqs_mom_vertex_n2_mtr_vertex}
 -
 G_0\lr{\activeidx{p^1_1}} \sum\limits_{v=1}^{m}
 \sum\limits_{q_1, \ldots, q_{2 v + 1}}
 \delta\lr{\activeidx{p_1^1 - q_1 - \ldots - q_{2 v + 1}}}
 V\lr{\activeidx{q_1, \ldots, q_{2 v + 1}}}
 \times \nonumber \\ \times
 \vev{
  \activeseq{\lrs{\activeidx{q_1, \ldots, q_{2 v + 1},} \, p^1_2}}
  \lrs{p^2_1, \ldots, p^2_{n_2}}
  \ldots
  \lrs{p^r_1, \ldots, p^r_{n_r}}
 }_{\activeidx{m - v}} ,
\end{eqnarray}
\begin{eqnarray}
 \vev{
  \activeseq{\lrs{p^1_1, p^1_2, \ldots, p^1_{n_1}}}
  \lrs{p^2_1,       \ldots, p^2_{n_2}}
  \ldots
  \lrs{p^r_1,       \ldots, p^r_{n_r}}
 }_{\activeseq{m}}
 = \nonumber
 \\ \label{sd_eqs_mom_vertex_newline1}
 =
 \frac{\delta\lr{\activeidx{p^1_1 + p^1_2}}}{V} \, G_0\lr{\activeidx{p^1_1}} \,
 \vev{
  \activeseq{\lrs{p^1_3, \ldots, p^1_{n_1}}}
  \lrs{p^2_1, \ldots, p^2_{n_2}}
  \ldots
  \lrs{p^r_1, \ldots, p^r_{n_r}}
 }_m
 +
 \\ \label{sd_eqs_mom_vertex_newline2}
 +
 \frac{\delta\lr{\activeidx{p^1_1 + p^1_{n_1}}}}{V} \, G_0\lr{\activeidx{p^1_1}} \,
 \vev{
  \activeseq{\lrs{p^1_2, p^1_3, \ldots, p^1_{n_1-1}}}
  \lrs{p^2_1,  \ldots, p^2_{n_2}}
  \ldots
  \lrs{p^r_1, \ldots, p^r_{n_r}}
 }_m
 +
 \\ \label{sd_eqs_mom_vertex_join}
 +
 \sum\limits_{A=4}^{n_1-2}
 \frac{\delta\lr{\activeidx{p^1_1 + p^1_A}}}{V} \, G_0\lr{\activeidx{p^1_1}} \,
 \vev{
  \activeseq{\lrs{p^1_2, \ldots, p^1_{A-1}}
  \lrs{p^1_{A+1}, \ldots, p^1_{n_1}}}
  \lrs{p^2_1, \ldots, p^2_{n_2}}
  \ldots
  \lrs{p^r_1, \ldots, p^r_{n_r}}
  }_m
 -
 \\ \label{sd_eqs_mom_vertex_vertex}
 -
 G_0\lr{\activeidx{p^1_1}} \sum\limits_{v=1}^{m}
 \sum\limits_{q_1, \ldots, q_{2 v + 1}}
 \delta\lr{\activeidx{p^1_1 - q_1 - \ldots - q_{2 v + 1}}}
 V\lr{\activeidx{q_1, \ldots, q_{2 v + 1}}}
 \times \nonumber \\ \times
 \vev{
  \activeseq{\lrs{\activeidx{q_1, \ldots, q_{2 v + 1},} \, p^1_2, p^1_3,  \ldots, p^1_{n_1}}}
  \lrs{p^2_1, \ldots, p^2_{n_2}}
  \ldots
  \lrs{p^r_1, \ldots, p^r_{n_r}}
  }_{\activeidx{m - v}} ,
\end{eqnarray}
\end{subequations}
where we have \iftoggle{activecolor}{again used \textcolor{red}{red} and \textcolor{violet}{violet} colors to highlight the individual momenta, order labels and momentum subsequences in correlators (\ref{corrs_expansion_def}) which are different on the right- and left-hand sides. We have also }{} introduced the bare propagator in momentum space
\begin{eqnarray}
\label{bare_propagator_def}
 G_0\lr{p} = \frac{1}{\Delta\lr{p} + m_0^2}, \quad m_0^2 \equiv \frac{\lambda}{4} .
\end{eqnarray}
\end{widetext}

\begin{figure}
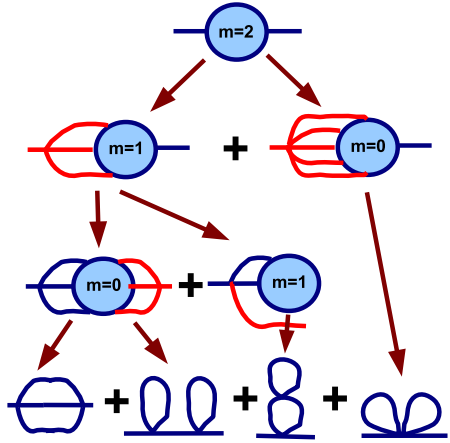

  \centering
  \includegraphics[width=\mfwa]{{{sd_structure1}}}
  \caption{An example of the recursive calculation of diagrams contributing to order $m = 2$ two-point correlator from Schwinger-Dyson equations (\ref{sd_eqs_mom_vertex}). Combinatorial diagram weights are not explicitly shown in order not to clutter the plot.}
  \label{fig:sd_structure1}
\end{figure}

It is easy to convince oneself that recursive solution of the Schwinger-Dyson equations (\ref{sd_eqs_mom_vertex}) generates the conventional weak-coupling expansion, with the auxiliary parameter $\xi$ formally serving as a coupling. For example, to find the order $m = 2$ contribution to the two-point correlator $\vev{\lrs{p_1, p_2}}$, we use the first equation in (\ref{sd_eqs_mom_vertex}) to express it in terms of four-point correlator of order $m = 1$ and the six-point correlator of order $m = 0$. The four-point correlator at order $m = 1$ can be related to two-point correlator at order $m = 1$ and the six-point correlator at $m = 0$ using the last equation in (\ref{sd_eqs_mom_vertex}). Finally, the two-point correlator at order $m = 1$ is again related to four-point correlator at $m = 0$ by virtue of the first equation in (\ref{sd_eqs_mom_vertex}). This chain of relations is schematically illustrated on Fig.~\ref{fig:sd_structure1}.

It is instructive to explicitly calculate the first perturbative correction to the two-point correlator, which can be readily found from the Schwinger-Dyson equations (\ref{sd_eqs_mom_vertex}) for the two-point correlator:
\begin{eqnarray}
\label{p0p1_1}
 \vev{\lrs{p_1 p_2}}_1 = G_0^2\lr{p_1} \frac{\delta\lr{p_1 + p_2}}{V} \Sigma_1\lr{p_1} ,
\end{eqnarray}
where
\begin{eqnarray}
\label{sigma_1}
 \Sigma_1\lr{p} = \nonumber \\ =
 -\frac{1}{V} \sum\limits_q \lr{V\lr{q, -q, p} + V\lr{p, q, -q}} G_0\lr{q}
 = \nonumber \\ =
 -\frac{2}{V} \sum\limits_q G_0\lr{q}
 \times \nonumber \\ \times
 \lr{\frac{\lambda}{4} + 2 \D\lr{p} + 2 \D\lr{q} - \D\lr{p-q}} .
\end{eqnarray}
In particular, $\Sigma_1\lr{p = 0} = -2$. If we interpret this quantity as the first-order correction to the self-energy, then at zero momentum this correction, being multiplied by $\lr{-\lambda/8}$ and $\xi = 1$, exactly cancels the bare mass term $\lambda/4$! Of course, this cancellation requires re-summation of the infinite chain of bubble diagrams, and at any finite order of the expansion all diagram weights are still infrared-finite.

\subsection{Metropolis algorithm for stochastic solution of linear equations}
\label{sec:diagmc:subsec:metropolis_general}

As discussed in the previous Subsection, Schwinger-Dyson equations in terms of disconnected field correlators always form an infinite system of \emph{linear equations}, which can be written in the following general form
\begin{eqnarray}
\label{sd_linear_general}
 \phi\lr{X} = b\lr{X} + \sum\limits_Y \, A\lr{X | Y} \phi\lr{Y},
\end{eqnarray}
where $A\lr{X | Y}$ is some theory-dependent linear operator, $b\lr{X}$ represents the ``source'' terms in the Schwinger-Dyson equations (typically, the contact delta-function term in the lowest-order equation, like in (\ref{matrix_sd_general_n2}) and (\ref{sd_eqs_mom_vertex}), and $X$ and $Y$ are generalized indices incorporating all arguments of field correlators which enter the Schwinger-Dyson equations. For the Schwinger-Dyson equations (\ref{sd_eqs_mom_vertex}) discussed in the previous Subsection, $X$ and $Y$ are sequences of momenta $\lrc{\lrc{p^1_1, \ldots, p^1_{n_1}}, \ldots, \lrc{p^r_1, \ldots, p^r_{n_r}}}$ which parameterize the multi-trace correlators (\ref{corrs_expansion_def}) with an additional order-counting integer variable $m$. Note that the set of $X$ indices is in most cases infinite, which excludes the application of standard numerical linear algebra methods to the equation (\ref{sd_linear_general}) (see however \cite{Kruczenski:16:1} for an attempt in this direction). Bringing the Schwinger-Dyson equations into the form (\ref{sd_linear_general}) can be subtle for field theories for which the free (quadratic) part of the action has flat directions or vanishes, but these subtleties are not relevant for the present work.

The solution $\phi\lr{X}$ of (\ref{sd_linear_general}) can be written as formal geometric series in $A$:
\begin{eqnarray}
\label{sd_linear_powerseries_explicit}
 \phi\lr{X} = \sum\limits_{n=0}^{+\infty} \sum\limits_{X_0} \ldots \sum\limits_{X_n} \delta\lr{X, X_n}
 \times \nonumber \\ \times
 A\lr{X_n | X_{n-1}} \ldots A\lr{X_1 | X_0} b\lr{X_0} .
\end{eqnarray}
If the diagrammatic expansion which is obtained by iterating Schwinger-Dyson equations is truncated at some finite order, there is only a finite number of terms in these series, which correspond to a finite number of diagrams contributing at a given expansion order.

The key idea of the DiagMC algorithm which we devise in this work is to evaluate the series (\ref{sd_linear_powerseries_explicit}) by stochastically sampling the sequences of generalized indices $\mathcal{S} = \lrc{X_n, \ldots, X_0}$ with arbitrary $n$ in (\ref{sd_linear_powerseries_explicit}) with probability
\begin{eqnarray}
\label{sd_linear_sampling_prob}
 w\lr{\mathcal{S}}
 =
 \frac{|A\lr{X_n | X_{n-1}}| \ldots |A\lr{X_1 | X_0}| |b\lr{X_0}|}{\N_w} ,
\end{eqnarray}
where $\N_w$ is the appropriate normalization factor.

The solution $\phi\lr{X}$ to the system (\ref{sd_linear_general}) can be found by histogramming the last generalized index $X_n$ in the sequence $\lrc{X_n, \ldots, X_0}$, where each occurrence of $X_n$ is weighted with the sign $\sigma\lr{X_n, \ldots, X_0} = \sign{A\lr{X_n | X_{n-1}}} \ldots \sign{A\lr{X_1 | X_0}} \sign{b\lr{X_0}}$. This sign reweighting puts certain limitations on the use of the method, since we are effectively sampling the series in which the linear operator $A\lr{X | Y}$ is replaced by $|A\lr{X | Y}|$ and which typically have a smaller radius of convergence. In the worst case, the expectation value of the reweighting sign $\sigma\lr{X_n, \ldots, X_0}$ can decrease exponentially or faster with $n$, thus limiting the maximal expansion order which can be sampled by the algorithm.

Let us also note that since for Schwinger-Dyson equations the generalized indices $X$ take infinitely many values, in practice histogramming of all possible values of $X$ is not possible, but also not necessary. E.g. if one is interested only in the momentum-space two-point correlator $\vev{\frac{1}{N} \tr\lr{\phi_{p_1} \, \phi_{p_2}}} = \vev{\lrs{p_1, p_2}}$ which corresponds to $X$ of the form $X = \lrc{\lrc{p_1, p_2}}$, one can histogram only the momenta $p_1$ and $p_2$ whenever $X$ contains only two of them. Thus no field correlators should be explicitly saved in computer memory, and the memory required by the algorithm does not directly scale with system volume (however, it does scale with the maximal expansion order, which might be correlated with the volume via required statistical error).

In order to simplify the notation in what follows, let us also define the quantities
\begin{eqnarray}
\label{n_def}
 \N\lr{Y} = \sum\limits_X |A\lr{X | Y}|, \quad \N_b = \sum\limits_X |b\lr{X}| .
\end{eqnarray}
In this work we propose to sample the sequences $\mathcal{S}$ with probability (\ref{sd_linear_sampling_prob}) by using the Metropolis-Hastings algorithm with the following updates:
\desclabel{Algorithm}\label{desc:transitions}\nopagebreak
\begin{description}
 \item[Add index:] With probability $p_+$ a new generalized index $X_{n+1}$ is added to the sequence $\lrc{X_n, \ldots, X_0}$, where the probability distribution of $X_{n+1}$ is $\pi\lr{X_{n+1} | X_n} = |A\lr{X_{n+1} | X_n}|/\N\lr{X_n}$. The sign variable $\sigma$ is multiplied by $\sign{A\lr{X_{n+1} | X_n}}$.
 \item[Remove index:] If the sequence contains more than one generalized index, with probability $1 - p_+$ the last generalized index $X_n$ is removed from the sequence, which thus changes from $\lrc{X_n, X_{n-1}, \ldots, X_0}$ to $\lrc{X_{n-1}, \ldots, X_0}$. The sign variable $\sigma$ is multiplied by $\sign{A\lr{X_n | X_{n-1}}}$.
 \item[Restart:] If the sequence contains only one generalized index $X_0$, it is replaced by $X_0'$ with probability $1 - p_+$. The probability distribution of $X_0'$ is $\pi\lr{X_0'} = |b\lr{X_0'}|/\N_b$. The sign variable $\sigma$ is set to $\sign{b\lr{X_0'}}$.
 \end{description}
These updates should be accepted or rejected with the probability \cite{Hastings:70:1}
\begin{eqnarray}
\label{meth_hast_acceptance_nobalance}
 \alpha\lr{\mathcal{S} \rightarrow \mathcal{S}'}
 = \min\lr{1, \frac{w\lr{\mathcal{S}'} \pi\lr{\mathcal{S' \rightarrow S}}}{w\lr{ \mathcal{S}} \pi\lr{\mathcal{S  \rightarrow S'}}}}.
\end{eqnarray}
Explicit calculation yields the following acceptance probabilities for the three updates introduced above:
\begin{eqnarray}
\label{acceptance_probabilities}
 \alpha_{add}     =  \min\lr{1, \frac{\N\lr{X_n} \lr{1 - p_+}}{p_+}} ,
 \nonumber \\
 \alpha_{remove}  =  \min\lr{1, \frac{p_+}{\N\lr{X_{n-1}} \lr{1 - p_{+}} } },
 \nonumber \\
 \alpha_{restart} = 1 .
\end{eqnarray}
After some algebraic manipulations the overall average acceptance rate can be expressed as
\begin{eqnarray}
\label{total_acceptance}
 \vev{\alpha}
 =
 \lr{1 - p_+} \frac{\N_b}{\N_w}
 + \nonumber \\ +
 2 \vev{\min\lr{ p_+, \lr{1 - p_+} \N\lr{X_n} }}  ,
\end{eqnarray}
where in the last line the expectation value is taken with respect to the stationary probability distribution of $X_n$. For the optimal performance of the algorithm it is desirable to choose $p_+$ (which is the only free parameter in our algorithm) in such a way that the acceptance rate is maximally close to unity. To this end we first perform Metropolis updates with some initial value of $p_+$, estimate the expectation values $\vev{\min\lr{ p_+', \lr{1 - p_+'} \N\lr{X_n} }}$ in (\ref{total_acceptance}) for several trial values $p_+'$ and then select the value for which the acceptance is maximal. This process is repeated several times until the values of the average acceptance and $p_+$ stabilize.

Sometimes the transformations $X_n \rightarrow X_{n+1}$ can be joined into groups of very similar transformations. For example, if we have to generate a random loop momentum when sampling the Feynman diagrams in momentum space, in the notation of Algorithm \ref{desc:transitions} generating every single possible momentum would be a separate transformation, but it is more convenient to describe them as a single group of transformations. Denoting possible outcomes of such transformations as $X_{n+1} \in \T\lr{X_n}$, where $\T\lr{X_n}$ labels the group of transformations, we can express the probability distribution of the new element $X_{n+1}$ in the \textbf{``Add index''} update in (\ref{desc:transitions}) as the product of the conditional probability $\pi\lr{X_{n+1} | \T\lr{X_n}}$ to generate the new element $X_{n+1}$ by one of the transformations in $\T\lr{X_n}$, times the overall probability $\pi\lr{\T | X_n}$ to perform any transformation from $\T$:
\begin{eqnarray}
\label{transformation_grouping}
 \pi\lr{X_{n+1} | X_n} = \pi\lr{X_{n+1} | \T\lr{X_n}} \pi\lr{\T | X_n},
 \nonumber \\
 \pi\lr{X_{n+1} | \T\lr{X_n}} = A\lr{X_{n+1} | X_n}/\N_{\T}\lr{X_n},
 \nonumber \\
 \N_{\T}\lr{X_n} = \sum_{X_{n+1} \in \T} A\lr{X_{n+1} | X_n},
 \nonumber \\
 \pi\lr{\T | X_n} = \N_{\T}\lr{X_n}/\N\lr{X_n},
 \nonumber \\
 \N\lr{X_n} = \sum_{\T\lr{X_n}} \N_{\T}\lr{X_n}.
\end{eqnarray}
Such grouping of transformations will be very convenient for a compact description of our DiagMC algorithm in what follows. Sometimes we will also omit the arguments of $\N_{\T}\lr{X_n}$ and $\T\lr{X_n}$ for the sake of brevity.

Finally, let us calculate the normalization factor $\N_w$ relating the (sign weighted) histogram of the last generalized index $X_n$ in the sequences $\mathcal{S} = \lrc{X_n, \ldots, X_0}$ and the solution $\phi\lr{X}$ of the linear equations (\ref{sd_linear_general}). Combining (\ref{sd_linear_sampling_prob}) and (\ref{sd_linear_general}), we obtain a linear equation for $\N_w$:
\begin{eqnarray}
\label{Nw_linear_equation}
 \N_w = \N_w \sum\limits_X w\lr{X}
 = \nonumber \\ =
 \N_w \sum\limits_{X,Y} |A\lr{X | Y}| w\lr{Y} + \sum\limits_X |b\lr{X}|
 = \nonumber \\ =
 \N_w \sum\limits_Y \N\lr{Y} w\lr{Y} + \N_b ,
\end{eqnarray}
where $w\lr{X}$ is the probability distribution of the last element $X_n$ in the sequences $\mathcal{S}$, with any sign $\sigma$. The solution of this equation can be written as
\begin{eqnarray}
\label{norm_factor_linear}
 \N_w = \frac{\N_b}{1 - \vev{\N\lr{X_n}}} ,
\end{eqnarray}
where again the expectation value is taken with respect to the stationary probability distribution of the Metropolis algorithm \ref{desc:transitions}.

The general implementation of the Metropolis algorithm \ref{desc:transitions} for which the user-defined matrix elements $A\lr{X | Y}$ and the source vector $b\lr{X}$ should be provided is available as a part of GitHub repository \cite{SDMetropolisGitHub}.

\subsection{DiagMC algorithm for the \texorpdfstring{large-$N$ $U\lr{N} \times U\lr{N}$}{large-N U(N) x U(N)} principal chiral model}
\label{sec:diagmc:subsec:diagmc_pcm}

In this Subsection we adapt the general Metropolis algorithm described in the previous Subsection~\ref{sec:diagmc:subsec:metropolis_general} for solving the linear Schwinger-Dyson equations (\ref{sd_eqs_mom_vertex}) obtained from the action (\ref{pcm_action}) of the large-$N$ principal chiral model (\ref{pcm_partition0}). The generalized indices $X$ will take values in the space of partitioned sequences of momenta
\begin{eqnarray}
\label{partitioned_sequences_def}
 X = \lrc{\lrc{p^1_1, \ldots, p^1_{n_1}}, \ldots, \lrc{p^r_1, \ldots, p^r_{n_r}}}_m ,
\end{eqnarray}
additionally labelled with an expansion order $m = 0, 1, 2, \ldots$. These sequences will be generated with the probability proportional to the correlators $\vev{\lrs{p^1_1, \ldots, p^1_{n_1}} \ldots \lrs{p^r_1, \ldots, p^r_{n_r}}}_m$ defined in (\ref{corrs_expansion_def}), up to the sign re-weighting with the sign variable $\sigma = \pm 1$, introduced in algorithm~\ref{desc:transitions}.

The system of equations (\ref{sd_eqs_mom_vertex}) already has the form (\ref{sd_linear_general}). The only non-zero elements of the source vector $b\lr{X}$ correspond the delta-function term (in the right-hand side of line (\ref{sd_eqs_mom_vertex_n2_contact})) in the first, inhomogeneous Schwinger-Dyson equation in (\ref{sd_eqs_mom_vertex}):
\begin{eqnarray}
\label{source_vector_components}
 b\lr{\lrc{\lrc{p, -p}}} = \delta_{m,0} G_0\lr{p}/V .
\end{eqnarray}
From equations (\ref{sd_eqs_mom_vertex}) we can also directly read off the following groups of nonzero elements of the $A\lr{X | Y}$ matrix:
\begin{widetext}
\begin{subequations}
\label{A_matrix_components}
\begin{eqnarray}
\label{A_matrix_components_newseq}
  A\lr{
   \lrc{
   \activeseq{\lrc{\activeidx{p^1_1, p^1_2}}},
   \lrc{p^2_1, \ldots, p^2_{n_2}},
      \ldots,
   \lrc{p^r_1, \ldots, p^r_{n_r}}
   }_m
   |
   \lrc{
    \lrc{p^2_1, \ldots, p^2_{n_2}},
    \ldots,
    \lrc{p^r_1, \ldots, p^r_{n_r}}
   }_m
  }
  = \nonumber \\ =
  \frac{G_0\lr{\activeidx{p^1_1}} \, \delta\lr{\activeidx{p^1_1 + p^1_2}}}{V} ,
  \quad \myremark{\text{(from the term (\ref{sd_eqs_mom_vertex_n2_mtr_newseq}))  }} \quad \quad 
\end{eqnarray}
\begin{eqnarray}
  \label{A_matrix_components_newline1}
  A\lr{
   \lrc{
   \activeseq{\lrc{\activeidx{p^1_1, p^1_2,} p^1_3, \ldots, p^1_{n_1}}},
      \ldots
   \lrc{p^r_1, \ldots, p^r_{n_r}}}_m
   |
   \lrc{
    \activeseq{\lrc{p^1_3, \ldots, p^1_{n_1}}},
    \ldots
    \lrc{p^r_1, \ldots, p^r_{n_r}}
   }_m
  }
  = \nonumber \\ =
  \frac{G_0\lr{\activeidx{p^1_1}}\, \delta\lr{\activeidx{p^1_1 + p^1_2}}}{V} ,
  \quad n_1 \geq 4 , 
  \quad \myremark{\text{(from the term (\ref{sd_eqs_mom_vertex_newline1}))  }} \quad \quad
\end{eqnarray}
\begin{eqnarray}
  \label{A_matrix_components_newline2}
  A\lr{
   \lrc{
   \activeseq{\lrc{\activeidx{p^1_1,} p^1_2, \ldots, p^1_{n_1 - 1}, \activeidx{p^1_{n_1}}}},
      \ldots
   \lrc{p^r_1, \ldots, p^r_{n_r}}}_m
   |
   \lrc{
    \activeseq{\lrc{p^1_2, \ldots, p^1_{n_1 - 1}}},
    \ldots,
    \lrc{p^r_1, \ldots, p^r_{n_r}}
   }_m
  }
  = \nonumber \\ =
  \frac{G_0\lr{\activeidx{p^1_1}} \, \delta\lr{\activeidx{p^1_1 + p^1_{n_1}}}}{V} ,
  \quad n_1 \geq 4 , 
  \quad \myremark{\text{(from the term (\ref{sd_eqs_mom_vertex_newline2}))  }} \quad \quad
\end{eqnarray}
\begin{eqnarray}
  \label{A_matrix_components_join}
  A\left(
   \lrc{
   \activeseq{\lrc{\activeidx{p^1_1,} \ldots, \activeidx{p^1_A,} \ldots, p^1_{n_1}}},
   \lrc{p^2_1, \ldots, p^2_{n_2}},
      \ldots
   \lrc{p^r_1, \ldots, p^r_{n_r}}
   }_m
   \right. \left. \right| \nonumber \\ \left| \right. \left.
   \lrc{
   \activeseq{\lrc{p^1_2, \ldots, p^1_{A-1}},
   \lrc{p^1_{A+1}, \ldots, p^1_{n_1}}},
   \lrc{p^2_1, \ldots, p^2_{n_2}},
   \ldots
   \lrc{p^r_1, \ldots, p^r_{n_r}}}_m
  \right)
  = \nonumber \\ =
  \frac{G_0\lr{\activeidx{p^1_1}}\, \delta\lr{\activeidx{p^1_1 + p^1_A}}}{V},
  \quad  A = 4, 6, \ldots, n_1 - 2 ,
  \quad n_1 \geq 6, 
  \quad \myremark{\text{(from the term (\ref{sd_eqs_mom_vertex_join}))  }} \quad \quad
\end{eqnarray}
\begin{eqnarray}
  \label{A_matrix_components_vertex}
   A\lr{
   \lrc{
   \activeseq{\lrc{\activeidx{p^1_1,} p^1_2, \ldots, p^1_{n_1}}},
   \ldots ,
   \lrc{p^r_1, \ldots, p^r_{n_r}}}_{\activeidx{m}}
   |
   \activeseq{\lrc{\lrc{\activeidx{q_1, \ldots, q_{2 v + 1},} p^1_2, \ldots, p^1_{n_1}}},
   \ldots ,
   \lrc{p^r_1, \ldots, p^r_{n_r}}}_{\activeidx{m - v}}
  }
  = \nonumber \\ =
  - G_0\lr{\activeidx{p^1_1}} \delta\lr{\activeidx{p^1_1 - q_1 - \ldots - q_{2 v + 1}}}
  V\lr{\activeidx{q_1, \ldots, q_{2 v + 1}}}   .
  \quad \myremark{\text{(from the terms (\ref{sd_eqs_mom_vertex_n2_vertex}), (\ref{sd_eqs_mom_vertex_n2_mtr_vertex}),(\ref{sd_eqs_mom_vertex_vertex}))  }} \quad \quad
\end{eqnarray}
\end{subequations}
\iftoggle{activecolor}{Here we have used \textcolor{red}{red} and \textcolor{violet}{violet} colors to highlight the individual momenta and order labels and momentum subsequences which are different for the first and the second indices of the $A\lr{X | Y }$ matrix.}{}
\end{widetext}

In the above representation we write down all nonzero elements of the matrix $A\lr{X | Y}$ for the fixed first index $X$. However, for the ``Add index'' update in the Algorithm~\ref{desc:transitions} we need to know all nonzero elements $A\lr{X_{n+1} | X_n}$ for the fixed second index $X_n$. In order to obtain a suitable representation of the matrix $A\lr{X | Y}$, we re-label the matrix elements (\ref{A_matrix_components}), assuming that the second index $Y$ in $A\lr{X | Y}$ has the most general form $Y = \lrc{\lrc{p^1_1, \ldots, p^1_{n_1}}, \ldots, \lrc{p^r_1, \ldots, p^r_{n_r}}}_m$, and express the momenta in the first index $X$ in terms of these momenta $p^a_i$. This gives
\begin{widetext}
\begin{subequations}
\label{A_matrix_components_transposed}
\begin{eqnarray}
\label{A_matrix_components_transposed_newseq}
  A\lr{
   \lrc{
   \activeseq{\lrc{\activeidx{p, -p}}},
   \lrc{p^1_1, \ldots, p^1_{n_1}},
      \ldots,
   \lrc{p^r_1, \ldots, p^r_{n_r}}
   }_m
   |
   \lrc{
    \lrc{p^1_1, \ldots, p^1_{n_1}},
    \ldots,
    \lrc{p^r_1, \ldots, p^r_{n_r}}
   }_m
  }
  = \nonumber \\ =
  \frac{G_0\lr{\activeidx{p}}}{V} ,  
  \quad \myremark{\text{(from the term (\ref{sd_eqs_mom_vertex_n2_mtr_newseq}))  }} \quad \quad
\end{eqnarray}
\begin{eqnarray}
  \label{A_matrix_components_transposed_newline1}
  A\lr{
   \lrc{
   \activeseq{\lrc{\activeidx{p, -p}, p^1_1, \ldots, p^1_{n_1}}},
      \ldots
   \lrc{p^r_1, \ldots, p^r_{n_r}}}_m
   |
   \lrc{
    \activeseq{\lrc{p^1_1, \ldots, p^1_{n_1}}},
    \ldots
    \lrc{p^r_1, \ldots, p^r_{n_r}}
   }_m
  }
  = \nonumber \\ =
  \frac{G_0\lr{\activeidx{p}}}{V} ,  
  \quad \myremark{\text{(from the term (\ref{sd_eqs_mom_vertex_newline1}))  }} \quad \quad
\end{eqnarray}
\begin{eqnarray}
  \label{A_matrix_components_transposed_newline2}
  A\lr{
   \lrc{
   \activeseq{\lrc{\activeidx{p,} p^1_1, \ldots, p^1_{n_1}, \activeidx{-p}}},
      \ldots
   \lrc{p^r_1, \ldots, p^r_{n_r}}}_m
   |
   \lrc{
    \activeseq{\lrc{p^1_1, \ldots, p^1_{n_1}}},
    \ldots,
    \lrc{p^r_1, \ldots, p^r_{n_r}}
   }_m
  }
  = \nonumber \\ =
  \frac{G_0\lr{\activeidx{p}}}{V} ,  
  \quad \myremark{\text{(from the term (\ref{sd_eqs_mom_vertex_newline2}))  }} \quad \quad
\end{eqnarray}
\begin{eqnarray}
  \label{A_matrix_components_transposed_join}
  A\left(
   \lrc{
   \activeseq{\lrc{\activeidx{p,} p^1_1, \ldots, p^1_{n_1}, \activeidx{-p,} p^2_1, \ldots, p^2_{n_2}}},
   \lrc{p^3_1, \ldots, p^3_{n_3}},
      \ldots
   \lrc{p^r_1, \ldots, p^r_{n_r}}
   }_m
   \right. \left. \right| \nonumber \\ \left| \right. \left.
   \lrc{
    \activeseq{\lrc{p^1_1, \ldots, p^1_{n_1}},
    \lrc{p^2_1, \ldots, p^2_{n_2}},}
    \lrc{p^3_1, \ldots, p^3_{n_3}},
    \ldots,
    \lrc{p^r_1, \ldots, p^r_{n_r}}
   }_m
  \right)
  = \nonumber \\ =
  \frac{G_0\lr{\activeidx{p}}}{V},
  \quad r \geq 2,  
  \quad \myremark{\text{(from the term (\ref{sd_eqs_mom_vertex_join}))  }} \quad \quad
\end{eqnarray}
\begin{eqnarray}
  \label{A_matrix_components_transposed_vertex}
   A\left(
   \lrc{
    \activeseq{\lrc{\activeidx{p^1_1 + p^1_2 + \ldots + p^1_{2 v + 1},} \, p^1_{2 v + 2}, \ldots, p^1_{n_1}}},
    \ldots ,
    \lrc{p^r_1, \ldots, p^r_{n_r}}}_{\activeidx{m + v}}
   \right. \left. \right| \nonumber \\ \left| \right. \left.
    \lrc{
    \activeseq{\lrc{\activeidx{p^1_1,} \ldots, p^1_{n_1}}},
    \ldots,
    \lrc{p^r_1, \ldots, p^r_{n_r}}
   }_{\activeidx{m}}
  \right)
  = \nonumber \\ =
  - G_0\lr{\activeidx{p^1_1 + p^1_2 + \ldots + p^1_{2 v + 1}}} \,
    V\lr{\activeidx{p^1_1, \ldots, p^1_{2 v + 1}}},
    \quad 2 v + 1 < n_1 ,
  \quad \myremark{\text{(from the terms (\ref{sd_eqs_mom_vertex_n2_vertex}), (\ref{sd_eqs_mom_vertex_n2_mtr_vertex}),(\ref{sd_eqs_mom_vertex_vertex}))  }} \quad
\end{eqnarray}
\end{subequations}
where the alphabetic labels of matrix elements are in one-to-one correspondence with alphabetic labels in (\ref{A_matrix_components}).
\end{widetext}
The matrix elements (\ref{A_matrix_components_transposed_newseq}) - (\ref{A_matrix_components_transposed_join}) form sets of elements parameterized by the momentum variable $p$. We denote the normalization factor $\mathcal{N}_{\mathcal{T}}\lr{X_n}$ in (\ref{transformation_grouping}) for these sets as
\begin{eqnarray}
\label{sigma0pcm}
 \Sigma_0 = \frac{1}{V}\sum_p G_0\lr{p} ,
\end{eqnarray}
where $\Sigma_0$ has an obvious interpretation of the one-loop self-energy. Correspondingly, the probability distribution of the new momenta $p$ is
\begin{eqnarray}
\label{pi0pcm}
 \pi_0\lr{p} \equiv G_0\lr{p}/\lr{V \Sigma_0} .
\end{eqnarray}

At the first sight it might seem that the Metropolis algorithm~(\ref{desc:transitions}) would require keeping in memory all the generalized indices $X_0, \ldots, X_n$ which enter the series (\ref{sd_linear_powerseries_explicit}). However, this is not necessary in practice, since the matrix $A\lr{X | Y}$ for the Schwinger-Dyson equations (\ref{sd_eqs_mom_vertex}) is very sparse, as one can see from the equations (\ref{A_matrix_components_transposed}). Also, only a few momenta in the generalized index $X_n = \lrc{\lrc{p^1_1, \ldots, p^1_{n_1}}, \ldots, \lrc{p^r_1, \ldots, p^r_{n_r}}}$ are different between the first and the second generalized indices of the matrix $A\lr{X_{n+1} | X_n}$. Thus only a few momenta will be changed when generating the new generalized index $X_{n+1}$ from $X_n$. Therefore it is sufficient to keep in computer memory only the topmost sequence $X_n$, as well as the sequence of transformations $X_n \rightarrow X_{n+1}$ which led to the current sequence, along with a few parameters which characterize these transformations. In fact, only the information which is necessary to recover $X_n$ from $X_{n+1}$ in the ``Remove index''  update in Algorithm~\ref{desc:transitions} should be stored.

We note also that the transformations $X_n \rightarrow X_{n+1}$ which correspond to matrix elements (\ref{A_matrix_components_transposed_newseq}), (\ref{A_matrix_components_newline1}), (\ref{A_matrix_components_newline2}), (\ref{A_matrix_components_transposed_vertex}) involve only the first momentum subsequence $\lrc{p^1_1, \ldots, p^1_{n_1}}$ in (\ref{partitioned_sequences_def}), and the transformation corresponding to (\ref{A_matrix_components_transposed_join}) involves also the second subsequence $\lrc{p^2_1, \ldots, p^2_{n_2}}$. This naturally endows the partitioned sequences (\ref{partitioned_sequences_def}) with the stack structure, in the algorithmic sense. Correspondingly, we will refer to the subsequence $\lrc{p^1_1, \ldots, p^1_{n_1}}$ as the ``topmost'' or ``first'' subsequence, the subsequence $\lrc{p^2_1, \ldots, p^2_{n_2}}$ - as the ``second'' sequence and so on.

Interestingly, the fact that the random sampling process operates on the stack implies the factorization of probabilities for subsequences in the stack \cite{Etessami:05:1,McKean:67:1,Buividovich:10:2}, which corresponds to factorization of single-trace correlators in the large-$N$ limit. If one proceeds keeping the $1/N^2$ terms in the equations (\ref{matrix_sd_general_n2_mtr}) and (\ref{matrix_sd_general}), our partitioned momentum sequences will no longer have stack structure. In particular, the fourth line of the Schwinger-Dyson equations (\ref{matrix_sd_general}) would correspond to a transformation where the topmost momentum subsequence is split into two subsequences, one of which is inserted into arbitrary subsequence in the partitioned sequence (\ref{partitioned_sequences_def}) \cite{Buividovich:16:3}. Obviously, the number of ways to do such a splitting and insertion grows polynomially both in the number of subsequences and their lengths. This polynomial growth of the number of terms in the Schwinger-Dyson equations results in the factorial growth of the number of non-planar Feynman diagrams with order \cite{Buividovich:16:3,Buividovich:11:1}. In contrast, the matrix elements (\ref{A_matrix_components_transposed}) do not depend on the number and lengths of subsequences, so that the algorithm samples a much smaller space of planar diagrams, the number of which grows only exponentially with expansion order. This is the reason why Diagrammatic Monte Carlo should be in general much more efficient in the large-$N$ limit.

Since the generalized indices $X$ take an infinite number of values, it is impossible but also not necessary to keep the matrix elements $A\lr{X | Y}$ in computer memory. Rather, they can be implemented as a function which, given the topmost index $X_n$ in the series (\ref{sd_linear_powerseries_explicit}), returns only nonzero values of $A\lr{X_{n+1} | X_n}$ and the corresponding set of Metropolis proposals $X_n \rightarrow X_{n+1}$. The \texttt{C} code of the Metropolis algorithm~\ref{desc:transitions} available at \cite{SDMetropolisGitHub} relies exactly on such a ``functional'' implementation of the matrix $A\lr{X_{n+1} | X_n}$.

\begin{widetext}
With all these considerations in mind, from (\ref{A_matrix_components_transposed}) we can finally deduce the following set of transformations $X_n \rightarrow X_{n+1}$ in the ``Add index'' update of Algorithm~\ref{desc:transitions}:
\desclabel{Transformation set}\label{desc:planar_pcm_transitions}
\begin{description}
  \item[Push a new pair of momenta to the stack,] from matrix elements (\ref{A_matrix_components_transposed_newseq}):
  \begin{itemize}
   \item A new subsequence containing a pair of momenta $\activeseq{\lrc{\activeidx{p, -p}}}$ is pushed to the top of the momentum stack. The probability distribution for $\activeidx{p}$ is $\pi_0\lr{\activeidx{p}}$, defined in (\ref{pi0pcm}).
   \item The order $m$ and the sign $\sigma$ do not change.
   \item The corresponding term in the Schwinger-Dyson equations is (\ref{sd_eqs_mom_vertex_n2_mtr_newseq}).
   \item The overall weight for these transformations, summed over all $\activeidx{p}$, is $\partsumI = \Sigma_0$, using the same notation as in (\ref{transformation_grouping}).
   \item To recover $X_n$ back from $X_{n+1}$, the topmost subsequence should be removed from the stack configuration $X_{n+1}$. No additional information should be saved for this backward step.
  \end{itemize}
  \item[Add new momenta to the topmost subsequence,] from matrix elements (\ref{A_matrix_components_transposed_newline1}), (\ref{A_matrix_components_transposed_newline2}):\nopagebreak
  \begin{itemize}
   \item A new pair of momenta is inserted into the topmost subsequence $\activeseq{\lrc{p^1_1, \ldots, p^1_{n_1}}}$, either a) at the beginning as $\activeseq{\lrc{\activeidx{p, -p,} \, p^1_1, \ldots, p^1_{n_1}}}$, or b) at the beginning and at the end as $\activeseq{\lrc{\activeidx{p,} \, p^1_1, \ldots, p^1_{n_1}, \activeidx{-p}}}$. The probability distribution for $\activeidx{p}$ is again $\pi_0\lr{\activeidx{p}}$.
   \item The order $m$ and the sign $\sigma$ do not change.
   \item The corresponding terms in the Schwinger-Dyson equations (\ref{sd_eqs_mom_vertex}) are (\ref{sd_eqs_mom_vertex_newline1}), (\ref{sd_eqs_mom_vertex_newline2}).
   \item The overall weights for each of the transformations a) and b) are $\partsum = \Sigma_0$.
   \item To recover $X_n$ back from $X_{n+1}$, either the two first momenta, or the first and the last momenta should be removed from the topmost subsequence in the stack. Correspondingly, one should remember whether the transformation a) or b) was chosen.
   \end{itemize}
   \item[Merge two topmost subsequences,] from matrix elements (\ref{A_matrix_components_transposed_join}):
   \begin{itemize}
    \item If there are two or more momentum subsequences in $X$, they are merged and interleaved with a pair of momenta $\activeidx{p, -p}$, where $\activeidx{p}$ is generated with probability distribution $\pi_0\lr{\activeidx{p}}$:
       \begin{eqnarray}
       \label{pcm_transformation_merge}
       \lrc{\activeseq{\lrc{p^1_1, \ldots, p^1_{n_1}}, \lrc{p^2_1, \ldots, p^2_{n_2}}}, \ldots }
       \rightarrow 
       \lrc{\activeseq{\lrc{\activeidx{p,} \, p^1_1, \ldots, p^1_{n_1}, \activeidx{-p,} p^2_1, \ldots p^2_{n_2}}}, \ldots } .
       \end{eqnarray}
    \item The order $m$ and the sign $\sigma$ do not change.
    \item The corresponding term in the Schwinger-Dyson equations (\ref{sd_eqs_mom_vertex}) is (\ref{sd_eqs_mom_vertex_join}).
    \item The overall weight for this transformation, summed over all $\activeidx{p}$, is $\partsum = \Sigma_0$.
    \item To recover $X_n$ back from $X_{n+1}$, one should split the topmost subsequence in the stack, which now has the form of the right-hand side of (\ref{pcm_transformation_merge}), at the position $n_1 + 1$, and remove the first momenta from the resulting subsequences. Correspondingly, the length $\activeidx{n_1}$ of the topmost subsequence in $X_n$ should be stored before the transformation $X_n \rightarrow X_{n+1}$.
   \end{itemize}
   \item[Create $\lr{2 v + 2}$-leg vertex,] from matrix elements (\ref{A_matrix_components_transposed_vertex}):
   \begin{itemize}
    \item If there are four or more momenta in the topmost momentum subsequence in $X_n$, replace the first $2 v + 1$, $v = 1 \ldots n_1/2-1$ momenta by a single momentum equal to their sum:
       \begin{eqnarray}
       \label{pcm_transformation_vertex}
        \lrc{\activeseq{\lrc{\activeidx{p^1_1, p^1_2, \ldots, p^1_{2 v + 1},} \, p^1_{2 v + 2}, \ldots, p^1_{n_1}}}, \ldots}
        \rightarrow 
        \lrc{\activeseq{\lrc{\activeidx{p^1_1 + p^1_2 + \ldots + p^1_{2 v + 1},} \, p^1_{2 v + 2}, \ldots, p^1_{n_1}}}, \ldots } .
       \end{eqnarray}
    \item The order $\activeidx{m}$ is increased by $v$, and the sign $\activeidx{\sigma}$ is multiplied by $-1$.
    \item The corresponding terms in the Schwinger-Dyson equations (\ref{sd_eqs_mom_vertex}) are (\ref{sd_eqs_mom_vertex_n2_vertex}), (\ref{sd_eqs_mom_vertex_n2_mtr_vertex}) and (\ref{sd_eqs_mom_vertex_vertex}).
    \item The weight of this transformation for each particular $v$ is
       \begin{eqnarray}
       \label{vertex_weight}
        A\lr{X_{n+1} | X_n}
        = 
        - G\lr{\activeidx{p^1_1 + \ldots + p^1_{2 v + 1}}} \, V\lr{\activeidx{p^1_1, \ldots, p^1_{2 v + 1}}} .
       \end{eqnarray}
    \item To recover $X_n$ back from $X_{n+1}$, one should replace the first momentum in the topmost subsequence in the stack by the momenta $\activeidx{p^1_1, \ldots, p^1_{2 v + 1}}$. To this end, the vertex order $\activeidx{v}$ and the $2 v$ momenta $\activeidx{p^1_1, \ldots, p^1_{2 v}}$ should be stored. The momentum $\activeidx{p^1_{2 v + 1}}$ can be recovered from momentum conservation constraint.
  \end{itemize}
  \item[Restart,] from the source vector (\ref{source_vector_components}):
  \begin{itemize}
  \item If the ``Restart'' update is selected in Algorithm~\ref{desc:transitions}, the new value of $X_0$ has a single subsequence with just two momenta: $X_0' = \lrc{\activeseq{\lrc{\activeidx{p, -p}}}}$, where $\activeidx{p}$ is distributed with probability $\pi_0\lr{\activeidx{p}}$.
  \item The order $\activeidx{m}$ is set to $\activeidx{m} = 0$, and the sign variable to $\activeidx{\sigma} = +1$. This state is also the initial state for the algorithm.
  \item The corresponding term in the Schwinger-Dyson equations (\ref{sd_eqs_mom_vertex}) is the term on the right-hand side of line (\ref{sd_eqs_mom_vertex_n2_contact}).
  \item The norm of the source vector is $\mathcal{N}_b = \sum\limits_{X_0'} b\lr{X_0'} = \Sigma_0$, which can be used to recover the overall normalization factor for field correlators from (\ref{norm_factor_linear}).
  \end{itemize}
\end{description}
\end{widetext}

The sequence of updates $X_n \rightarrow X_{n+1}$ in the Metropolis algorithm~\ref{desc:transitions} with transformations \ref{desc:planar_pcm_transitions} can be visually interpreted as a step-by-step process of drawing planar, in general disconnected Feynman diagrams with arbitrary number of external legs. The transformations~\ref{desc:planar_pcm_transitions} are used in the \textbf{``Add index''} update and correspond to adding a randomly chosen new element to the diagram. The first three transformations in the set~\ref{desc:planar_pcm_transitions} correspond to drawing free propagator lines with different positions of endpoints. The last transformation in the set~\ref{desc:planar_pcm_transitions} draws an interaction vertex with $2 v + 2$ legs by joining $2 v + 1$ external lines into a single external line. It is also this transformation which turns external lines into internal ones. The \textbf{``Remove index''} update in the Metropolis Algorithm \ref{desc:transitions} erases the diagram element which was drawn last, and can be thought of as an ``Undo'' operation. The initial state of the algorithm is just a single free propagator line, which is the simplest Feynman diagram one can ever draw. Simultaneous generation of random momenta distributed with the probability $\pi_0\lr{p}$ proportional to the free propagator $G_0\lr{p}$ implements the Monte-Carlo integration over internal loop momenta in Feynman diagrams. In this way the diagram weights are represented as products of propagators with freely generated momenta times vertex functions and propagators which contain some linear combinations thereof. Let us also note that the momentum conservation is automatically valid for all the transformations in (\ref{desc:planar_pcm_transitions}), which always lead to zero total momentum for any momentum subsequence in the stack.

Computationally, the most expensive operation in our DiagMC algorithm is the calculation of vertex functions $V\lr{p^1_1, \ldots, p^1_{2 v + 1}}$ for all $v = 1 \ldots n_1/2 - 1$, which is necessary to obtain acceptance probabilities for all possible ``Create vertex'' transformations in the set~\ref{desc:planar_pcm_transitions}. From (\ref{vertex_func_def}) one can infer that for each $v$ the calculation of the vertex function involves $\sim v^2$ floating-point operations. As a result, the CPU time needed to calculate all vertex functions scales as $n_1^3$. This CPU time is dominating the performance of the algorithm in the regime sufficiently close to the continuum limit, where subsequences with rather large lengths $n_1$ can appear. It is not inconceivable that this scaling can be made milder by carefully optimizing the calculation of vertex functions and re-using some data from previous Metropolis updates. We leave such developments for future work.

After calculation of vertex functions, the next most expensive and frequently used part of the algorithm is the generation of random momenta with probability $\pi_0\lr{p}$. To this end one can implement any method for generating continuous variables with definite probability distribution, for example, the biased Metropolis algorithm \cite{Bazavov:05:1}. Note that this allows to work directly in the infinite volume limit, which is one of the attractive features of DiagMC approach in general \cite{Prokofev:0802.2923}.

In this work we have still chosen to work with finite but sufficiently large spatial lattice size $L_0 \times L_1$, which is much larger than the correlation length for all values of $\lambda$ which we have used. In this case we generate the momenta $p$ by randomly choosing among a finite number of discrete values with given (in general, unequal) probabilities. By creating an indexed ``lookup table'' for the discrete values of momenta this procedure can be performed in CPU time which scales only logarithmically with volume.

The only reason for working in a finite volume is that this simplifies the histogramming of correlators of $g_x$ variables (original $U\lr{N}$-valued variables in the partition function (\ref{pcm_partition0})), see Subsection~\ref{sec:pcm_numres:subsec:observables} for a more detailed discussion. The performance of the algorithm, however, practically does not depend on the lattice size as long as it is much larger than the correlation length. Rather, the dynamically generated mass gap in the principal chiral model (\ref{pcm_partition0}) effectively sets the infrared scale for the algorithm. Needless to say, finite lattice size $L_0$ in the Euclidean time direction $\mu = 0$ is also necessary to study the theory at finite temperature (see Subsection \ref{sec:pcm_numres:subsec:finitetemp}).

The non-convergence of the re-summed perturbation theory of Section~\ref{sec:lattpt} for principal chiral model on finite-size one-dimensional lattice suggests that the series obtained on finite-size lattices might have worse convergence properties than in the infinite-volume limit. We expect that these problems will appear only at very high orders of perturbation theory. Indeed, replacing continuum integrals over Brillouin zone in the infrared-finite weak-coupling expansion of Section~\ref{sec:lattpt} by discrete sums of the form $\frac{1}{L^2} \sum\limits_p$ with $p_{\mu} = \frac{2 \pi m_{\mu}}{L}$, $m_{\mu} = 0 \ldots L - 1$ results in corrections to the continuum integrals which decrease with $L$ approximately as $e^{-m L}$, but increase with diagram order. For example, for a chain of $l$ one-loop diagrams, such corrections depend on order $l$ and lattice size $L$ as $\lr{1 + a e^{-m L}}^l$, with some constant $a$. Thus if the expansion is limited to some maximal order $M$ which also limits the maximal number of loop momenta, finite-volume corrections can always be made smaller by using sufficiently large lattice volume $V$. In Subsection~\ref{sec:pcm_numres:subsec:zerotemp} we explicitly demonstrate the smallness of finite-volume corrections by comparing the results of simulations at $108 \times 108$ and $256 \times 256$ lattices.

It seems also that our DiagMC algorithm does not allow for a straightforward parallelization. The only potential for parallelization which we currently see is to relegate the generation of random momenta with probability distribution $\pi_0\lr{p}$ to a separate MPI or OpenMP thread. However, as our algorithm can operate directly in the infinite-volume limit, parallelization might not be really necessary. Having a multi-CPU computer at hand, one can speed up simulations by gathering statistics independently on different CPUs.

In the next Section~\ref{sec:pcm_numres} we discuss in detail the numerical results which we have obtained using the algorithm~\ref{desc:transitions} with transformation set~\ref{desc:planar_pcm_transitions}, as well as the details of data collection and processing. The production code with the transformations \ref{desc:planar_pcm_transitions} in the Metropolis algorithm \ref{desc:transitions} is available as a part of GitHub repository \cite{SDMetropolisGitHub}.

\section{Numerical results of the DiagMC simulations of the \texorpdfstring{large-$N$ $U\lr{N} \times U\lr{N}$}{large-N U(N) x U(N)} principal chiral model}
\label{sec:pcm_numres}

\subsection{Correlators of \texorpdfstring{$g_x$}{gx} variables in terms of \texorpdfstring{$\phi_p$}{phip} fields}
\label{sec:pcm_numres:subsec:observables}


While the DiagMC algorithm described in Subsection~\ref{sec:diagmc:subsec:diagmc_pcm} stochastically samples the correlators of $\phi_p$ fields in momentum space, physically one is typically interested in correlators of $U\lr{N}$-valued $g_x$ variables which enter the partition function (\ref{pcm_partition0}).

The first non-trivial correlator is the expectation value $\vev{\frac{1}{N} \tr g_x}$ of the trace of $g_x$. It should vanish if the $U\lr{N} \times U\lr{N}$ symmetry of the principal chiral model (\ref{pcm_partition0}) is not broken. By virtue of the Mermin-Wagner theorem which prohibits spontaneous symmetry breaking in $D = 2$ dimensions, this should be the case at any finite value of $N$. Exact solution of the $O\lr{N}$ sigma model at $N \rightarrow \infty$ suggests also that spontaneous symmetry breaking does not happen for nonlinear sigma models in $D = 2$ in the large-$N$ limit.

However, the infrared-finite weak-coupling expansion sampled by our DiagMC algorithm is built around the vacuum state with $g_x = 1$, $\vev{\frac{1}{N} \tr g_x} = 1$, which explicitly breaks the $U\lr{N} \times U\lr{N}$ symmetry of the principal chiral model. To ensure that our DiagMC algorithm produces physically consistent results, it is thus important to check that the expectation value $\vev{\frac{1}{N} \tr g_x}$ approaches zero as the maximal order of the weak-coupling expansion $M$ is extrapolated to infinity. We demonstrate this on Fig.~\ref{fig:Gx_convergence} below.

Using the definition (\ref{cayley_map}), it is straightforward to expand the expectation value $\vev{\frac{1}{N} \tr g_x}$ in powers of the fields $\phi_x$. Since our perturbative series (\ref{expansion_general}) was obtained from the same expansion, it is natural to extend our formal power counting scheme based on the auxiliary parameter $\xi$ also to correlators of $g_x$ variables, which leads to
\begin{eqnarray}
\label{gx_vev_space}
 \vev{\frac{1}{N} \tr g_x} = \frac{1}{V} \sum\limits_x \vev{\frac{1}{N} \tr g_x}
 = \nonumber \\ =
 1 + 2 \sum\limits_{k=1}^{+\infty} \lr{-\frac{\lambda}{8}}^k \, \xi^k \, \frac{1}{V} \sum\limits_x \vev{\frac{1}{N} \tr \phi_x^{2 k}} ,
\end{eqnarray}
where in the first line we have used translational invariance and $\xi$ should be set to $\xi = 1$ at the end of the calculation.

Yet another observable which we consider in this work is the two-point correlator $\vev{\frac{1}{N} \tr\lr{g^{\dag}_x \, g_y}}$ of $g_x$ variables. This is the simplest nontrivial correlator from which one can infer the mean energy, the mass gap and the static screening length in the large-$N$ principal chiral model (\ref{pcm_partition0}). Similarly to (\ref{gx_vev_space}), we can use the definition (\ref{cayley_map}) to express this two-point correlator in terms of $\phi$ fields as
\begin{eqnarray}
\label{gxgy_vev_space}
 \vev{\frac{1}{N} \tr\lr{g^{\dag}_x \, g_y}}
 =
 2 \vev{\frac{1}{N} \tr g_x} - 1
 + \nonumber \\ +
 4 \sum\limits_{k,l=1}^{+\infty}
 \lr{-1}^{\frac{k-l}{2}} \, \lr{\frac{\lambda}{8}}^{\frac{k+l}{2}} \, \xi^{\frac{k+l}{2}} \,
 \vev{\frac{1}{N} \tr\lr{\phi_x^k \phi_y^l}}
 = \nonumber \\ =
 \lr{2 \vev{\frac{1}{N} \tr g_x} - 1}
 + \nonumber \\ +
 4 \sum\limits_{k=1}^{+\infty} \lr{-\frac{\lambda}{8}}^k \, \xi^k \,
 \sum\limits_{l=1}^{2 k - 1} \lr{-1}^l \vev{\frac{1}{N} \tr\lr{\phi_x^l \phi_y^{2 k - l}} } .
\end{eqnarray}

In order to perform a controllable extrapolation of the expansion order to infinity, we now limit our expansion in powers of $\xi$ to some maximal order $M$, which will serve as extrapolation parameter. Furthermore, we use the momentum-space field correlators (\ref{corrs_expansion_def}) which are sampled in our DiagMC simulations and explicitly take into account the momentum conservation, which implies translational invariance for the correlators of $g_x$ variables. This allows us to express the expectation values (\ref{gx_vev_space}) and (\ref{gxgy_vev_space}) as
\begin{eqnarray}
\label{gx_vev_momentum}
 \vev{\frac{1}{N} \tr g_x}_M
 =
 1
 + \nonumber \\ +
 2 \sum\limits_{k=1}^M
 \sum\limits_{m=0}^{M - k}
 \lr{-\frac{\lambda}{8}}^{k+m} \, \xi^{k+m} \,
 \times \nonumber \\ \times
 \sum\limits_{p_1, \ldots, p_{2 k}}
 \vev{\lrs{p_1, \ldots, p_{2 k}}}_m .
\end{eqnarray}
\begin{eqnarray}
\label{gxgy_vev_momentum}
 \vev{\frac{1}{N} \tr\lr{g^{\dag}_x g_0}}_M
 =
 2 \vev{\frac{1}{N} \tr g_x}_M - 1
 + \nonumber \\ +
 4 \sum\limits_{k=1}^M
   \sum\limits_{m=0}^{M - k}
 \lr{-\lambda/8}^{k+m} \, \xi^{k+m} \,
 \times \nonumber \\ \times
 \sum\limits_{p_1, \ldots, p_{2 k}}
 \Gamma\lr{x; \, p_1, \ldots, p_{2 k}}
 \vev{\lrs{p_1, \ldots, p_{2 k}}}_m
 \nonumber \\
 \Gamma\lr{x; \, p_1, \ldots, p_{2 k}}
 = \nonumber \\ =
 \sum\limits_{l=1}^{2 k - 1} \lr{-1}^l \cos\lr{\lr{p_1 + \ldots + p_l} x} .
\end{eqnarray}
In these expressions, we use the subscript $M$ for the correlators of $g_x$ variables to denote that our weak-coupling expansion (as defined in Section~\ref{sec:lattpt}) was truncated at order $M$.

While the algorithm~\ref{desc:transitions} with the transformation set~\ref{desc:planar_pcm_transitions} can in principle sample correlators $\vev{\lrs{p_1, \ldots, p_{2 k}}}_m$ to arbitrarily high orders $m$, we have found that in practice one can achieve better precision by sampling only field correlators $\vev{\lrs{p_1, \ldots, p_{2 k}}}_m$ with $k+m \leq M$. Indeed, only these field correlators contribute to the expectation values (\ref{gx_vev_momentum}) and (\ref{gxgy_vev_momentum}). In order to limit Monte-Carlo sampling to $k+m \leq M$, we set to zero all matrix elements $A\lr{X_{n+1} | X_n}$ for transformations which lead to $k+m > M$ in the Metropolis algorithm \ref{desc:transitions}. In our simulations we have been able to measure the first $M = 12$ coefficients of the infrared-finite weak-coupling expansion of Section~\ref{sec:lattpt} with reasonably small statistical error. Note also that once the field correlators $\vev{\lrs{p_1, \ldots, p_{2 k}}}_m$ are measured for all values of $k + m < M_{max}$ up to some maximal value $M_{max}$ of $M$, we can calculate at once the truncated correlators $\vev{\frac{1}{N} \tr g_x}_M$ and $\vev{\frac{1}{N} \tr\lr{g^{\dag}_x g_0}}_M$ for all $M \leq M_{max}$. This is very convenient in practice for the extrapolation of numerical results to the infinite expansion order $M \rightarrow \infty$.

In order to measure the field correlators (\ref{gx_vev_momentum}) and (\ref{gxgy_vev_momentum}), it is of course not necessary to histogram the sequences (\ref{partitioned_sequences_def}) with multiple momentum variables, which would require extremely large amount of RAM memory. The expectation value (\ref{gx_vev_momentum}) involves only the sum $\sum\limits_{p_1 \ldots p_{2 k}} \vev{\lrs{p_1, \ldots, p_{2 k}}}_m$, which is proportional to the total probability to get \emph{any} momentum sequence $\lrc{\lrc{p_1, \ldots, p_{2 k}}}_m$ of length $2 k$ with order $m$. Thus we only need to histogram the non-negative integer variables $k$ and $m$, which can only take values with $k + m \leq M$. This requires a histogram with $M \lr{M + 1}/2$ bins, which takes a negligibly small amount of RAM memory for realistic values of $M$ ($M \leq M_{max} = 12$ in our case).

Likewise, in order to measure the two-point correlator (\ref{gxgy_vev_momentum}) at some fixed distance $x$, we have to histogram only the variables $k$ and $m$, additionally weighting each momentum sequence $\lrc{\lrc{p_1, \ldots, p_{2 k}}}$ with the factor $\Gamma\lr{x; \, p_1, \ldots, p_{2 k}}$. Thus in order to measure the two-point correlator (\ref{gxgy_vev_momentum}) we again need the histogram with $M \lr{M + 1}/2$ bins for every value of $x$ we are interested in. In order to avoid frequent calculation of the function $\Gamma\lr{x; \, p_1, \ldots, p_{2 k}}$ which in practice takes significant amount of CPU time, we have calculated the two-point correlator $\sum_p e^{i p x} \vev{\frac{1}{N} \tr\lr{g^{\dag}_x g_0}}_M$ in momentum space. This Fourier transformation turns the function of $\Gamma\lr{x; \, p_1, \ldots, p_{2 k}}$ into the collection of $\delta$-functions
\begin{eqnarray}
\label{Gamma_func_mspace}
 \Gamma\lr{p; \, p_1, \ldots, p_{2 k}}
 = \nonumber \\ =
 \sum\limits_{l=1}^{2 k - 1} \lr{-1}^l \delta\lr{p - p_1 - \ldots - p_l}
\end{eqnarray}
which can be more easily incorporated into histogramming process (in deriving (\ref{Gamma_func_mspace}) we have taken into account the invariance of correlators under the replacement $x_{\mu} \rightarrow L_{\mu} - x_{\mu}$). After the momentum-space correlator $\sum_p e^{i p x} \vev{\frac{1}{N} \tr\lr{g^{\dag}_x g_0}}_M$ is measured, it can be easily transformed back to coordinate space by an inverse Fourier transform. The implementation of this faster and simpler histogramming procedure was the main reason for working with finite lattice size in our simulations.

In order to translate the sign-weighted probabilities obtained by histogramming $k$ and $m$ into the physical values of correlators (\ref{corrs_expansion_def}), we have to multiply them with the normalization factor (\ref{norm_factor_linear}). While equation (\ref{norm_factor_linear}) suggests that this normalization factor can be calculated by measuring the expectation value $\vev{\mathcal{N}\lr{X_n}}$ over the Monte-Carlo history, in practice we have found that $\mathcal{N}\lr{X_n}$ is a very noisy observable which is likely to have a heavy-tailed probability distribution. Thus directly using the equation (\ref{norm_factor_linear}) to obtain the normalization factor $\mathcal{N}_w$ results in large statistical errors. We have used a more convenient practical strategy and fixed $\mathcal{N}_w$ to be the ratio of the weighted probability in the histogram bin with $k=1$, $m=1$ and the first perturbative correction to the two-point function $\sum_{p_0, p_1} \vev{p_0 p_1}_1$, which we have explicitly calculated using (\ref{p0p1_1}). As a cross-check, we made sure that using the value of $\mathcal{N}_w$ calculated in this way leads to the correct estimate of the next perturbative correction $\sum_{p_0, p_1} \vev{p_0 p_1}_2$, which we have also explicitly calculated.

\subsection{Simulation setup, sign problem, and restoration of \texorpdfstring{$U\lr{N} \times U\lr{N}$}{U(N) x U(N)} symmetry}
\label{sec:pcm_numres:subsec:setup_misc}

In our simulations we perform two scans over parameters of the principal chiral model (\ref{pcm_partition0}). In the first scan, discussed in Subsection~\ref{sec:pcm_numres:subsec:zerotemp}, we fix the lattice size to $L_0 \times L_1 = 108 \times 108$ and study the dependence of physical observables on the t'Hooft coupling $\lambda$ in the range $\lambda = 3.012 \ldots 5.0$, which includes the scaling region of the principal chiral model (\ref{pcm_partition0}) where the continuum physics sets in, as well as the point of the large-$N$ phase transition \cite{Gross:80:1} between the weak- and the strong-coupling phases at $\lambda_c = 3.27$ \cite{Rossi:94:2}. To check the finite-volume effects, we have also performed a single simulation with $L_0 \times L_1 = 256 \times 256$ lattice at $\lambda = 3.1$. In the second scan, discussed in Subsection~\ref{sec:pcm_numres:subsec:finitetemp}, we fix the t'Hooft coupling to $\lambda = 3.012$ and study the temperature dependence of physical observables by varying the temporal lattice size $L_0$ in the range $L_0 = 4 \ldots 90$. As discussed above, we work on finite-volume lattices only to simplify the histogramming procedure, and the speed of our DiagMC algorithm depends on the lattice size at most logarithmically.

We perform histogramming of observables every time when our DiagMC algorithm reaches a configuration with only one momentum subsequence in the stack. One can also explicitly take into account the large-$N$ factorization (\ref{factorization_largeN}) and sample from multiple subsequences in the stack, however, this does not give a large advantage since in this case the samples appear to be rather strongly correlated. In order to avoid potential problems with autocorrelations, for each parameter set we run around $6 \cdot 10^4$ simulations of $2 \cdot 10^8$ Metropolis updates each with statistically independent seeds for the \texttt{ranlux} random number generator \cite{Luscher:93:1}. The outcome of each of these simulations is treated as statistically independent data point, and the statistical error for physical observables is calculated as the standard error for this data set. The total number of Metropolis updates for each data set is thus around $10^{13}$, which takes around $10^4$ CPU-hours (single core). The speed of our algorithm can be probably increased by an order of magnitude by carefully optimizing the calculation of vertex functions and generation of random momenta. Let us also note that achieving a comparable precision for the large-$N$ extrapolation of the results of conventional Monte-Carlo simulations takes a comparable amount of CPU time \cite{Buividovich:17:4}.

In each of statistically independent simulations, during the first $5 \cdot 10^4$ Metropolis updates we tune the probability $p_+$ of the \textbf{``Add index''} update in the Metropolis algorithm~\ref{desc:transitions} in order to maximize the average acceptance rate. The optimal value of $p_+$ and the resulting acceptance rate take values in the range $p_+ \simeq 0.4 \ldots 0.6$,  $\vev{\alpha} \simeq 0.5 \ldots 0.65$ and depend rather weakly on the t'Hooft coupling and temperature. The average length of the sequence $\lrc{X_n, \ldots, X_0}$ of generalized indices in the Metropolis algorithm~\ref{desc:transitions} is in the range between $\vev{n} = 17.3$ (for $\lambda = 3.012$) and $\vev{n} = 18.1$ (for $\lambda = 5.0$). The average number of subsequences in the momentum stack on which the transformations~\ref{desc:planar_pcm_transitions} operate is in the range between $\vev{r} = 1.2$ (for $\lambda = 5.0$) and $\vev{r} = 1.4$ (for $\lambda = 3.012$). The dependence of both $\vev{n}$ and $\vev{r}$ on temperature is much weaker, and is completely within the ranges given above.

\begin{figure}[h!tpb]
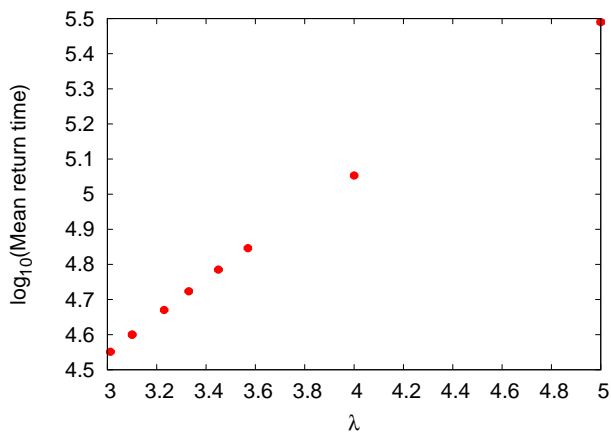

  \centering
  \includegraphics[width=0.33\textwidth,angle=-90]{{{mean_return_time_summary_lambda}}}
  \caption{The dependence of the average number of Monte-Carlo steps between the \textbf{``Restart''} updates in the Metropolis algorithm~\ref{desc:transitions} on the t'Hooft coupling $\lambda$.}
  \label{fig:return_time}
\end{figure}

As the most pessimistic estimate for the autocorrelation time of our DiagMC algorithm, we can use the average number of Monte-Carlo steps between the \textbf{``Restart''} updates in the Metropolis algorithm~\ref{desc:transitions}. Indeed, the \textbf{``Restart''} update completely resets the state of the algorithm, and the sequence of generalized indices generated after the \textbf{``Restart''} update is completely statistically independent from the sequence which was generated before that. On Fig.~\ref{fig:return_time} we illustrate the dependence of this ``mean return time'' on the t'Hooft coupling $\lambda$. It appears that the \textbf{``Restart''} updates are very rare for all values of $\lambda$. On average, they are separated by $\sim 10^4 \ldots 10^5$ Monte-Carlo steps. The mean return time strongly increases towards larger values of the t'Hooft coupling $\lambda$. Together with the $\lambda$ dependence of $\vev{n}$, this observation shows that for larger $\lambda$ higher orders in the infrared-finite weak-coupling expansion of Section~\ref{sec:lattpt} become more important, and the algorithm spends more time sampling correlators $\vev{p_1 \ldots p_{2 k}}_m$ with larger values of $k + m$. Correspondingly, returns to the initial state of the algorithm with $k = 1$, $m = 0$ are less frequent. In the scan over finite temperatures, the mean return time practically does not depend on temperature, and only at the highest temperatures which correspond to $L_0 = 4$ and $L_0 = 8$ it exhibits a quick increase from approximately $3.5 \cdot 10^4$ to $5.3 \cdot 10^4$.

\begin{figure}[h!tpb]
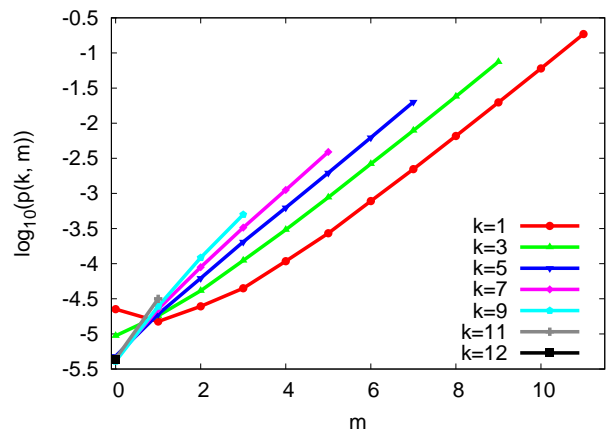

  \centering
  \includegraphics[width=0.33\textwidth,angle=-90]{{{visit_frequency}}}
  \caption{The probability $p\lr{k, m}$ to obtain the topmost momentum subsequence with $2 k$ momenta and order counter $m$ for $\lambda = 3.1$.}
  \label{fig:visit_frequency}
\end{figure}

On Fig.~\ref{fig:visit_frequency} we illustrate the probability distribution of the length $k$ of the topmost subsequence in the stack and the order counter $m$ in our DiagMC algorithm for $\lambda = 3.1$. The largest fraction of time (almost $20\%$) is spent on sampling the two-point correlator $\vev{p_0 p_1}_m$ with the maximal value $m = M_{max}-1$ ( $=11$ in our case). The least probable configuration is $m = 0$, $k = 12$, which is the disconnected correlator with the maximal number of external legs which contributes to expectation values (\ref{gx_vev_momentum}) and (\ref{gxgy_vev_momentum}). In total, the algorithm spends most time sampling the higher-order diagrams. This is a completely reasonable behaviour, as Monte-Carlo integration over loop momenta at high values of $m$ requires a lot of statistics, and also sign cancellations are most important for higher-order Feynman diagrams.

Let us now present our numerical results. We start by demonstrating that the expectation value $\vev{\frac{1}{N} \tr g_x}_M$ indeed tends to zero as the maximal order $M$ in the truncated weak-coupling expansion (\ref{gx_vev_momentum}) becomes larger. This is an indication that in the limit of large $M$ the global $U\lr{N} \times U\lr{N}$ symmetry of the principal chiral model (\ref{pcm_partition0}) is restored.

\begin{figure}[h!tpb]
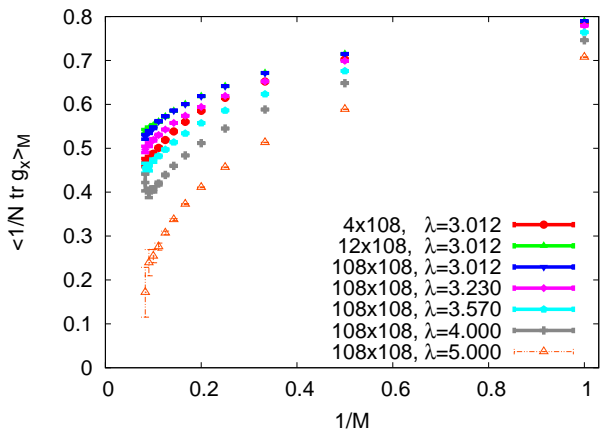

  \centering
  \includegraphics[width=0.33\textwidth,angle=-90]{{{Gx_convergence}}}\\
  \caption{Convergence of the expectation value $\vev{\frac{1}{N} \tr g_x}_M$ of the single field variable to zero as the truncation order $M$ in the weak-coupling expansion (\ref{gx_vev_momentum}) is increased.}
  \label{fig:Gx_convergence}
\end{figure}

On Fig.~\ref{fig:Gx_convergence} we show the dependence of $\vev{\frac{1}{N} \tr g_x}_M$ on $1/M$ for several values of t'Hooft coupling and several temperatures. In all cases $\vev{\frac{1}{N} \tr g_x}_M$ exhibits a nonlinear but monotonically decreasing dependence on $1/M$. For the largest t'Hooft couplings $\lambda = 5.0, \, 4.0$ and for the highest temperature $1/T = L_0 = 4$ the tendency of $\vev{\frac{1}{N} \tr g_x}_M$ to vanish is rather clear. In contrast, it seems that at small $\lambda$ the dependence on $M$ should become very nonlinear at small $1/M$ for $\vev{\frac{1}{N} \tr g_x}_M$ to extrapolate to zero. Indeed, the slope of $\vev{\frac{1}{N} \tr g_x}_M$ becomes larger towards smaller values of $1/M$. These observations support the expectation that $\vev{\frac{1}{N} \tr g_x}_M$ vanishes at $M \rightarrow \infty$.

\begin{figure}[h!tpb]
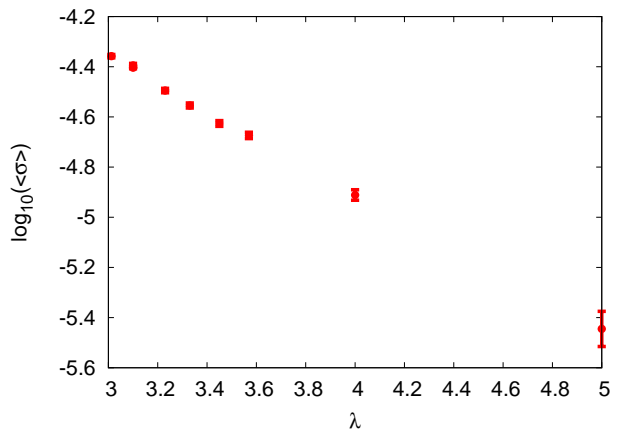

  \centering
  \includegraphics[width=0.33\textwidth,angle=-90]{{{mean_sign_summary_lambda}}}
  \caption{Expectation value of the reweighting sign $\sigma$ in the Metropolis algorithm~\ref{desc:transitions} as a function of the t'Hooft coupling $\lambda$.}
  \label{fig:sign_summary}
\end{figure}

As already mentioned in Subsection~\ref{sec:diagmc:subsec:diagmc_pcm}, not all coefficients $A\lr{X_{n+1} | X_n}$ are positive for the transformation set~\ref{desc:planar_pcm_transitions}. We thus have to use the sign reweighting, as discussed in Subsection~\ref{sec:diagmc:subsec:metropolis_general}. There are two potential sources of the sign problem which might hinder our DiagMC simulations: sign cancellations within the expectation values of the correlators $\vev{\lrs{p_1, \ldots, p_{2 k}}}_m$ of $\phi$ variables, and sign cancellations within the observables (\ref{gx_vev_momentum}) and (\ref{gxgy_vev_momentum}) themselves. In order to quantify sign cancellations in the correlators $\vev{p_1, \ldots, p_{2 k}}_m$, on Fig.~\ref{fig:sign_summary} we plot the expectation value of the sign variable $\sigma$ in the Metropolis algorithm~\ref{desc:transitions} as a function of the t'Hooft coupling $\lambda$. For all values of $\lambda$ the mean sign $\vev{\sigma}$ appears to be rather small, around $10^{-4} \ldots 10^{-5}$, thus sign problem is indeed important in our simulations. A somewhat encouraging feature is that the sign problem becomes milder towards the weak-coupling limit in which the theory approaches the continuum limit. However, from Fig.~\ref{fig:Gx_convergence}, as well as from the exactly solvable example of the 2D $O\lr{N}$ sigma model (see Subsection~\ref{sec:solvable_examples:subsec:2D}) we can conclude that at the same time the infrared-finite weak-coupling expansion of Section~\ref{sec:lattpt} converges less rapidly at small $\lambda$, so that more terms in the series should be sampled. The mean sign $\vev{\sigma}$ also depends rather weakly on temperature, and exhibits a rapid drop from $\vev{\sigma} \sim 5 \cdot 10^{-5}$ to $\vev{\sigma} \sim 3 \cdot 10^{-5}$ only at highest temperatures ($L_0 = 4$ and $L_0 = 8$). On Fig.~\ref{fig:mean_sign_vs_order} in the next Subsection we will also illustrate the sign problem for a particular physical observable (mean link).

\subsection{Two-point correlators at zero temperature}
\label{sec:pcm_numres:subsec:zerotemp}

In this Subsection we consider the simplest nontrivial two-point correlation function $\vev{\frac{1}{N} \tr\lr{g^{\dag}_x g_0}}$ of $U\lr{N}$-valued fields. From this correlator one can extract such important physical properties of the principal chiral model (\ref{pcm_partition0}) as the energy density and the mass gap in the principal chiral model (\ref{pcm_partition0}). It has been also extensively studied using conventional Monte-Carlo simulations of both $SU\lr{N} \times SU\lr{N}$ and $U\lr{N} \times U\lr{N}$ principal chiral models at several large values of $N$ \cite{Vicari:94:1,Rossi:94:2,Rossi:94:1,Buividovich:17:4}. Weak-coupling expansion of $\vev{\frac{1}{N} \tr\lr{g^{\dag}_x g_0}}$ was given up to three first orders in $\lambda$ at arbitrary values of $N$ in \cite{Rossi:94:1}. In this Subsection we will compare the large-$N$ extrapolations of these results with the results of our DiagMC simulations. This allows us to compare the performance of our algorithm with conventional Monte-Carlo simulations, as well as the convergence of our infrared-finite weak-coupling expansion with that of the standard lattice perturbation theory.

\begin{figure*}[h!tpb]
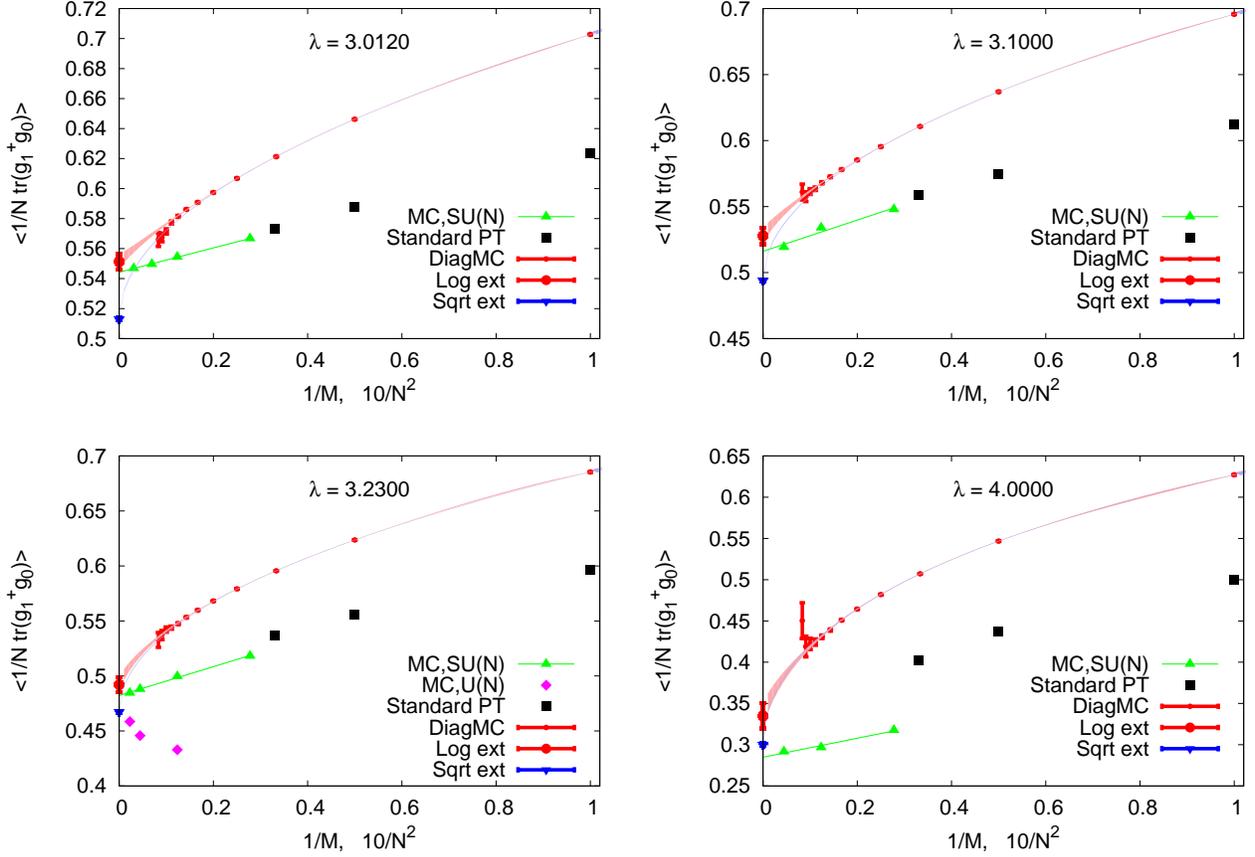

  \centering
  \includegraphics[width=\mfwa,angle=-90]{{{link_convergence_d2_t108_s108_l3.0120}}}
  \includegraphics[width=\mfwa,angle=-90]{{{link_convergence_d2_t108_s108_l3.1000}}}\\
  \includegraphics[width=\mfwa,angle=-90]{{{link_convergence_d2_t108_s108_l3.2300}}}
  \includegraphics[width=\mfwa,angle=-90]{{{link_convergence_d2_t108_s108_l4.0000}}}\\
  \caption{A comparison of the mean link values $\vev{\frac{1}{N} \tr\lr{g^{\dag}_1 g_0}}$ as obtained from DiagMC simulations of the infrared-finite weak-coupling expansion (\ref{expansion_general}), from standard Monte-Carlo simulations at finite $N$ (``MC,SU(N)'' and ``MC,U(N)''), and from the three lowest orders of the standard lattice perturbation theory (``Standard PT''). Perturbative results are plotted as the functions of the inverse expansion truncation order $1/M$. DiagMC data is extrapolated to $1/M \rightarrow 0$ by fitting the $1/M$ dependence to the functions (\ref{mlink_extrapolation_log}) (red strip) and (\ref{mlink_extrapolation_sqrt}) (blue strip). Strip width as well as the error bars on the extrapolated values (``Log ext'' and ``Sqrt ext'') are the confidence intervals for the extrapolating functions. The results of standard Monte-Carlo simulations are plotted as functions of $10/N^2$, and extrapolated to $1/N^2 \rightarrow 0$ using linear fits for $SU\lr{N}$ data (solid green lines).}
  \label{fig:pcm_link_convergence}
\end{figure*}

A particularly simple and important observable is the mean link $\vev{\frac{1}{N} \tr\lr{g^{\dag}_1 g_0}}$, which is also related to the energy density of the principal chiral model. On Fig.~\ref{fig:pcm_link_convergence} we illustrate the convergence of the truncated series (\ref{gxgy_vev_momentum}) for the mean link by plotting $\vev{\frac{1}{N} \tr\lr{g^{\dag}_1 g_0}}_M$ as a function of inverse truncation order $1/M$. The dependence on $1/M$ is nonlinear with the curvature increasing towards larger $M$, which requires some non-linear extrapolation procedure to reach the limit $M \rightarrow \infty$ ($1/M \rightarrow 0$). We have found that the dependence of the truncated series for the mean link on $1/M$ is well described by the following two extrapolating functions:
\begin{eqnarray}
 \label{mlink_extrapolation_log}
 \vev{\frac{1}{N} \tr\lr{g^{\dag}_1 g_0}}_M = A - B \frac{\log\lr{M}}{M} + \frac{C}{M} , \\
\label{mlink_extrapolation_sqrt}
 \vev{\frac{1}{N} \tr\lr{g^{\dag}_1 g_0}}_M = A + \frac{B}{\sqrt{M}} + \frac{C}{M} .
\end{eqnarray}
We perform extrapolations by fitting the numerical data with these functions using the $\chi^2$ weight which includes the data covariance matrix to account for correlations between numerical values at different $M$. The fit parameters are $A$, $B$ and $C$, with $A$ obviously corresponding to the result of extrapolation to $M \rightarrow \infty$. Upon fitting, both extrapolating functions (\ref{mlink_extrapolation_log}) and (\ref{mlink_extrapolation_sqrt}) lead to roughly the same values of $\chi^2$ over the number of degrees of freedom in the fit. These extrapolations are shown on Fig.~\ref{fig:pcm_link_convergence} as red and blue strips for the functions (\ref{mlink_extrapolation_log}) and (\ref{mlink_extrapolation_sqrt}), respectively. The strip width corresponds to the confidence interval set by the statistical uncertainty in the optimal values of $A$, $B$ and $C$. The difference between the two extrapolations (\ref{mlink_extrapolation_log}) and (\ref{mlink_extrapolation_sqrt}) can be used as an estimate of a systematic error of our extrapolation procedure in the absence of other prior information which helps to choose between the two fits (see below).

To compare our data with the results of conventional Monte-Carlo simulations, on Fig.~\ref{fig:pcm_link_convergence} we also show the mean link values obtained in Monte-Carlo simulations of both $SU\lr{N} \times SU\lr{N}$ and $U\lr{N} \times U\lr{N}$ principal chiral models as functions of $1/N^2$ (rescaled as $10/N^2$ for better readability of the plots). These data is taken from \cite{Vicari:94:1,Rossi:94:2,Rossi:94:1} for t'Hooft couplings $\lambda = 3.100, \, 3.230, \, 4.000$ and $N = 6, \, 9, \, 15, \, 21$ and from \cite{Buividovich:17:4} for $\lambda = 3.012$ and $N = 6, \, 9, \, 12, \, 18$. The finite-$N$ data is extrapolated to the large-$N$ limit using the linear fits of the form $\vev{\frac{1}{N} \tr\lr{g^{\dag}_1 g_0}} = A + B/N^2$, which are shown on Fig.~\ref{fig:pcm_link_convergence} as straight solid green lines. We have to note that the Monte-Carlo data for $\lambda = 3.100, \, 3.230, \, 4.000$ shows noticeable deviations from the expected $A + B/N^2$ behavior which significantly exceed the very small statistical errors in these data. Thus our $N \rightarrow \infty$ extrapolation might be also not completely accurate. Also the Monte-Carlo data for the finite-$N$ $U\lr{N} \times U\lr{N}$ principal chiral model, taken from \cite{Rossi:94:2}, does not seem to follow the $1/N^2$ scaling of finite-$N$ corrections.

For the three values $\lambda = 3.0120, \, 3.100$ and $\lambda = 3.230$ the extrapolation of the finite-$N$ data for the $SU\lr{N} \times SU\lr{N}$ principal chiral model clearly approaches the result of the $M \rightarrow \infty$ extrapolation with the extrapolating function (\ref{mlink_extrapolation_log}). While the differences of the $N \rightarrow \infty$ and $M \rightarrow \infty$ extrapolations lie within the statistical uncertainty of the $M \rightarrow \infty$ extrapolation of the DiagMC data, extrapolating function (\ref{mlink_extrapolation_log}) has a small but noticeable tendency for over-estimating the mean link. On the other hand, the second extrapolating function (\ref{mlink_extrapolation_sqrt}) underestimates the mean link rather significantly, and we thus conclude that the first extrapolating function (\ref{mlink_extrapolation_log}) yields more realistic results in the $M \rightarrow \infty$ limit.

\begin{figure}[h!tpb]
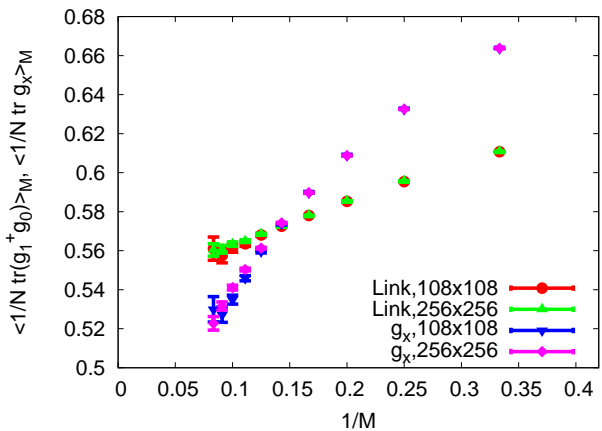

  \centering
  \includegraphics[width=\mfwa,angle=-90]{{{link_volume_comparison}}}\\
  \caption{A comparison of truncated expansions (\ref{gx_vev_momentum}) and (\ref{gxgy_vev_momentum}) for the mean link $\vev{\frac{1}{N} \tr\lr{g_1^{\dag} g_0}}$ and the single-field expectation value $\vev{\frac{1}{N} \tr g_x}$ for lattice sizes $L_0 \times L_1 = 108 \times 108$ and $L_0 \times L_1 = 256 \times 256$. CPU time is roughly the same for both simulations.}
  \label{fig:link_volume_comparison}
\end{figure}

To illustrate the ability of our DiagMC algorithm to work at arbitrarily large volumes without significant slow-down, on Fig.~\ref{fig:link_volume_comparison} we also compare the convergence of the truncated expansions (\ref{gx_vev_momentum}) and (\ref{gxgy_vev_momentum}) for the mean link $\vev{\frac{1}{N} \tr\lr{g_1^{\dag} g_0}}$ and the single-field expectation value $\vev{\frac{1}{N} \tr g_x}$ for lattice sizes $L_0 \times L_1 = 108 \times 108$ and $L_0 \times L_1 = 256 \times 256$. Both simulations took approximately the same CPU time, and had the same strength of the sign problem.

In order to compare the convergence of our massive lattice perturbation theory constructed in Section~\ref{sec:lattpt} with the convergence of the standard lattice perturbation theory constructed using massless propagators, on Fig.~\ref{fig:pcm_link_convergence} we also show the one-loop, two-loop and three-loop ($M = 1, \, 2, \, 3$, respectively) results of \cite{Rossi:94:1} for the mean link at $N \rightarrow \infty$. It appears that conventional perturbation theory yields the results which are, at least for the first three orders, closer to the physical result and also seem to converge faster. Thus the price which we have to pay for the absence of infrared divergences is the slower convergence of our infrared-finite weak-coupling expansion.

For the largest value $\lambda = 4.00$ which we consider our massive perturbation theory with both extrapolations (\ref{mlink_extrapolation_log}) and (\ref{mlink_extrapolation_sqrt}) yields the result which is significantly larger than the large-$N$ extrapolation of the Monte-Carlo data. Also the standard lattice perturbation theory \cite{Rossi:94:1} seems to converge to a larger value. This disagreement between weak-coupling expansions and the physical result is an indication of the large-$N$ transition between the weak-coupling and the strong-coupling regimes \cite{Gross:80:1}, which in the principal chiral model (\ref{pcm_partition0}) happens at $\lambda_c = 3.27$ \cite{Rossi:94:2}. After this transition, the physical values should be rather calculated using the strong-coupling expansion.

\begin{figure}[h!tpb]
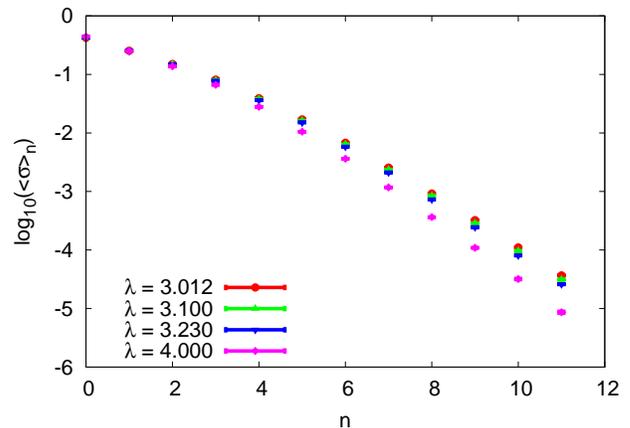

  \centering
  \includegraphics[width=\mfwa,angle=-90]{{{mean_sign_vs_order}}}\\
  \caption{Average sign $\vev{\sigma}_n$ of the infrared-finite weak-coupling expansion diagrams which contribute to the link expectation value (\ref{gxgy_vev_momentum}) at order $n = k + m$ as a function of $n$ for different values of $\lambda$.}
  \label{fig:mean_sign_vs_order}
\end{figure}

As we have already discussed in Subsection~\ref{sec:pcm_numres:subsec:setup_misc} above, our simulations are limited to around $M \sim 10^1$ first orders of the weak-coupling expansion due to more and more important sign cancellations between different Feynman diagrams. To illustrate this sign problem in more detail, we have considered the sign cancellations within contributions of order $\lr{-\frac{\lambda}{8}}^{m+k}$ with fixed $m+k = n$ to the mean link $\vev{\frac{1}{N} \tr\lr{g^{\dag}_1 g_0}}_M$ given by (\ref{gxgy_vev_momentum}). We have characterized these sign cancellations by the ratio $\vev{\sigma}_n = \vev{\frac{I_n^+ - I_n^-}{I_n^+ + I_n^-}}$, where $I_n^+$ and $I_n^-$ are the sums of the absolute values of diagrams which contribute at order $n = k + m$ to the mean link (\ref{gxgy_vev_momentum}) with positive and negative re-weighting signs $\sigma$, respectively, and $\vev{\ldots}$ denotes averaging over Monte-Carlo history. On Fig.~\ref{fig:mean_sign_vs_order} we illustrate that $\vev{\sigma}_n$ decays seemingly exponentially with $n$ for several values of the t'Hooft coupling $\lambda$. Since our algorithm spends most time sampling diagrams with large $n \sim M$, this explains the overall strength of the sign problem which we have previously illustrated on Fig.~\ref{fig:sign_summary}.

\begin{figure}[h!tpb]
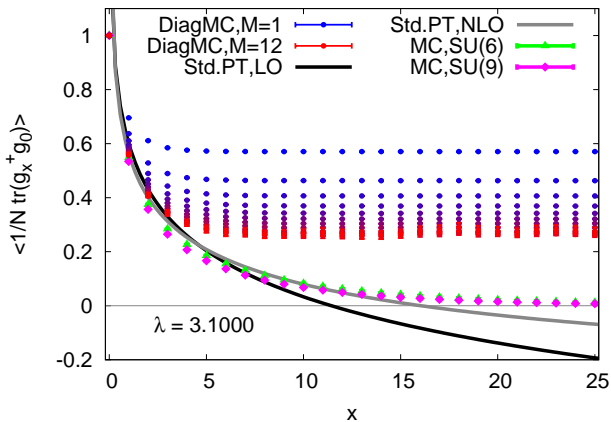

  \centering
  \includegraphics[width=\mfwa,angle=-90]{{{correlator_d2_t108_s108_l3.1000}}}\\
  \caption{Coordinate dependence of the two-point correlator $\vev{\frac{1}{N} \tr\lr{g^{\dag}_x g_0}}_M$ given by (\ref{gxgy_vev_momentum}) at various truncation orders $M$ of the infrared-finite weak-coupling expansion at $\lambda = 3.1$. The values of $M$ are encoded in color, from pure blue for $M = 1$ to pure red for $M = 12$. For comparison, we also show the data from the standard Monte-Carlo simulations with $N=6$ and $N = 9$, and the first two orders of the standard perturbative expansion of $\vev{\frac{1}{N} \tr\lr{g^{\dag}_x g_0}}$.}
  \label{fig:pcm_correlator}
\end{figure}

Finally, on Fig.~\ref{fig:pcm_correlator} we compare the $x$ dependence of the two-point correlator $\vev{\frac{1}{N} \tr\lr{g^{\dag}_x g_0}}_M$ at $\lambda = 3.1$ with Monte-Carlo data obtained in \cite{Buividovich:17:4} at $N = 6$ and $N = 9$. For DiagMC results the truncation order $M$ is encoded in color, from pure blue for $M = 1$ to pure red for $M = 12$. While at distances of a few lattice spacings the DiagMC data clearly approaches the Monte-Carlo data, at large distances the convergence of our massive perturbation theory becomes slower. Since the expression (\ref{gxgy_vev_momentum}) for $\vev{\frac{1}{N} \tr\lr{g^{\dag}_x g_0}}_M$ contains the constant $x$-independent contribution, it is not surprising that it decays down to some finite ``plateau'' value, rather than exponentially decaying to zero as the full correlators for $SU\lr{6}$ and $SU\lr{9}$ do. The height of this ``plateau'' is, however, monotonically decreasing with $M$. It seems that one has to go to much higher values of $M$ than we use to really see how the exponential decay of the $\vev{\frac{1}{N} \tr\lr{g^{\dag}_x g_0}}$ emerges.

It is also interesting to compare the correlators obtained from our infrared-finite weak-coupling expansion with the results of the standard lattice perturbation theory, for which the first two orders in coordinate space were given in \cite{Rossi:94:1}. These results are shown on Fig.~\ref{fig:pcm_correlator} as solid black (for the first order) and grey (for the second order) lines. As in the case of mean plaquette, we observe that at short distances the standard perturbative expansion is closer to the first-principle Monte-Carlo data than those from our infrared-finite weak-coupling expansion. The results of standard perturbative expansion show, however, a completely different behavior at large distances. At distance $x \sim 10$ they become negative, and further grow logarithmically with distance. In this sense in the far infrared they behave very differently from the exponentially decaying physical result. Furthermore, the second-order result of \cite{Rossi:94:1} starts growing logarithmically towards positive infinity at $x \gtrsim 3 \cdot 10^3$. While such large distances are certainly not interesting for numerical analysis, this discussion illustrates the infrared divergences of the standard lattice perturbation theory which become only stronger at higher orders.

\subsection{Finite-temperature phase transition}
\label{sec:pcm_numres:subsec:finitetemp}

\begin{figure}[h!tpb]
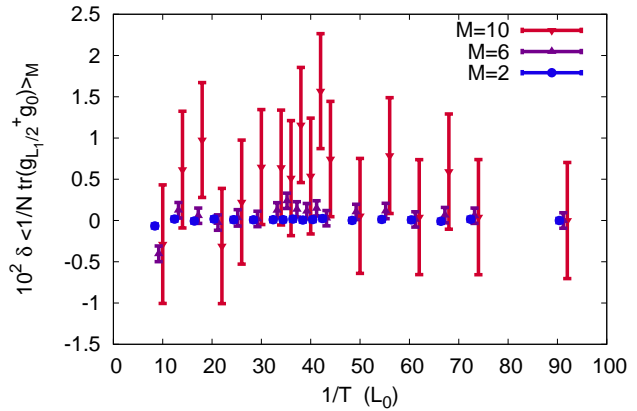

  \centering
  \includegraphics[width=\mfwa,angle=-90]{{{finite_t_scan}}}\\
  \caption{Temperature dependence of the correlator $\vev{\frac{1}{N} \tr\lr{g^{\dag}_x g_0}}_M$ at $x = \lr{0, L_1/2}$ (midpoint), $\lambda = 3.012$ and different truncation orders $M$ of the infrared-finite weak-coupling expansion. The zero-temperature values at $L_0 = 108$ and corresponding truncation order $M$ are subtracted.}
  \label{fig:finite_t_scan}
\end{figure}

As a further application of our DiagMC algorithm, in this Subsection we consider the principal chiral model (\ref{pcm_partition0}) at finite temperature. Not much is known about the finite-temperature physics of this model. As for other asymptotically free quantum field theories, on general grounds one can expect that at sufficiently high temperature the $SU\lr{N}$-singlet degrees of freedom are effectively ``deconfined''. However, for the principal chiral models with $N > 2$ this expectation has never been explicitly tested. One of the subtle points which make such tests difficult is the absence of a local order parameter for this ``deconfinement'' transition in the principal chiral model. From general physical intuition in such situations one expects either a first-order phase transition or a crossover. Understanding the ``deconfinement transition'' in the principal chiral model can be important for understanding the physics of one-dimensional compactifications of this model, which have been actively used in recent years for constructing resurgent trans-series \cite{Unsal:14:1}. An explicit analytic calculation \cite{Cubero:15:1} of the correlation functions in the large-$N$ $SU\lr{N} \times SU\lr{N}$ principal chiral model based on the Bethe ansatz techniques, unfortunately, does not allow so far to study the thermodynamics of this model.

In a Monte-Carlo study \cite{Buividovich:17:4} co-authored by one of us the numerical results for the energy density and the correlation length at different temperatures are presented for the principal chiral model (\ref{pcm_partition0}) at quite large but finite $N$, for the same t'Hooft coupling $\lambda = 3.012$ and the spatial lattice size $L_1 = 108$ as in the present study. The study \cite{Buividovich:17:4} indicates a weak enhancement of correlation length at some critical temperature, which very slowly tends to become sharper towards larger $N$. No jump or hysteresis of the free energy indicative of a first-order phase transition was observed.

In order to get further insights into the nature of this finite-temperature phase transition or crossover in the principal chiral model, we have performed DiagMC simulations at several different temperatures, which correspond to different lattice sizes $L_0 = 4 \ldots 108$ in the Euclidean time direction at fixed t'Hooft coupling $\lambda = 3.012$ (and hence fixed lattice spacing, which only depends on $\lambda$). As a physical observable which is most sensitive to long-range correlations of $g_x$ fields we have chosen the value of the two-point correlator $\vev{\frac{1}{N} \tr\lr{g^{\dag}_x g_0}}_M$ at $x = \lr{0, L_1/2}$, which is the point most distant from the source at $\lr{0, 0}$ on a periodic lattice of spatial size $L_1$. This point is also the ``midpoint'' of the correlator where it takes the minimal value. To get a clearer signal, we subtract the zero-temperature data from the finite-temperature data. The result is shown on Fig.~\ref{fig:finite_t_scan} for the truncation orders $M = 2, \, 6, \, 10$. The spatial lattice size is again $L_1 = 108$.

As the truncation order $M$ is increased, the difference between the finite-temperature and zero-temperature correlator seems to grow. However, the statistical errors also grow with $M$ and for most values of $L_0$ the differences lie within statistical errors. However, for the smallest values $L_0 \simeq 4, \, 8$ the difference clearly decreases, indicating the decrease of correlation length at very high temperatures which is probably related to Debye screening. Furthermore, for $L_0$ roughly between $35$ and $45$ the difference seems to grow with $M$ beyond statistical uncertainty, resulting in a sort of peak structure around $L_0 \simeq 40$. This enhancement of correlations was also reported in \cite{Buividovich:17:4} at the same value of $L_0$. It can be interpreted is an indication of a weak finite-temperature phase transition. Since in DiagMC simulations the correlator $\vev{\frac{1}{N} \tr\lr{g^{\dag}_x g_0}}_M$ quickly grows with $M$, from our data we cannot exclude the scenario of finite-order phase transition at which the correlation length would diverge.

\section{Extension to gauge theories}
\label{sec:gaugeext}

In this Section we discuss possible extension of the approaches outlined in this work to non-Abelian gauge theories at finite density, which is the ultimate motivation of this work. The most straightforward extension would be to construct the infrared-finite weak-coupling expansion of Section~\ref{sec:lattpt} for pure $U\lr{N}$ lattice gauge theory with the partition function
\begin{eqnarray}
\label{lgt_partition0}
 \mathcal{Z} = \int\limits_{U\lr{N}} dg_{x,\mu}
 \nonumber \\
 \expa{
  -\frac{N}{4 \lambda} \sum\limits_{x,\mu,\nu}
   \re\tr\lr{g_{x,\mu} g_{x+\hat{\mu},\nu} g^{\dag}_{x+\hat{\nu},\mu} g^{\dag}_{x,\nu}}
  } .
\end{eqnarray}
Using the Cayley map parameterization (\ref{cayley_map}) and the $U\lr{N}$ integration measure (\ref{cayley_jacobian}) in terms of $N \times N$ Hermitian matrices $A_{x,\mu}$, one can calculate the few first orders of the expansion of the action in powers of the t'Hooft coupling $\lambda$:
\begin{eqnarray}
\label{lgt_partition_cayley}
 \mathcal{Z} = \int\limits dA_{x,\mu}
 \exp\left(
  -N \sum_{x,\mu}\, \alpha^{2} \Tr A_{x,\mu}^2
  - \right. \nonumber \\ \left.  -
  \frac{N \alpha^2}{\lambda} \sum_{x,\mu,\nu}\, \Tr\lr{F_{x,\mu\nu}^2}
  + O\lr{\nabla^2, \alpha^4}
  \right) ,
\end{eqnarray}
where $F_{x,\mu\nu} = \nabla_{\mu} A_{x,\nu} - \nabla_{\nu} A_{x, \mu} - 2 i \alpha \lrs{A_{x,\mu}, A_{x,\nu}}$ is the lattice field strength tensor constructed from the non-compact field $A_{x,\mu}$ and $\nabla_{\mu}$ is the forward lattice derivative. Again we have to choose $\alpha^2 \sim \lambda$ to arrive at the canonical normalization of the kinetic terms in the action. Higher-order terms in the action of the $A_{x,\mu}$ gauge fields are either suppressed by higher powers of $\alpha$ or contain higher lattice derivatives which are suppressed in the naive continuum limit. Again we see that the ``gluon'' field $A_{x,\mu}$ has acquired the mass term proportional to the t'Hooft coupling $\lambda$. The action (\ref{lgt_partition_cayley}) is of course still invariant under gauge transformations $g_{x,\mu} \rightarrow \Omega_x g_{x,\mu} \Omega^{\dag}_{x+\hat{\mu}}$, which, however, now have quite a non-trivial nonlinear form in terms of $A_{x,\mu}$ variables. In particular, they are different from the conventional continuum gauge transformations $A_{x,\mu} \rightarrow A_{x,\mu} + \nabla_{\mu} \phi_x - i \lrs{A_{x,\mu}, \phi_x}$. This is why the nonzero gluon mass in (\ref{lgt_partition_cayley}) does not contradict the gauge invariance of the theory. It is interesting here to draw parallels with the gluon mass which one can obtain from non-perturbative solution of (truncated) Schwinger-Dyson equations and Ward identities of continuum theory \cite{Cornwall:82:1}.

The inclusion of finite-density quarks would require sampling of both fermionic and bosonic correlators. Since fermionic propagators at finite density are in general complex-valued, this would require additional re-weighting and make the sign problem worse. For perturbative expansions with truncation error which scales exponentially $\epsilon_M \sim \alpha^M$, $|\alpha|<1$ with the truncation order $M$ this still allows to obtain the result in a computational time which scales as a polynomial of the required error, provided one can sample Feynman diagrams in a time which grows only exponentially with expansion order \cite{Rossi:1703.10141}. In the large-$N$ limit, the number of Feynman diagrams grows exponentially with their order, thus even the direct summation will satisfy this requirement. However, for our infrared-finite weak-coupling expansion (\ref{expansion_general}), the truncation error scales as $\epsilon_M \sim M^{-\gamma}$, $\gamma > 0$ (see (\ref{mlink_extrapolation_log}), (\ref{mlink_extrapolation_sqrt})), thus computational time will grow faster then polynomial in the required error, similarly to what happens for conventional Monte-Carlo simulations with sign problem. An important direction for further work is thus to find the re-summation strategy which would lead to exponentially fast convergence of the series, with bold diagrammatic expansions being one of the options.

\begin{figure}[h!tpb]
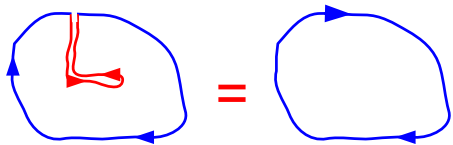

  \centering
  \includegraphics[width=\mfwa]{{{zigzag_symmetry}}}
  \caption{Zigzag symmetry of Wilson loops in gauge theory.}
  \label{fig:zigzag_symmetry}
\end{figure}

A less straightforward extension of this work to non-Abelian gauge theories might be based on the lattice strong-coupling expansion, combined with hopping expansion if dynamical quarks are included and the Veneziano large-$N$ limit is taken. Inside the convergence radius of the strong-coupling and hopping expansions, their truncation error should scale down exponentially with truncation order. At the same time, in the large-$N$ limit the number of strong-coupling expansion diagrams should grow not faster than exponentially with order \cite{Zuber:NPB1985}. Furthermore, for hopping expansion adding chemical potential does not introduce any additional sign cancellations between the diagrams, but merely modifies the hopping coefficients in the time direction \cite{deForcrand:11:1,deForcrand:14:1,Unger:14:1,Unger:17:1}. Thus even despite the sign cancellations which happen between strong-coupling expansion diagrams at higher orders, DiagMC algorithms based on strong-coupling expansion can be expected to solve, at least formally, the computational complexity problem (in the terminology of \cite{Rossi:1703.10141}) for finite-density gauge theories in the Veneziano large-$N$ limit.

The Metropolis algorithm~\ref{desc:transitions} allows to systematically sample the strong-coupling expansion for large-$N$ non-Abelian gauge theories. To this end it should be applied to solve the gauge-invariant Schwinger-Dyson equations for lattice Wilson loops $\vev{\frac{1}{N} \tr\lr{g_{x_1, \mu_1} g_{x_2, \mu_2} \ldots g_{x_n, \mu_n}  } }$ (Migdal-Makeenko loop equations \cite{Migdal:81:1,Eguchi:82:1}). These equations are very similar in structure to equations (\ref{matrix_sd_general}) in Subsection~\ref{sec:diagmc:subsec:sdeqs}.

In fact, we have already tried to solve the Migdal-Makeenko loop equations in large-$N$ pure gauge theory using the algorithm~\ref{desc:transitions}, sampling the loops on the lattice with probabilities proportional to Wilson loops $\vev{\frac{1}{N} \tr\lr{g_{x_1, \mu_1} g_{x_2, \mu_2} \ldots g_{x_n, \mu_n}  } }$. Unfortunately, this straightforward attempt did not allow to go beyond the first few orders of strong-coupling expansion, which could anyway be computed manually. The reason is the ``zigzag symmetry'' \cite{Polyakov:97:1,Drukker:99:1} between Wilson loops which differ by the insertion of arbitrary path traversed forward and immediately backward (see Fig.~\ref{fig:zigzag_symmetry}), which stems from unitarity of link matrices $g_{x,\mu} g_{x+\mu,-\mu} \equiv g_{x, \mu} g^{\dag}_{x,\mu} = 1$. Most of the computer time goes into sampling numerous redundant loops with multiple ``zigzag'' insertions, and nontrivial loops which contribute to higher orders of strong-coupling expansion are entropically disfavoured. Possible solution to this problem is to rewrite the Migdal-Makeenko loop equations in terms of ``irreducible'' loops without zigzag insertions, which makes the equation structure significantly more nontrivial. We hope that this ``gauge-fixing'' in loop space would allow to go to sufficiently high orders of strong-coupling expansion. Work in this direction is in progress.

Another challenge is to extend this approach to finite $N$, where the combinatorial structures involved in the strong-coupling expansion coefficients become more complicated \cite{Unger:14:1,Samuel:80:1}, and $1/N$ expansion might not work properly \cite{Samuel:80:1}. Importantly, the baryon bound states only exist at finite $N$. Nevertheless, even the finite-density gauge theory in the large-$N$ Veneziano limit might exhibit some universal features for densities below the baryon condensation threshold and is therefore physically interesting \cite{Kiritsis:14:1,Gursoy:1612.00899}. Needless to say, large-$N$ limit is also important in the context of AdS/CFT holographic duality between gauge theories and gravity.

\section{Conclusions}
\label{sec:conclusions}

In this work we have explored the advantages and disadvantages of the Diagrammatic Monte-Carlo (DiagMC) approach for studying asymptotically free non-Abelian lattice field theories in the weak-coupling regime. To this end we have combined two largely independent tools: the Metropolis algorithm for solving infinitely-dimensional Schwinger-Dyson equations, and the partial resummation of lattice perturbation theory which leads to massive bare propagators.

While our primary motivation are non-Abelian lattice gauge theories, eventually coupled to finite-density fermions, in this exploratory study we have considered an example of the large-$N$ $U\lr{N} \times U\lr{N}$ principal chiral model. This model has significantly simpler weak-coupling expansion, while having most of the non-perturbative features of the large-$N$ pure gauge theories on the lattice.

The construction of Metropolis algorithm~\ref{desc:transitions} from Schwinger-Dyson equations is very general, and can be applied to practically any quantum field theory. Its main conceptual advantages are:
\begin{itemize}
  \item The applicability of algorithm to \textbf{any expansion}, even \textbf{without explicit diagrammatic rules}. In particular, it can be used to sample both weak- and strong-coupling expansions in lattice gauge theory.
  \item The ability to work directly in the \textbf{infinite-volume~limit}. Of course, this does not imply the absence of slowing down towards the continuum limit, since the number of expansion coefficients which should be sampled to reach a given precision might grow with correlation length (see e.g. Subsection~\ref{sec:solvable_examples:subsec:2D}), and sign problem might make this growth faster (see Subsection~\ref{sec:pcm_numres:subsec:setup_misc}).
  \item The ability to work directly in the \textbf{large-$N$ limit}, where the number of diagrams contributing to either strong- or weak-coupling expansions becomes significantly smaller. Systematic calculation of the coefficients of $1/N$-expansion is also possible \cite{Buividovich:16:3}. Large-$N$ limit is not directly accessible in most other DiagMC simulations of non-Abelian gauge theories \cite{deForcrand:11:1,deForcrand:14:1,Vairinhos:14:1,Unger:14:1,Gattringer:1609.00124,Unger:17:1}.
\end{itemize}
Our algorithm is particularly suitable for exploring the critical behavior in large-$N$ quantum field theories with manifestly positive weights of Feynman diagrams, for example, the planar $\phi^4$ model with negative coupling constant or the Weingarten model of random planar surfaces on the lattice \cite{Weingarten:80:1}.

The resummation strategy used in this work also seems to be quite general for non-Abelian field theories. Its particular advantages are:
\begin{itemize}
  \item \textbf{The absence of infrared divergences} which are characteristic for the standard weak-coupling expansion of massless quantum field theories. While these divergences cancel in physical observables, they should be regulated at intermediate steps, which enhances the sign problem upon removing the infrared regulator due to cancellations between numerically large summands of opposite signs. In more conventional approaches to lattice perturbation theory such as Numerical Stochastic Perturbation Theory (NSPT) \cite{Pineda:14:1,Pineda:15:1} dealing with infrared divergences also requires additional infrared regulators.
  \item \textbf{The absence of factorial divergences} in the weak-coupling expansion in the large-$N$ limit, which arise at sufficiently high orders in conventional lattice perturbation theory. The general arguments of Section~\ref{sec:lattpt}, however, does not exclude the possibility of exponential divergences. Our arguments in favour of convergence of our infrared-finite weak-coupling expansion are only based on the numerical evidence - for up to $M = 500$ orders for the large-$N$ $O\lr{N}$ sigma model, and for up to $M = 12$ orders for the principal chiral model.
\end{itemize}
Of course, these advantageous features come at a certain price, and in some respects the resummed perturbative expansion behaves worse than the bare one:
\begin{itemize}
    \item Sampling our weak-coupling expansion using a DiagMC algorithm leads to the \textbf{sign~problem} due to in general non-positive weights of Feynman diagrams, even though the conventional Monte-Carlo simulations do not have any sign problem. Furthermore, for lattice gauge theory this sign problem is expected to become worse if finite-density fermions are added.
    \item \textbf{The convergence is slower} than for the conventional lattice perturbation theory, see Figs.~\ref{fig:pcm_link_convergence}~and~\ref{fig:pcm_correlator}.
    \item The expansion starts from the \textbf{wrong~vacuum~state} with explicitly broken $U\lr{N} \times U\lr{N}$ symmetry of the principal chiral model. While this symmetry seems to be gradually restored at high orders, at finite expansion order this explicit symmetry breaking results in constant contribution to the correlators of $g_x$ variables, which in reality are exponentially decaying.
\end{itemize}

Summarizing these advantages and disadvantages, it seems that the DiagMC approach based on the weak-coupling expansion is comparable in terms of computational time to conventional simulation methods as applied to non-Abelian lattice field theories in the large-$N$ limit. In particular, for Fig.~\ref{fig:pcm_link_convergence} the finite-$N$ data from conventional Monte-Carlo and the DiagMC data at infinite $N$ were produced within the comparable amount of CPU time. It is also conceivable, for example, that applying Numerical Stochastic Perturbation Theory (NSPT) \cite{Pineda:14:1,Pineda:15:1} to finite-$N$ principal chiral model and then extrapolating to large-volume and large-$N$ limits would lead to the comparable quality of numerical data within the comparable CPU time. It remains to be seen how efficient the sampling of strong-coupling expansion might be.

Probably the most promising direction for further improvement is the Bold Diagrammatic Monte-Carlo (BDMC) approach, which can efficiently capture non-perturbative information by combining perturbative expansion with a self-consistent mean-field-like approximation \cite{Pollet:1012.5808,Prokofev:1110.3747,Davody:1307.7699,Pollet:1701.02680}. Unfortunately, currently most of BDMC algorithms for bosonic fields are based on certain truncations of Schwinger-Dyson equations even for the simplest case of scalar field theory with $\phi^4$ interactions \cite{Davody:1307.7699,Pollet:1701.02680}.

One possibility to construct the bold diagrammatic expansion would be to work with Schwinger-Dyson equations for $n$-particle-irreducible field correlators \cite{Berges:hep-ph/0401172}. Another possibility, more specific for non-Abelian field theories, is to introduce the matrix-valued Lagrange multiplier enforcing the unitarity constraint $g^{\dag}_x g_x = 1$, and integrate out the $g_x$ fields, which would automatically ensure the $U\lr{N} \times U\lr{N}$-invariant vacuum state with $\vev{\frac{1}{N} \tr g_x} = 0$. For the large-$N$ $O\lr{N}$ sigma model, a similar treatment immediately yields the exact non-perturbative solution, with the mass being equal to the saddle-point value of the Lagrange multiplier field. In fact, this mean-field solution resums our double series (\ref{on_series_2D}) with coupling and logs of coupling into a single exponential $\expa{-\frac{4 \pi}{\lambda}}$, which is the only term in the true trans-series for the mass gap in the large-$N$ $O\lr{N}$ sigma model. One could thus expect that if such approach works also for the principal chiral model, it might re-sum many terms in our expansion into exponentials, thus turning it into the true trans-series of the form (\ref{transseries_example}) with optimal convergence properties and reducing the sign problem due to cancellations between Feynman diagrams. Implementing this approach in practice turns out to be non-trivial, and would be explored in future work.

\begin{acknowledgments} This work was supported by the S.~Kowalevskaja award from the
Alexander von Humboldt Foundation. The authors are grateful to the Mainz Institute for Theoretical Physics (MITP) for its hospitality and its partial support during the completion of this work. The calculations were performed on ``iDataCool'' at Regensburg University, on the ITEP cluster in Moscow and on the LRZ cluster in Garching. We acknowledge valuable discussions with S.~Valgushev, F.~Werner, B.~Svistunov, T.~Sulejmanpasic, L.~{von~Smekal}, N.~Prokof'ev, G.~Dunne, O.~Costin, A.~Cherman and G.~Bali.
\end{acknowledgments}

%

\appendix

\section{Vertex functions in the small momentum limit}
\label{apdx:vertex_functions_ir}

The vertex functions (\ref{vertex_func_def}) can be drastically simplified if all the momenta in their arguments are much smaller than the lattice cutoff scale. In this limit the lattice Laplacian $\D\lr{p}$ behaves as $\D\lr{p} \approx p^2$, and the vertex function $V(q_1,\cdots,q_{2n+1})$ can be represented as
\begin{widetext}
\begin{eqnarray}
\label{vertapdx_original}
 V(q_1,\cdots,q_{2n+1})
 = 
 \sum_{k=1}^{2n+1} (-1)^{k-1}
 \sum_{m=0}^{k-1}
 \D(q_{m+1}+\cdots+q_{m+2n+2-k})
 = \nonumber \\ =
 \sum_{k=1}^{2n+1} (-1)^{k-1} \sum_{m=0}^{k-1}
 \lr{q_{m+1}+\cdots+q_{m+2n+2-k}}^2
 = \nonumber \\ =
 \sum_{k=1}^{2n+1} (-1)^{k-1} \sum_{m=0}^{k-1}
 \sum_{i=1}^{2n+2-k} q_{m+i}^2
 + 
 2 \sum_{k=1}^{2n+1} (-1)^{k-1} \sum_{m=0}^{k-1} \sum_{i=1}^{2n+2-k}\sum_{j=1}^{i-1}q_{m+i}q_{m+j}
\end{eqnarray}
The first term in the last line of the above equation can be simplified as
\begin{eqnarray}
\label{vertapdx_first_term}
\sum_{k=1}^{2n+1} (-1)^{k-1} \sum_{m=0}^{k-1}
\sum_{i=1}^{2n+2-k} q_{m+i}^2
 = 
 \sum_{r=1}^{2n} q_{2r}^2\;\sum_{k=1}^{2n+1} (-1)^{k-1} \sum_{m=0}^{k-1} \sum_{l=0}^{2r-1}
\delta_{m,l}\;\theta(2n+2-k-2r+l+1)
 + \nonumber\\ + \sum_{r=0}^{2n} q_{2r+1}^2\;\sum_{k=1}^{2n+1} (-1)^{k-1} \sum_{m=0}^{k-1} \sum_{l=0}^{2r}
\delta_{m,l}\;\theta(2n+2-k-2r+l)
 = \nonumber\\ =
 \sum_{r=1}^{2n} q_{2r}^2\sum_{l=0}^{2r-1}\;\sum_{k=l+1}^{2n+2-2r+l} (-1)^{k-1}+
\sum_{r=0}^{2n} q_{2r+1}^2 \sum_{l=0}^{2r}\;\sum_{k=l+1}^{2n+1-2r+l} (-1)^{k-1}
 = 
 \sum_{r=0}^{2n} q_{2r+1}^2 ,
\end{eqnarray}
\end{widetext}
where $\theta$ is the unit step function. Similar algebra gives for the second term in the last line of (\ref{vertapdx_original}):
\begin{eqnarray}
\label{vertapdx_second_term}
 2 \sum_{k=1}^{2n+1} (-1)^{k-1} \sum_{m=0}^{k-1}
\sum_{i=1}^{2n+2-k}\sum_{j=1}^{i-1}q_{m+i}q_{m+j}
 = \nonumber \\ =
 2 \sum_{r=0}^{2n}\sum_{s=0}^{r-1} q_{2r+1} q_{2s+1}
\end{eqnarray}
Combining (\ref{vertapdx_first_term}) and (\ref{vertapdx_second_term}), we get
\begin{eqnarray}
\label{vertapdx_final}
 V(q_1, \ldots, q_{2n+1}) = \lr{q_1 + q_3+\cdots+q_{2n+1}}^2 ,
\end{eqnarray}
which is the formula (\ref{vertex_func_IR}) in the main text.

\section{Integration measures on \texorpdfstring{$S_N$}{Sn} and \texorpdfstring{$U\lr{N}$}{U(N)} manifolds}
\label{apdx:integration_measures}

\subsection{Cayley map for \texorpdfstring{$U\lr{N}$}{U(N)} group}
\label{apdx:integration_measures:subsec:cayley}

With the Cayley mapping $g = \frac{1 + i \phi}{1 - i \phi} = \frac{2}{1 - i \phi} - 1$, the group-invariant distance element $ds^2 = \tr\lr{d g \, d g^{\dag}}$ on $U\lr{N}$ can be expressed as
\begin{eqnarray}
\label{UN_cayley_ds}
 ds^2
 = \nonumber \\ =
 \tr\lr{
 \frac{- 2 i }{1 - i \phi} d\phi \frac{1}{1 - i \phi}
 \frac{  2 i }{1 + i \phi} d\phi \frac{1}{1 + i \phi}
 }
 = \nonumber \\ =
 4 \tr\lr{\frac{1}{1 + \phi^2} d\phi \frac{1}{1 + \phi^2} d\phi}
   ,
\end{eqnarray}
The invariance of $ds^2$ under shifts $g \rightarrow u g$, $u \in U\lr{N}$ implies, in particular, the invariance of the measure under similarity transformations $g \rightarrow u g u^{\dag}$ which allow to diagonalize $g$ and hence the matrix $\phi$ in (\ref{UN_cayley_ds}). We thus assume that the matrix $\phi_{ij}$ has diagonal form $\phi_{ij} = \phi_i \delta_{ij}$. The distance element $ds^2$ can be then written as a convolution of the coordinate differentials $d\phi_{ij}$ with the diagonal metrics:
\begin{eqnarray}
\label{UN_cayley_ds_metrics}
 ds^2 = g_{ij \, kl} d\phi_{ij} d\phi_{kl}
 = \nonumber \\ =
 \sum\limits_i \frac{d\phi_{ii}^2}{1 + \phi_i^2}
 +
 2 \sum\limits_{j > i} \frac{\lr{d \Re \phi_{ij}}^2 + \lr{d \Im \phi_{ij}}^2}{\lr{1 + \phi_i^2} \lr{1 + \phi_j^2}}  ,
\end{eqnarray}
where we have taken into account that the off-diagonal components $\phi_{ij}$ are not all independent, but rather satisfy $\phi_{ij} = \bar{\phi}_{ji}$. We can now express the integration measure, up to an irrelevant normalization factor, as $\int dg = \int d\phi \sqrt{g}$, where $g$ is the determinant of the metric tensor which is equal to the product of all the diagonal metric elements appearing in (\ref{UN_cayley_ds_metrics}):
\begin{eqnarray}
\label{UN_cayley_ds_measure1}
 g = \lr{\prod\limits_{j > i} f_i^2 f_j^2} \prod\limits_{i} f_i = \prod\limits_{i,j} f_i f_j
 = \nonumber \\ =
 \lr{\prod\limits_i f_i}^{2 N} = \det{1 + \phi^2}^{-2 N} ,
\end{eqnarray}
where we have denoted $f_i = \lr{1 + \phi_i^2}^{-1}$. Combining everything together, we arrive at
\begin{eqnarray}
\label{UN_cayley_ds_measure}
 \int\limits_{U\lr{N}} dg
 =
 \int\limits_{\mathbb{H}_{N \times N}} d\phi \, \det{1 + \phi^2}^{- N} ,
\end{eqnarray}
where on the right hand side the integration goes over all $N \times N$ Hermitian matrices. This expression is identical to (\ref{cayley_jacobian}) in the main text, up to the trivial rescaling $\phi \rightarrow \alpha \phi$.

\subsection{Exponential map for \texorpdfstring{$U\lr{N}$}{U(N)} group}
\label{apdx:integration_measures:subsec:expmap}

Proceeding in the same way as for the Cayley map, for the exponential map $g = e^{i \phi}$ we can relate the differentials of $\phi$ and $g$ as
\begin{eqnarray}
\label{expmap_differential}
 dg = \int_{0}^{1} dz \, e^{i z \phi} d\phi e^{i \lr{1 - z} \phi} .
\end{eqnarray}
The group-invariant distance element takes the form
\begin{eqnarray}
\label{exmap_ds}
 ds^2 = \tr\lr{d g \, d g^{\dag}}
 = \nonumber \\ =
 \int\limits_{0}^{1} \int\limits_{0}^{1} dz_1 dz_2 \,
 \tr\lr{e^{i \lr{z_2 - z_1} \phi} \, d\phi \, e^{-i \lr{z_2 - z_1} \phi} d\phi}  .
\end{eqnarray}
Again, we can now use the invariance of the measure under similarity transformations $g \rightarrow u g u^{\dag}$, $\phi \rightarrow u g u^{\dag}$, $u \in U\lr{N}$ to diagonalize $g$ and $\phi$, and express the distance element (\ref{exmap_ds}) as
\begin{eqnarray}
\label{exmap_ds_diagonal}
\sum\limits_{i,j}
 \int\limits_{0}^{1} \int\limits_{0}^{1} dz_1 dz_2 \,
 e^{i \lr{z_2 - z_1} \lr{\phi_i - \phi_j}} |d \phi_{i j}|^2
 =
 \sum\limits_{i} d\phi_{ii}^2
 + \nonumber \\ +
 2 \sum\limits_{i > j}
 \int\limits_{0}^{1} \int\limits_{0}^{1} dz_1 dz_2 \,
 \cos\lr{\lr{z_2 - z_1} \lr{\phi_i - \phi_j}} |d \phi_{i j}|^2
 = \nonumber \\ =
 \sum\limits_{i} d\phi_{ii}^2
 +
 8 \sum\limits_{i > j}
 \frac{\sin^2\lr{\lr{\phi_i - \phi_j}/2}}{\lr{\phi_i - \phi_j}^2} |d \phi_{i j}|^2 .  \hspace{1.5cm}
\end{eqnarray}
We see that the metric tensor is again diagonal with our choice of group coordinates, and we finally get for the integration measure (up to an overall normalization factor)
\begin{eqnarray}
\label{expmap_measure}
 \int\limits_{U\lr{N}} dg
 =
 \int\limits_{\mathbb{M}} d\phi
 \prod\limits_{i > j}
 \frac{\sin^2\lr{\lr{\phi_i - \phi_j}/2}}{\lr{\phi_i - \phi_j}^2} ,
\end{eqnarray}
where $\mathbb{M}$ is some closed domain in the $N^2$-dimensional space $\mathbb{H}_{N \times N}$ of the Hermitian $N \times N$ matrices. The $U\lr{N}$ group manifold is uniquely covered by the integration in (\ref{expmap_measure}) if $\mathbb{M}$ is bounded by the $N^2-1$-dimensional manifolds at which at least two eigenvalues of $\phi$ coincide. E.g. in the case of $U\lr{2}$ group, $\mathbb{M}$ is the direct product of the $3$-dimensional ball and the one-dimensional circle $S_1$.

In order to compare with the formula (\ref{cayley_jacobian}) in the main text, it is instructive to expand the Jacobian (\ref{expmap_measure}) in powers of $\phi$:
\begin{eqnarray}
\label{expmap_measure_expanded}
\int\limits_{\mathbb{M}} d\phi
 \expa{\sum\limits_{i, j}
 \log\lr{\frac{\sin^2\lr{\lr{\phi_i - \phi_j}/2}}{\lr{\phi_i - \phi_j}^2}}
 }
 = \nonumber \\ =
 \int\limits_{\mathbb{M}} d\phi \expa{- \sum\limits_{i, j} \frac{\lr{\phi_i - \phi_j}^2}{12} + O\lr{\lambda^4}  }
 = \nonumber \\ =
\int\limits_{\mathbb{M}} d\phi \expa{- \frac{N \tr\phi^2 - \tr\phi \, \tr\phi}{6} + O\lr{\phi^4}  } ,
\end{eqnarray}
which illustrates the appearance of double-trace terms already in the lowest order of the expansion.

\subsection{Stereographic map for the sphere \texorpdfstring{$S_N$}{Sn}}
\label{apdx:integration_measures:subsec:stereo}

In order to find the integration measure on the $N$-sphere $S_N$ parameterized by the coordinates $\phi_i$ as $n_0 = \lr{1 - \phi^2}\lr{1 + \phi^2}^{-1}$, $n_i = 2 \phi_i \, \lr{1 + \phi^2}^{-1}$, $\phi^2 \equiv \sum_i \phi_i^2$, we again start by first expressing the $O\lr{N}$-invariant distance element $ds^2 = d n_a \, d n_a$ on $S_N$ in terms of the coordinates $\phi_i$:
\begin{eqnarray}
\label{sn_ds}
 ds^2 = d n_0^2 + d n_i^2
 =
 \frac{16 \lr{\phi_i d\phi_i}^2}{\lr{1 + \phi^2}^4}
 + \nonumber \\ +
 \lr{
 \frac{2 d\phi_i}{1 + \phi^2}
 -
 \frac{4 \phi_j d\phi_j \phi_i}{\lr{1 + \phi^2}^2}
 }^2
 =
 \frac{4 \, d\phi_i^2}{\lr{1 + \phi^2}^2} .
\end{eqnarray}
The metric tensor is again diagonal in the stereographic coordinates $\phi_i$ (which is not surprising, as stereographic mapping is conformal), and the integration measure (up to a normalization constant) can be written in terms of the square root of the metric determinant as
\begin{eqnarray}
\label{sn_measure}
 \int\limits_{S_N} dn
  =
 \int\limits_{\mathbb{R}^{N-1}} d^{N-1}\phi \, \lr{1 + \phi^2}^{-\lr{N-1}} ,
\end{eqnarray}
where the integration is over the whole $N-1$ dimensional real space $\mathbb{R}^{N-1}$. In the large-$N$ limit one can also replace the power of $-\lr{N-1}$ in (\ref{sn_measure}) by $-N$, which leads to the expression (\ref{stereographic_jacobian}) in the main text after the trivial re-scaling $\phi \rightarrow \sqrt{\lambda}/2 \, \phi$.

\end{document}